\documentclass[twocolumn,trackchanges]{aastex63}

 \hypersetup{linkcolor=blue,citecolor=blue,filecolor=cyan,urlcolor=blue}

\usepackage{array}
\usepackage{booktabs}
\usepackage{multirow}
\usepackage{makecell}
\usepackage{hyperref}
\usepackage[hyphenbreaks]{breakurl}
\usepackage{balance}
\usepackage{amsmath}
\usepackage{threeparttable}
\usepackage{lineno}

\shorttitle{Discoveries of two protoclusters at $z=2.24$}

\shortauthors{Shi et al.}

\begin{document}

\title{Spectroscopic Confirmation of two Extremely Massive Protoclusters BOSS1244 and BOSS1542 at $z=2.24$}

\correspondingauthor{Dongdong Shi}
\email{ddshi@pmo.ac.cn}

\author[0000-0002-4314-5686]{Dong~Dong Shi}
\affiliation{Purple Mountain Observatory, Chinese Academy of Sciences, 10 Yuan Hua Road, Nanjing, Jiangsu 210023, China}
\affiliation{School of Astronomy and Space Sciences, University of Science and Technology of China, Hefei 230026, China}
\affiliation{Steward Observatory, University of Arizona, 933 North Cherry Avenue, Tucson, AZ 85721, USA}


\author{Zheng Cai}
\affiliation{Department of Astronomy and Tsinghua Center for Astrophysics, Tsinghua University, Beijing 100084, China}

\author{Xiaohui Fan}
\affiliation{Steward Observatory, University of Arizona, 933 North Cherry Avenue, Tucson, AZ 85721, USA}

\author{Xian~Zhong Zheng}
\affiliation{Purple Mountain Observatory, Chinese Academy of Sciences, 10 Yuan Hua Road, Nanjing, Jiangsu 210023, China}
\affiliation{School of Astronomy and Space Sciences, University of Science and Technology of China, Hefei 230026, China}

\author{Yun-Hsin Huang}
\affiliation{Steward Observatory, University of Arizona, 933 North Cherry Avenue, Tucson, AZ 85721, USA}

\author{Jiachuan Xu}
\affiliation{Steward Observatory, University of Arizona, 933 North Cherry Avenue, Tucson, AZ 85721, USA}





\begin{abstract}

We present spectroscopic confirmation of two new massive galaxy protoclusters at $z=2.24\pm0.02$, BOSS1244 and BOSS1542, traced by groups of Coherently Strong Ly$\alpha$ Absorption (CoSLA) systems imprinted in the absorption spectra of a number of quasars from the Sloan Digital Sky Survey III (SDSS~III) and identified as overdensities of narrowband-selected H$\alpha$ emitters (HAEs).  Using MMT/MMIRS and LBT/LUCI near-infrared (NIR) spectroscopy, we confirm 46 and 36 HAEs in the BOSS1244 ($\sim55\,$arcmin$^{2}$) and BOSS1542 ($\sim61\,$arcmin$^{2}$) fields, respectively. BOSS1244 displays a South-West (SW) component at $z=2.230\pm0.002$  and another North-East (NE) component at $z=2.246\pm0.001$  with the line-of-sight velocity dispersions of $405\pm202$\,km\,s$^{-1}$ and  $377\pm99$\,km\,s$^{-1}$, respectively. Interestingly,  we find that the SW region of BOSS1244 contains two substructures in redshift space, likely merging to form a larger system. In contrast, BOSS1542 exhibits an extended filamentary structure with a low velocity dispersion of $247\pm32$\,km\,s$^{-1}$ at $z=2.241\pm0.001$, providing a  direct confirmation of a large-scale cosmic web in the early Universe. The galaxy overdensities $\delta_{\rm g}$ on the scale of 15\,cMpc are $22.9\pm4.9$, $10.9\pm2.5$, and $20.5\pm3.9$ for the BOSS1244 SW, BOSS1244 NE, and BOSS1542 filament, respectively. They are the most overdense galaxy protoclusters ($\delta_{\rm g}>20$) discovered to date at $z>2$. These systems are expected to become virialized at $z\sim0$ with a total mass of $M_{\rm SW}=(1.59\pm0.20)\times10^{15}$\,$M_{\sun}$, $M_{\rm NE} =(0.83\pm0.11)\times10^{15}$\,$M_{\sun}$ and $M_{\rm filament}=(1.42\pm0.18)\times10^{15}$\,$M_{\sun}$, respectively. 
Our results suggest that the dense substructures of BOSS1244 and BOSS1542 will eventually evolve into the Coma-type galaxy clusters or even larger. Together with BOSS1441 described in \cite{Cai2017a}, these extremely massive overdensities at $z=2-3$ exhibit different morphologies, indicating that they are in different  assembly  stages in the formation of early galaxy clusters.  Furthermore, there are two quasar pairs in BOSS1441, one quasar pair in BOSS1244 and BOSS1542, CoSLAs detected in these quasar pairs can be used to trace the extremely massive large-scale structures of the Universe.

\end{abstract}



\section{Introduction} \label{sec:intro}

In a cold dark matter universe dominated by a cosmological constant ($\Lambda$CDM), 
theories of structure formation predict that galaxy formation preferentially occurs 
along large-scale filamentary or sheet-like overdense structures in the early universe. 
The intersections of such filaments host ``protoclusters'' \citep{vanAlbada1961,Peebles1970,Sunyaev1972}, 
which evolve into viralized massive galaxy clusters at the present epoch  \citep[e.g.,][]{Bond1996,Cen2000,Muldrew2015,Overzier2016}. 
Protoclusters provide ideal laboratories to study galaxy properties in dense 
environments and the environment dependence of galaxy formation and evolution 
in the early universe. 
Galaxies in dense environments appear 
to be more massive, with lower specific Star Formation Rates (SFRs), and their growth in the early universe is accelerated in the sense that  protocluster galaxies formed 
most of their stars earlier than field galaxies \citep{Hatch2011}.

Present-day massive galaxy clusters are dominated by spheroidal galaxies with 
low star formation activities while those in the general fields are mostly still actively forming stars \citep{Skibba2009,Collins2009,Santos2014,Santos2015}.
These galaxy clusters are key to tracing
the formation of the most massive dark matter halos, 
galaxies and supermassive black holes (SMBHs) \citep{Springel2005}. 
Cluster galaxies show a tight ``red sequence'' and obey the 
``morphology-density'' relation, indicating the impact of dense environments 
on the star formation activities of the inhabitants \citep{Visvanathan1977,
Dressler1980,Bower1992,Goto2003}. To understand the physical processes 
that drive both the mass build-up in galaxies and the quenching of star formation, 
we need to investigate galaxies and their surrounding gas within and around the 
precursors of present-day massive galaxy clusters$-$protoclusters at $z>2$. The 
transition period before protocluster member galaxies began to quench and evolved to the massive 
clusters currently observed is a crucial phase to study their physical properties and the 
mechanisms driving their evolution \citep{Kartaltepe2019}.

In the last decades,  protoclusters at $z>2$ have been successfully 
discovered and several techniques for tracing the overdense regions have 
been developed. 
These include performing ``blind'' deep surveys (of H$\alpha$ emitters (HAEs), Ly$\alpha$ emitters (LAEs), Lyman 
break galaxies (LBGs) and photo-z selected galaxies) \citep{Steidel1998,Steidel2000,Steidel2003,Shimasaku2003,Steidel2005,
Ouchi2005,Toshikawa2012,Chiang2014,Fevre2015,Planck2015,Toshikawa2016,
Jiang2018,Lemaux2018,Shi2020} and targeting rare massive halo tracers, e.g., quasars, radio 
galaxies, Ly$\alpha$ blobs (LABs) and submilimeter galaxies \citep[SMGs;][]{Pascarelle1996,
Pentericci2000,Kurk2000,Kurk2004a,Venemans2002,
Venemans2004,Venemans2005,Venemans2007,Daddi2009,Kuiper2011,Hatch2011,
Hatch2014,Hayashi2012,Cooke2014,Cooke2016,Husband2016,Casasola2018}. 
The former is limited by their relatively small survey volumes, while the 
latter may suffer from the strong selection biases and small duty cycles \citep{Cai2016,Cai2017a}. 
A promising selection technique for protoclusters is through the group of gas absorption systems 
along multiple sightlights to background quasars or galaxies \citep{Lee2014,Stark2015,Cai2016,Miller2019}.

\cite{Cai2016} demonstrated that the extremely massive 
overdensities at $z=2-3$ traced by groups of Coherently Strong intergalactic Ly$\alpha$ 
absorption (CoSLA). This approach utilizes the largest library of quasar spectra, such as those from the SDSS-III Baryon Oscillations Spectroscopic Survey (BOSS),  to locate extremely rare, strong H\,{\small I} absorption from the IGM and select candidates for the most massive protoclusters \citep[e.g.,][]{Liang2021}. 
\cite{Cai2016}  used cosmological simulations to show that the correlation between IGM Ly$\alpha$ optical depth and matter overdensities peaks on the scale of $10-30$\,$h^{-1}$ comoving Mpc (cMpc), finding that the strongest IGM Ly$\alpha$ absorption systems trace 4$\sigma$ extreme tail of mass 
overdensities on 15\,$h^{-1}$\,cMpc  \citep{Lee2018, Mukae2020}. This technique is referred as MApping the Most Massive Overdensity Through Hydrogen (MAMMOTH). 
Using MAMMOTH technique, the massive BOSS1441 overdensity at $z=2.32\pm0.02$ is selected from the early data release of SDSS-III BOSS. The LAE overdensity in BOSS1441 is $10.8\pm2.6$ on a 15  cMpc scale, which could collapse to a massive cluster with $\geq10^{15}$\,$M_{\sun}$ at the present day \citep{Cai2017a}.  Furthermore, an ultra-luminous Enormous Ly$\alpha$ nebulae (also known as MAMMOTH-1) with a size of $\sim 442$\,kpc at $z=2.32$ is discovered at the density peak of BOSS1441 \citep{Cai2017b}, which is used to trace the densest and most active regions of galaxy and cluster formation \citep[e.g.,][]{Cantalupo2014, Arrigoni2015, Borisova2016, Cai2018, Cai2019}.

\begin{figure*}[ht]
\setlength{\abovecaptionskip}{0pt}
\begin{center}
\includegraphics[trim=2mm 0mm 2mm 4mm,clip,height=0.45\textwidth]{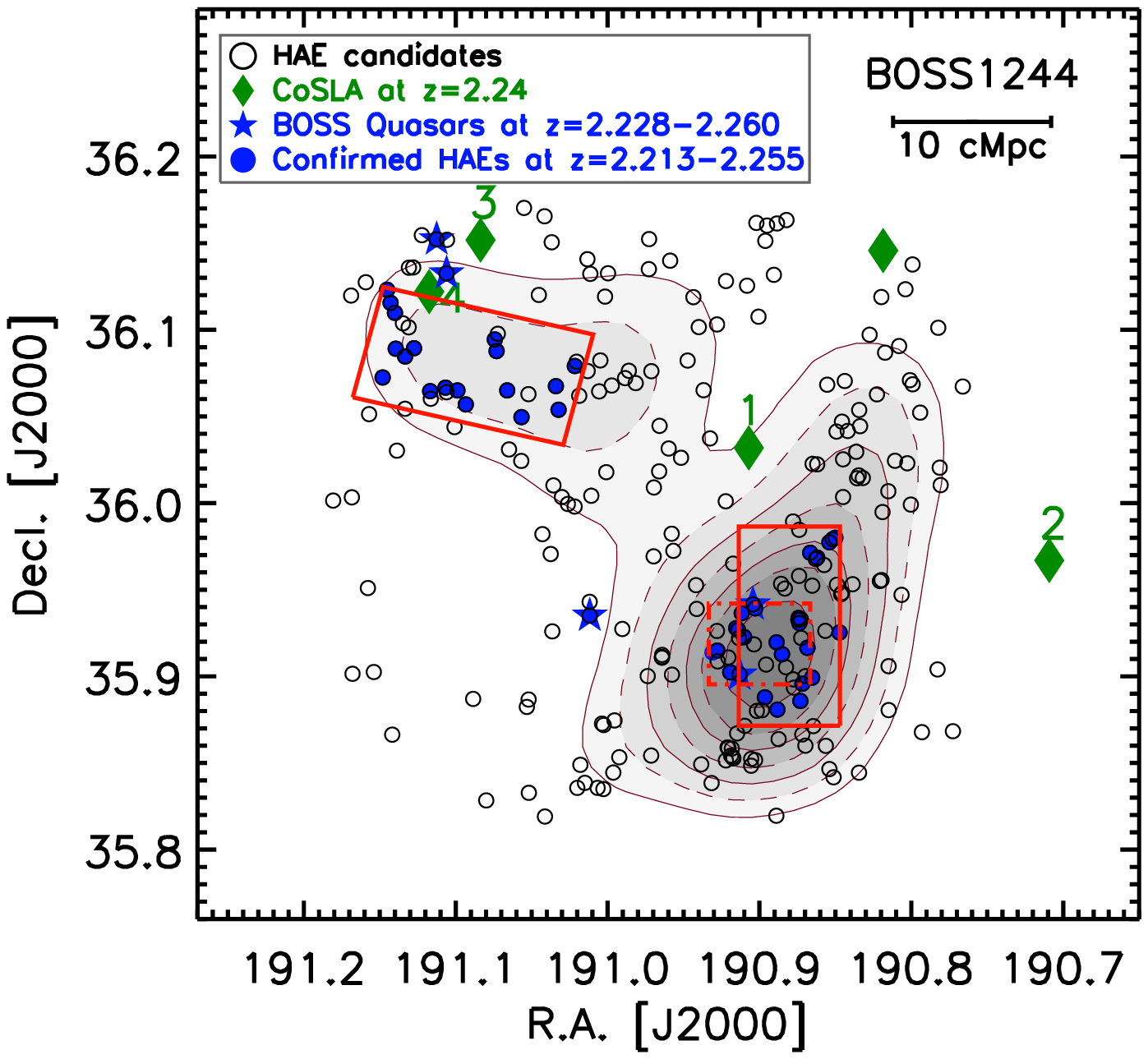}
\includegraphics[trim=2mm 0mm 2mm 4mm,clip,height=0.45\textwidth]{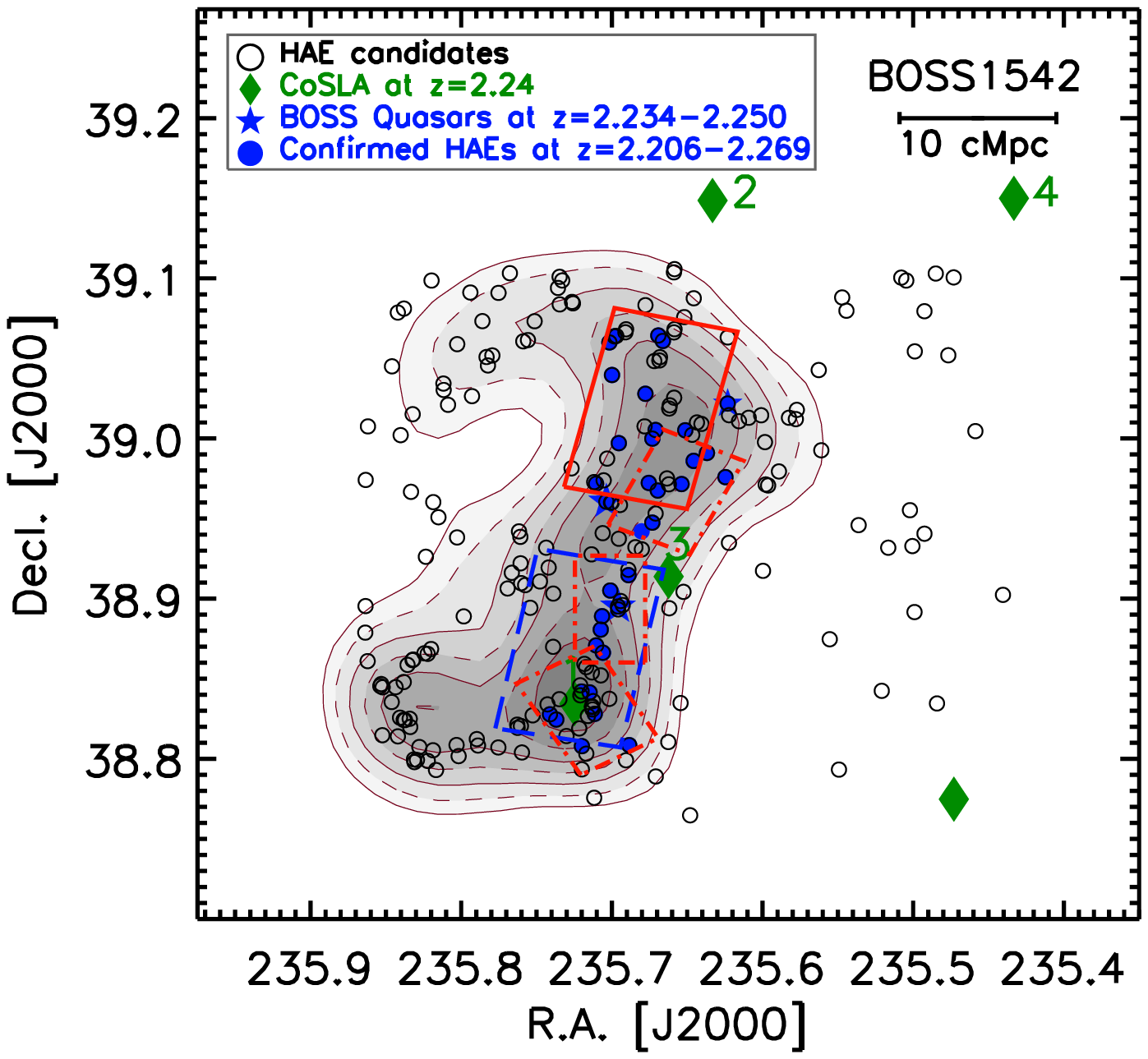}
\caption{Slitmask regions of MMT/MMIRS (red boxes: $4\arcmin\times6\farcm9$) and LBT/LUCI (red dot-dashed boxes: $4\farcm0\times2\farcm8$) in BOSS1244 (left) and BOSS1542 (right). BOSS1244-mask1 and BOSS1244-mask2 are located in the SW and NE regions, respectively. The dashed blue box in the right panel refers to BOSS1542-mask1, which was not observed due to the bad weather conditions.  These structures are traced by groups of Ly$\alpha$ absorption systems (green diamonds) and BOSS quasars (blue stars) at the redshift of $z=2.24\pm0.02$. The black circles are the selected HAE candidates and the filled circles show the spectroscopically confirmed HAEs through MMT/MMIRS and LBT/LUCI NIR spectroscopy. The solid and dashed lines represent contours of the density map of HAEs.  A galaxy number density of 0.2 per arcmin$^2$ is adopted as the contour interval and the inner density peak reaches $\sim2\,$\,arcmin$^{-2}$.} 
\label{fig:fig111}
\end{center}
\end{figure*}

Two more CoSLA candidates selected using the MAMMOTH technique, BOSS1244 and BOSS1542, have been confirmed to be extremely massive overdensities within 20\,$h^{-1}$\,cMpc at $z=2.24\pm0.02$ using the H$\alpha$ Emitter (HAE) candidates identified from near-infrared (NIR) narrow-band imaging \citep{Zheng2020}.  Currently, only a few protoclusters at $z\ge2$ are identified by NIR spectroscopy of HAEs. For example, the well-studied PKS~1138 at $z=2.16$ and USS~1558 at $z=2.53$ protoclusters associated with radio galaxy environments will evolve into the massive galaxy clusters with masses of $\sim10^{15}\,M_{\sun}$ at $z\sim0$ \citep{Kurk2004a,Hayashi2012,Shimakawa2014,Shimakawa2018a,Shimakawa2018b}.  \cite{Tanaka2011} reported a protocluster associated with the radio galaxy 4C~23.56 at $z=2.48$ had $\sim5-6$ times excess of HAEs from the general field, which might evolve into a galaxy cluster with the present-day mass of $>10^{14}\,M_{\sun}$. \cite{Darvish2020} confirmed a protocluster CC2.2 at $z=2.23$ with a present-day mass of $9.2\times10^{14}\,M_{\sun}$ through the NIR spectroscopy of HAEs.  Recently, \cite{Koyama2021} presented a $Planck$-selected protocluster at $z=2.16$ associated with an overdensity of HAEs, six HAEs at $z=2.150-2.164$ were identified through spectroscopy.  Although they do not calculate the present-day mass, we estimate that the fate of the protocluster may be a Virgo-type galaxy cluster ($\sim 10^{14}\,M_{\sun}$) at present day based on the volume they gave.

In this paper, we present NIR spectroscopic followup observations of these HAE candidates, quantitatively showing that BOSS1244 and BOSS1542 are indeed extremely overdense, and will collapse to two extremely massive clusters (like Coma cluster or even larger) at $z\sim0$.  We also use these identified HAEs to analyze dynamical properties (e.g., velocity dispersion, dynamical mass) and evolutionary stages of two overdensities.  Furthermore, we estimate the total mass
of the two overdensities to the present-day based on the galaxy overdensities.  The physical properties of HAEs in BOSS1244 and BOSS1542 will be presented in future work (Shi, D.D. et al. in preparation). The spectroscopic observations and data reduction are descried in Section~\ref{sec:observe}. In Section~\ref{sec:analyse}, 
we present the main analyses and results, and discuss the implications for the extremely overdense regions in 
Section~\ref{sec:discuss}.  Our  conclusions are summarized in Section~\ref{sec:conclusion}.
Throughout this paper, we adopt the cosmological parameters of $\Omega_{\rm M} = 0.3$, $\Omega_{\rm \Lambda} =0.7$ and $ H\rm_0 = 70$\,km\,s$^{-1}$\,Mpc$^{-1}$ and magnitudes are presented in the AB system unless otherwise specified. At $z=2.24$, 1$\arcmin$ corresponds to 0.495\,physical Mpc (pMpc) and 1.602\,comoving Mpc (cMpc), respectively.

\section{Observations and Data Reductions} \label{sec:observe}

\cite{Zheng2020}  carried out  deep NIR imaging observations with Canada$-$France$-$Hawaii Telescope (CFHT) through the narrowband (NB) $H_{\rm 2}S1$ and broadband (BB) $K_{\rm s}$ filters to identify emission-line objects. 
The NB technique selects objects with an excess of emission at $\lambda=2.13\,\mu$m. The emission-line may be [O\,{\small III}] from emitters at $z\simeq3.25$, [O\,{\small II}] at $z\simeq4.71$ or Pa$\alpha$/Pa$\beta$ at $z\simeq0.14$/0.66 and [S\,{\small III}] at 
$z\simeq1.23$/1.35 \citep[e.g.,][]{Geach2008, Sobral2012}. In total, 244/223 line emitters 
are selected with rest-frame $EW \textgreater 45 $\,\AA \,\,and $H_{2}S1 \textless 22.5$\,mag over the effective area of 417/399 arcmin$^{2}$ in BOSS1244/BOSS1542.  
As shown in \cite{Zheng2020}, about 80\% of these emitters are HAEs 
at $z=2.24\pm0.02$, and the H$\alpha$ luminosity functions (LF) can be 
derived in a statistical manner. Their results show that the shape of the 
H$\alpha$ LF of BOSS1244 agrees well with that of the general fields, while 
in BOSS1542, the LF exhibits a prominent excess at the high end, likely 
caused by the enhanced star formation or AGN activity.
We perform NIR spectroscopic observations to confirm the extremely overdense 
nature of two regions and better quantify their overdensities.    

\begin{deluxetable*}{ccccccccc}
\tablenum{1}
\tablecaption{MMT/MMIRS and LBT/LUCI NIR spectroscopic observations for the BOSS1244 and BOSS1542 fields.   \label{tab:tab1}} 
\tablewidth{0pt}
\tablehead{
\colhead{Field/SlitMask} & N$_{\rm obj}$ & \colhead{R.A.}  & \colhead{Decl.} & \colhead{P.A.} & \colhead{Grism+Filter} & \colhead{Exp.time} & \colhead{ObsDate} & \colhead{Seeing} \\
 \colhead{} & \colhead{} & \colhead{(J2000.0)} & \colhead{(J2000.0)}  & \colhead{(deg)} & \colhead{} & \colhead{(second)} & \colhead{}  & \colhead{Average ($\arcsec$)}
}
\decimalcolnumbers
\startdata
 BOSS1244/mask1 & 19/28 & $12^{h} 43^{m} 31.28^{s}$ & $+35\degr 55\arcmin 44\farcs38$ &  $180$ & $K3000+K\rm spec$ & 12,024 & 2017/06/14,18  &  1.16  \\
 BOSS1244/mask2 & 18/18 & $12^{h} 44^{m} 21.26^{s}$ & $+36\degr 04\arcmin 44\farcs62$ &  $-104$ & $K3000+K\rm spec$ & 12,024 & 2017/06/11,12,17 & 1.05  \\
 BOSS1542/mask1 & 0/26 & $15^{h} 42^{m} 53.20^{s}$ & $+38\degr 52\arcmin 06\farcs50$ &  $-11$ & $K3000+K\rm spec$ & \dots & \dots & \dots  \\
 BOSS1542/mask2 & 14/23 & $15^{h} 42^{m} 41.78^{s}$ & $+39\degr 01\arcmin 07\farcs62$ &  $-13$ & $K3000+K\rm spec$ & 7,236 & 2017/05/14 & 0.83  \\
 \hline
 \hline
 BOSS1244/mask  & 8/18 & $12^{h} 43^{m} 34.76^{s}$ & $+35\degr 55\arcmin 07\farcs33$ &  $90$ & $HK\rm spec$$+HK\rm spec$ & 5,256  & 2017/04/13  &  $\le1.2$  \\
 BOSS1542/mask1 & 8/14 & $15^{h} 42^{m} 51.80^{s}$  & $+38\degr 49\arcmin 49\farcs37$ &  $30$ & $HK\rm spec$$+HK\rm spec$ & 7,668 & 2017/04/12  &  $\le1.2$ \\
 BOSS1542/mask2 & 7/9 & $15^{h} 42^{m} 48.18^{s}$ & $+38\degr 53\arcmin 35\farcs39$ &  $0$ & $HK\rm spec$$+HK\rm spec$ & 6,480 & 2017/04/13 & $\le1.2$  \\
 BOSS1542/mask3 & 7/13 & $15^{h} 42^{m} 37.66^{s}$ & $+38\degr 57\arcmin 59\farcs07$ &  $-25$ & $HK\rm spec$$+HK\rm spec$ & 5,508 & 2017/04/13  & $\le1.2$ \\
\enddata
\tablecomments{ Above double lines are the MMT/MMIRS observations, below the double lines are the observations in LBT/LUCI. The number to the left of slash in column (2) is the number of successfully confirmed galaxies, and the number to the right of slash in column (2) is total number of HAE candidates in the slitmask. No observations in MMT/MMIRS BOSS1542/mask1 due to bad weather.}
\end{deluxetable*}

\subsection{MMT/MMIRS Spectroscopy} \label{sec:mmirs}

Spectroscopic observations of the HAE candidates in the density peak regions of BOSS1244 and BOSS1542 are carried out  using the Multiple Mirror Telescope (MMT) and Magellan Infrared Spectrograph \citep[MMIRS;][]{McLeod2012}, mounted on the MMT telescope (PI: Zheng, X.Z.) in 2017.   
MMIRS is a near infrared (NIR) imager with an imaging Field of View (FOV) $6\farcm9\times6\farcm9$ and multi-object spectrograph (MOS) over  $4\arcmin\times6\farcm9$.\footnote{\url{http://hopper.si.edu/wiki/mmti/MMTI/MMIRS/ObsManual}} We use the ``xfitmask" program\footnote{\url{http://hopper.si.edu/wiki/mmti/MMTI/MMIRS/ObsManual/MMIRS+Mask+Making}} to design our slit masks.  There are two masks designed in each field.  The red rectangles in Figure~\ref{fig:fig111} are the observed slit mask regions of two fields and the dashed blue rectangle in BOSS1542 is the unobserved slit mask region due to the bad weather.  A slit width of 1\farcs0  and a slit length of 7\farcs0 are adopted for observing our science targets, and the low noise gain (0.95) is used.  The targets are prioritized based on their  $K_{\rm s}$ magnitudes with 0\farcs7 aperture: we rank the highest, medium and lowest priorities to objects with $K_{\rm s}$ magnitudes in the range of 18.9$-$22.8, 22.8$-$24.3, 24.3$-$25.0 mag, respectively.  
The $K3000$ grism and $K\rm spec$ filter are used to take spectra over the wavelength coverage of 1.90$-$2.45\,$\micron$ \citep{Chilingarian2015}.  With these configurations, the resolution of $1\farcs0$ slit width in the $K3000$ grism and $K\rm spec$ filter corresponds to $R=1200$. We dither along the slit between individual 300\,s exposures. 
Three out of four masks were successfully observed under the average seeing conditions of  $0\farcs83-1\farcs16$ with the total integration time of 8.69 hrs in Semester 2017A. Four to five alignment stars with $13.5 < H < 16\,$mag (Vega) were chosen in our masks.  In addition, we observed A0V stars in each mask at a similar airmass in order to derive the spectral response function and remove atmospheric absorption pattern from spectra. Details of MMIRS slit masks are summarized in Table~\ref{tab:tab1}.  

\begin{figure}[ht]
\setlength{\abovecaptionskip}{0pt}
\begin{center}
\includegraphics[trim=16mm 2mm 0mm 11mm,clip,height=0.4\textwidth]{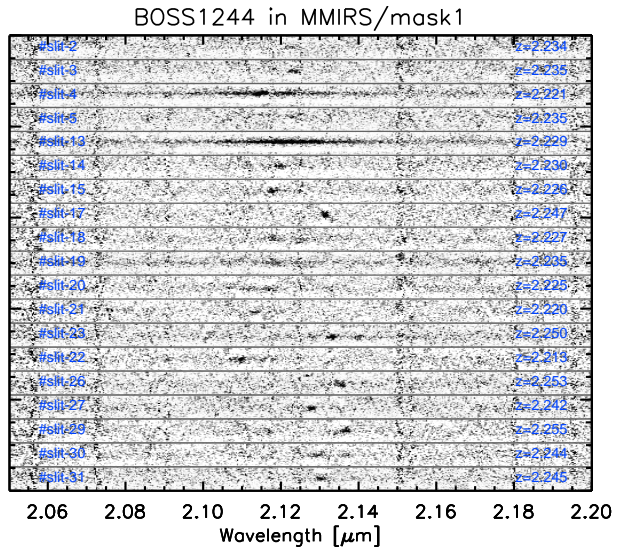}
\caption{2D spectra of 19 HAEs in BOSS1244 observed via MMT/MMIRS slit mask1. Skyline residuals are marked as vertical lines with a higher noise level. The slit and spectrum number are labeled. Almost all spectra (excluding slit-23 and slit-30) have only one emission line, and their continua are barely detected. The spectrum of slit-13 is consistent with a quasar with a broad H$\alpha$ emission line. } 
\label{fig:fig1}
\end{center}
\end{figure}

We use the standard MMIRS data reduction pipeline\footnote{\url{https://bitbucket.org/chil\_sai/mmirs-pipeline}} to process our MMIRS data \citep{Chilingarian2015}.  
Four point dither pattern ($ABA'B': +1\farcs8-1\farcs4,+1\farcs4-1\farcs8$) mode was adopted in our MOS mask observations.  
The major steps of data reduction include nonlinearity correction, dark subtraction, spectral tracing, flat-fielding, wavelength calibration, sky subtraction and telluric correction.  Two-dimensional (2D) spectra were extracted from the original frames without resampling after tracing and distortion mapping, and we set per-slit normalization for flat-fielding due to the imperfect illumination of the detector plane.  
We use airglow $OH$ lines for wavelength calibration given our faint targets and long exposure time ($300\,$s), 
but internal arc frames will be used if  the $OH$ based computation fails. 
 The sky subtraction is done using a technique modified from the original one given in \cite{Kelson2003}. 
For telluric correction, the pipeline computes the empirical atmosphere transmission function by the ratio of the observed telluric standard 
star spectrum and a synthetic stellar atmosphere of star. The empirical transmission function is corrected for the airmass difference between the observations of telluric standard and 
the science target. 
The 1D spectra are extracted from the reduced 2D spectra at every slit. 
The 2D/1D spectra with sky subtracted and telluric corrected 
are obtained. 

The NB excess fluxes are used to perform the absolute flux calibration 
for our spectral lines considering no slit stars are included in each slit mask. 
The object fluxes are calculated using the photometric magnitude at 
$H_{2}S1$- and $K_{\rm s}$-band with the equation (2) described in \cite{Zheng2020}.
We then convolve NB filter with the 1D spectra of Gaussian fitting to calculate total electrons, and correct the scaling factors ($\sim 2-5\times10^{-22}$) of 1D spectra at every target.  
We also use alignment stars (marked with ``BOX") to check the absolute flux calibration, and the scaling factors are $\sim5\times10^{-22}$ in the $H_{2}S1$ and $K_{\rm s}$ bands, which is $1-3$ times larger than the aforementioned method. This may be mainly due to the large slit width ($4\arcsec$) of alignment stars so that more light is collected. Therefore, we use NB excess flux to take the absolute flux calibration for our final calibrated spectra. 


\begin{figure}[ht]
\setlength{\abovecaptionskip}{0pt}
\begin{center}
\includegraphics[trim=16mm 2mm 0mm 13mm,clip,height=0.4\textwidth]{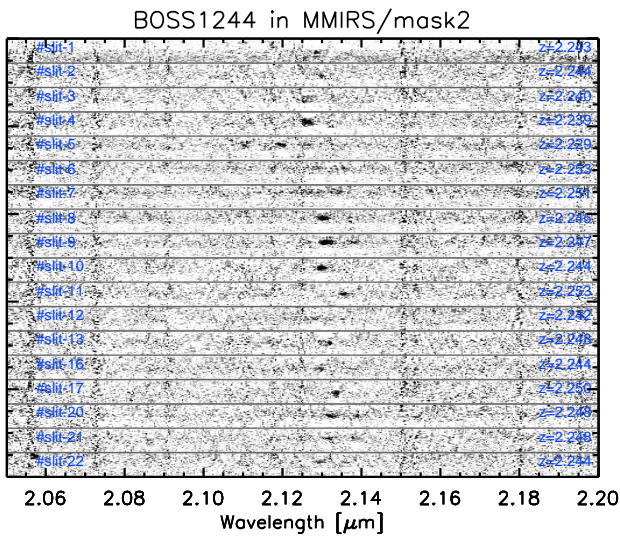}
\caption{2D spectra of 18 HAEs in BOSS1244 observed via MMT/MMIRS slit mask2. Skyline residuals are marked as vertical lines with a higher noise level. The slit and spectrum number are labeled. The spectra of slit 9, 20 and 21 show multiple lines. The rest exhibit only a single emission line.  Continuum is  barely visible in all spectra.} 
\label{fig:fig2}
\end{center}
\end{figure}

In total 46 HAE targets in BOSS1244 (two masks) and 23 HAE targets in BOSS1542 (one mask) are covered in the MMT/MMIRS observations.  We use a single Gaussian function to fit the emission line and measure the observed wavelength through H$\alpha$ line (the rest frame H$\alpha$ emission line is 6564.61\AA \footnote{\url{http://classic.sdss.org/dr7/algorithms/linestable.html}} in vacuum) to derive the redshift of HAEs.  If multiple components are detected, we will use multiple Gaussians in multiple regions to fit them. The equation $z=\lambda_{\rm obs}/\lambda_{\rm rest}-1$ is used to compute the redshift, where $\lambda_{\rm rest}$ is the rest-frame wavelength and $\lambda_{\rm obs}$ is the observe-frame wavelength.

\begin{figure}[ht]
\setlength{\abovecaptionskip}{0pt}
\begin{center}
\includegraphics[trim=16mm 2mm 0mm 28mm,clip,height=0.37\textwidth]{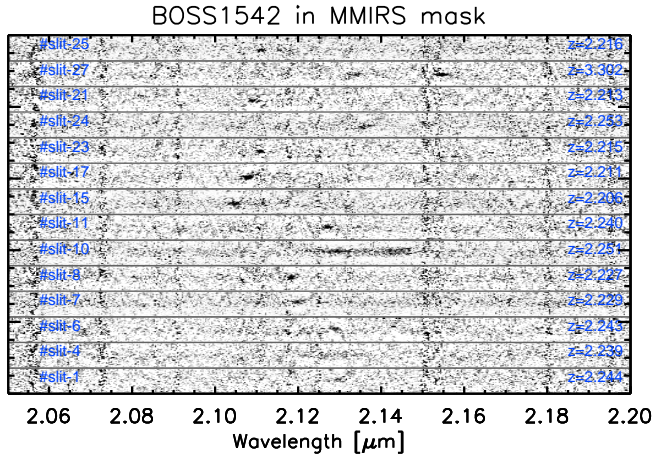}
\caption{2D spectra of 14 HAEs in BOSS1542 observed via MMT/MMIRS slit mask2. Skyline residuals are visible as vertical lines with a higher noise level. The slit and spectrum ID are indicated. We note that the spectrum of slit-27 from an [O\,{\small III}] emitter at $z=3.302$.} 
\label{fig:fig3}
\end{center}
\end{figure}

\subsection{LBT/LUCI Spectroscopy} \label{sec:luci}

To increase the sample size of spectroscopically-confirmed HAEs, three masks (36 targets) in BOSS1542 and one mask (18 targets) in BOSS1244 were observed using the LBT Utility Camera in the Infrared (LUCI) mounted on the Large Binocular Telescope (LBT) in semester 2017A (PI: Fan, X.). The relatively small dot-dashed red boxes in Figure~\ref{fig:fig111} are the observed slit mask regions of two fields.  LUCI\footnote{\url{https://www.lsw.uni-heidelberg.de/users/jheidt/LBT_links/LUCI_UserMan.pdf}} is able to provide imaging, longslit spectroscopy, and MOS spectroscopy over a FOV of four 
square arcminutes. We choose MOS spectroscopy and the slit masks are designed through the LMS\footnote{\url{https://sites.google.com/a/lbto.org/luci/preparing-to-observe/mask-preparation}} software.  Six alignment stars are used to correct telescope pointing and instrument rotation angle. The N1.8 camera with $4\farcm0\times2\farcm8$ FOV, $HK \rm spec$ grating with low resolution ($R=1900$) and $HK \rm spec$ filter are selected.  
Slits of 1\farcs0$\times$8\farcs0, 0\farcs5 $\times$0\farcs5 and 4\farcs0$\times$4\farcs0 are used for our targets, alignment stars, guide stars, respectively.  These observations were observed under the good seeing ($<1\farcs2$) conditions and each exposure takes 240 seconds. The total integration time in every mask is listed in Table~\ref{tab:tab1}.

Our LBT data were reduced using $Flame$ \citep{Belli2018}, a flexible data 
reduction pipeline written in Interactive Data Language (IDL) for NIR and optical spectroscopic data.     
We carried out data reduction following the reduction procedure given in the manual\footnote{\url{https://github.com/siriobelli/flame/blob/master/docs/flame_usermanual.pdf}} of $Flame$. We briefly describe the key steps below. Firstly, we set the inputs, initialize and create data structure. The reduction includes diagnostics of the observing conditions, calibrations on each of the science frames 
(including cosmic rays, bad pixels, dark frames and flat-fields), 
slit identification and cutout extraction, wavelength calibration, 
illumination correction, sky subtraction, and the extraction of 1D spectra 
from reduced and combined frames. More details about the pipeline can be found 
in the \cite{Belli2018}. We also used NB excess flux to derive the absolute 
flux calibration. The method of calculating the redshifts of HAE candidates 
is the same as MMT/MMIRS, which is described in Section~\ref{sec:mmirs}.

\section{Results and Analysis} \label{sec:analyse}

\subsection{Spectroscopic Confirmation of HAE Candidates} \label{sec:cmd}

We detect emission lines in 37 of 46 HAE candidates in BOSS1244 with spectra obtained 
from MMT/MMIRS. All of them are confirmed to be HAEs in redshift range of $2.213<z<2.255$.  
Note that most spectra show only one emission line and weak or no continuum. 
We further check these objects using the $BzK$ diagram from \cite{Daddi2004}, 
finding that they all fall into the region occupied by galaxies at $z>1.4$. 
We thus confirm that the detected lines are H$\alpha$ and these objects are HAEs.

Our MMT/MMIRS observations in  BOSS1244 gave an overall success rate of 80\% (37 in 46) in identifying HAEs. 
The detection rate is  $\sim$68\% (19 in 28) and 100\% (all 18) in mask1 and mask2 of BOSS1244, respectively. 
The difference in detection rate is largely caused by observational conditions: mask2 was taken under a better   condition than 
mask1 (see Table~\ref{tab:tab1}), although their integration times are the same and targets' line fluxes are similar.  
There may be some main reasons for the undetected targets in mask1:  
(a) some are too faint (NB$>$22.0\,mag) to be detected; 
(b) large seeing ($\sim$ 1\farcs2) conditions smear the signals of faint lines below the detection limit; 
(c) some bright targets may not be real emission line galaxies.

\begin{figure*}[ht]
\setlength{\abovecaptionskip}{0pt}
\begin{center}
\includegraphics[trim=6mm 55mm 0mm 10mm,clip,height=0.6\textwidth]{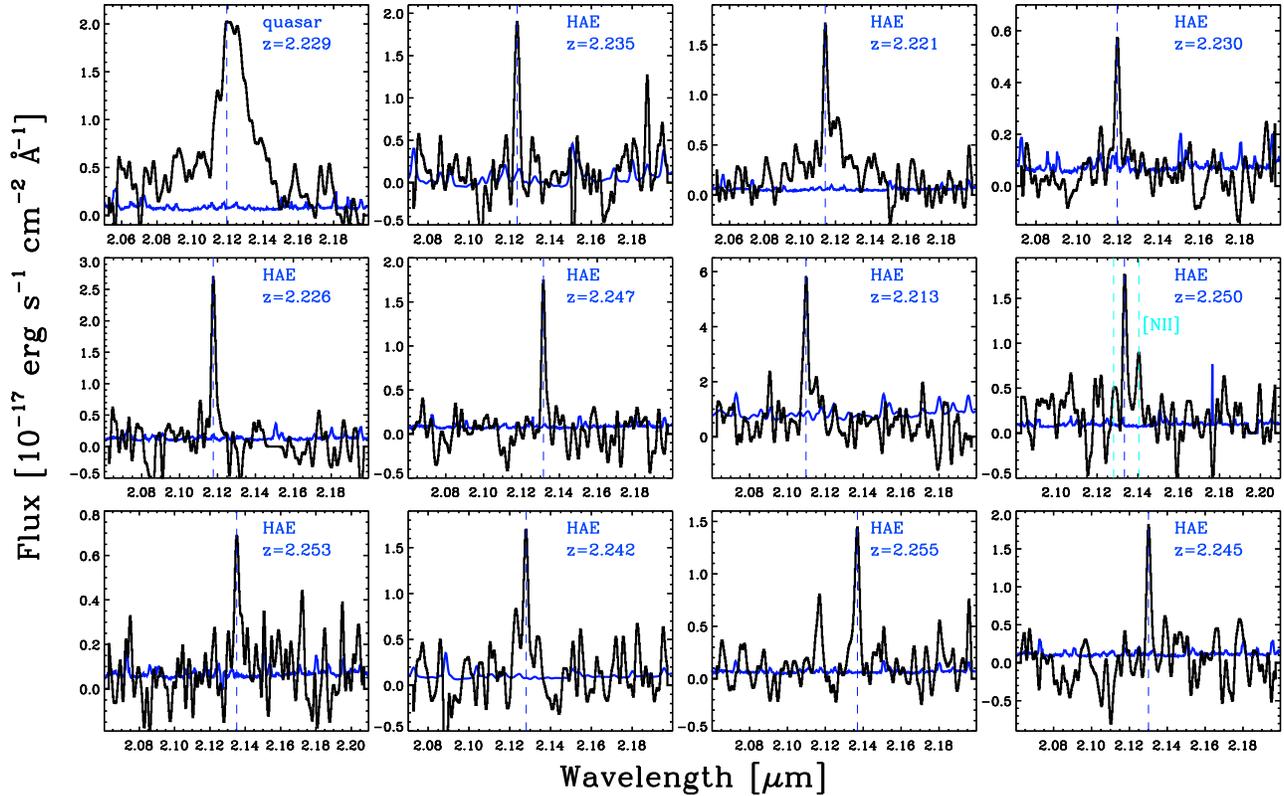}

\caption{MMT/MMIRS 1D spectra of 12 HAEs obtained from the BOSS1244 slitmasks, including one BOSS quasar.  The blue line in every plot is the sky line and the vertical blue dashed line is the H$\alpha$ emission. The vertical cyan dashed line is the [N\small II] emission.}
\label{fig:fig4}
\end{center}
\end{figure*}

\begin{figure*}[ht]
\setlength{\abovecaptionskip}{-90pt}
\begin{center}

\includegraphics[trim=6mm 55mm 0mm 10mm,clip,height=0.6\textwidth]{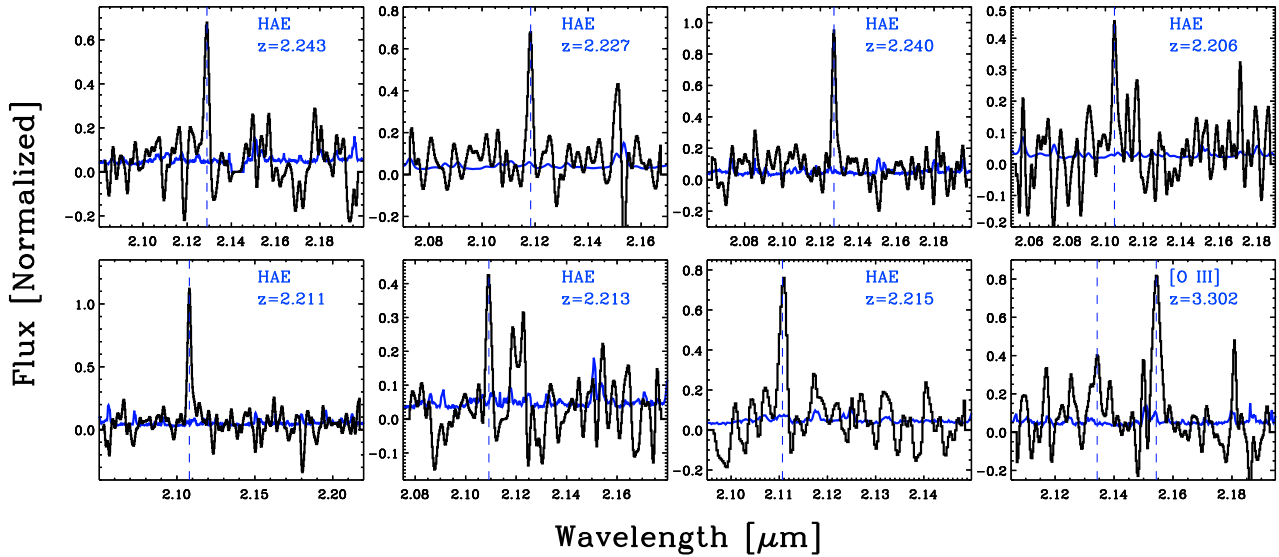}

\caption{MMT/MMIRS 1D spectra of eight HAEs in BOSS1542. The bottom-right panel shows the spectrum of an [O \small III] emitter at z=3.302. The blue curves represent the sky emission. The vertical dashed lines mark the emission lines H$\alpha$ or [O\,{\small III}]$\lambda\lambda4960,5008\rm\AA$ emission.} 
\label{fig:fig5}
\end{center}
\end{figure*}

\begin{figure*}[ht]
\setlength{\abovecaptionskip}{8pt}
\begin{center}
\includegraphics[trim=0mm 0mm 0mm 0mm,height=0.72\textwidth]{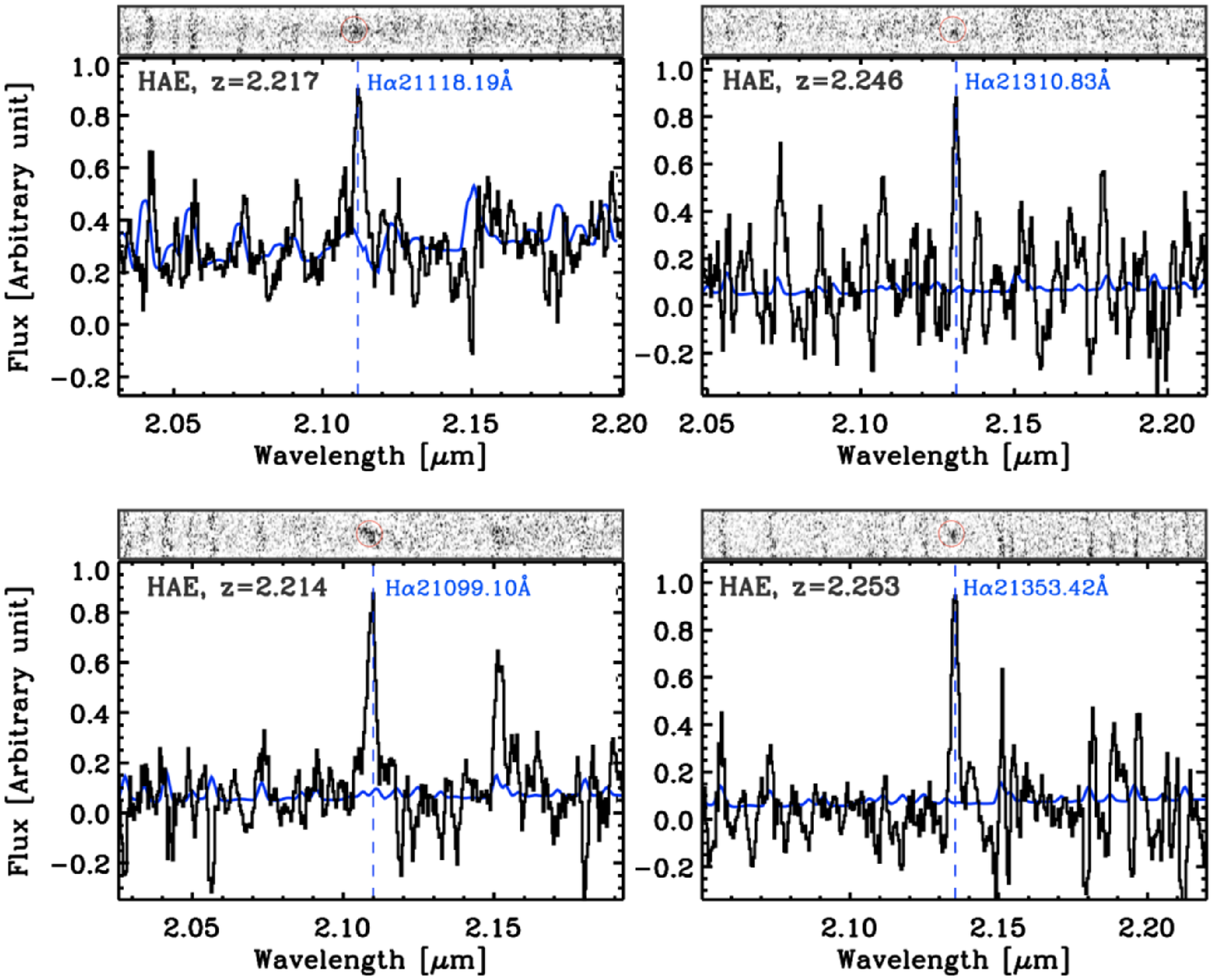}
\caption{LBT/LUCI 1D and 2D spectra of four HAEs in BOSS1244. The blue line in each panel is the sky line and the vertical blue dashed line is the H$\alpha$ emission.} 
\label{fig:fig6}
\end{center}
\end{figure*}

The reduced 2D spectra of HAEs in BOSS1244 are shown in Figure~\ref{fig:fig1} and Figure~\ref{fig:fig2}.  In slit mask1, slit-13 is the spectrum of a quasar with a broad emission line of full width at half maximum (FWHM)$=$4031\,km\,s$^{-1}$. This quasar is also included in SDSS Data Release 14 Quasar catalog (DR14Q).
From these spectra, only five targets (two in mask1 and three in mask2) show  H$\alpha$ resolved from [N\,{\small II}] line. 
We used multiple Gaussian functions to fit them simultaneously if more than two emission lines are resolved. Figure~\ref{fig:fig4} shows
the H$\alpha$ lines of some HAEs from our observations.

Moreover, we obtained one mask in BOSS1244 with LBT/LUCI. Eight out of 18 (44\%) HAE candidates have  H$\alpha$ emission line detected.  
Of them,  three HAEs are overlapped with MMT/MMIRS mask1, and have consistent redshifts. Figure~\ref{fig:fig6} presents the extracted 1D and 2D LUCI spectra for four objects.  
Altogether,  46 galaxies (including 41 HAEs and 5 quasars) at $2.213<z<2.255$ in 
BOSS1244 are identified from our NIR observations. These spectroscopically confirmed HAEs are listed in Table~\ref{tab:tabA1}.

For BOSS1542, we obtained one mask (23 targets in total) spectroscopic observation with MMT/MMIRS and three masks (36 targets in total) with LBT/LUCI. Using the method described before, 14 HAEs at $2.206<z<2.253$ are confirmed through 
MMT/MMIRS spectroscopic observations. The detection rate is 14/23 (61\%), lower 
than that of the previous  
two masks in BOSS1244. We note that the mask in BOSS1542 has a shorter 
integration time although it was observed under a better condition. 
In addition, we identify slit-27 as an [O\,{\small III}] emitter at $z=3.302$, because [O\,{\small III}]$\lambda\lambda4960,5008\rm\AA$\,\,lines are resolved.
Excluding the slit-27 target, the detection rate of HAEs is 13/23 (57\%). This low detection rate in BOSS1542 is mainly due to the shorter exposure time (shown in Table~\ref{tab:tab1}).  
Similarly, we identify 22 of 36 targets  at $2.215<z<2.269$ as HAEs using spectra obtained with LBT/LUCI, giving a success rate of 61\%.
Three HAEs observed with LBT/LUCI in BOSS1542 are quasars included in SDSS DR14Q. In total, 36 galaxies (including 33 HAEs and 3 quasars) at $2.206<z<2.269$ and one [O\,{\small III}] emitter at $z=3.302$  in BOSS1542 are confirmed by our NIR spectroscopic observations. These spectroscopically confirmed HAEs are listed in Table~\ref{tab:tabA2}.

 For HAEs, we calculate their redshifts based on the best-fit Gaussian profiles to the H$\alpha$ emission line and other emission lines (e.g., [O\,{\small III}] and [N\,{\small II}]). Figure~\ref{fig:fig8} shows the histogram of spectroscopic redshifts in BOSS1244 and in BOSS1542, respectively.  As discussed below, our spectroscopic observations confirm  BOSS1244 and BOSS1542 as extreme galaxy overdensities, indicating that HAEs are an effective tracer of  the overdense region of large-scale structures. We will present the physical properties of HAEs in the extremely overdense environments in a subsequent paper (Shi, D.D. et al. in preparation).

\begin{figure*}[ht]
\setlength{\abovecaptionskip}{8pt}
\begin{center}
\includegraphics[trim=0mm 0mm 0mm 0mm,height=1.1\textwidth]{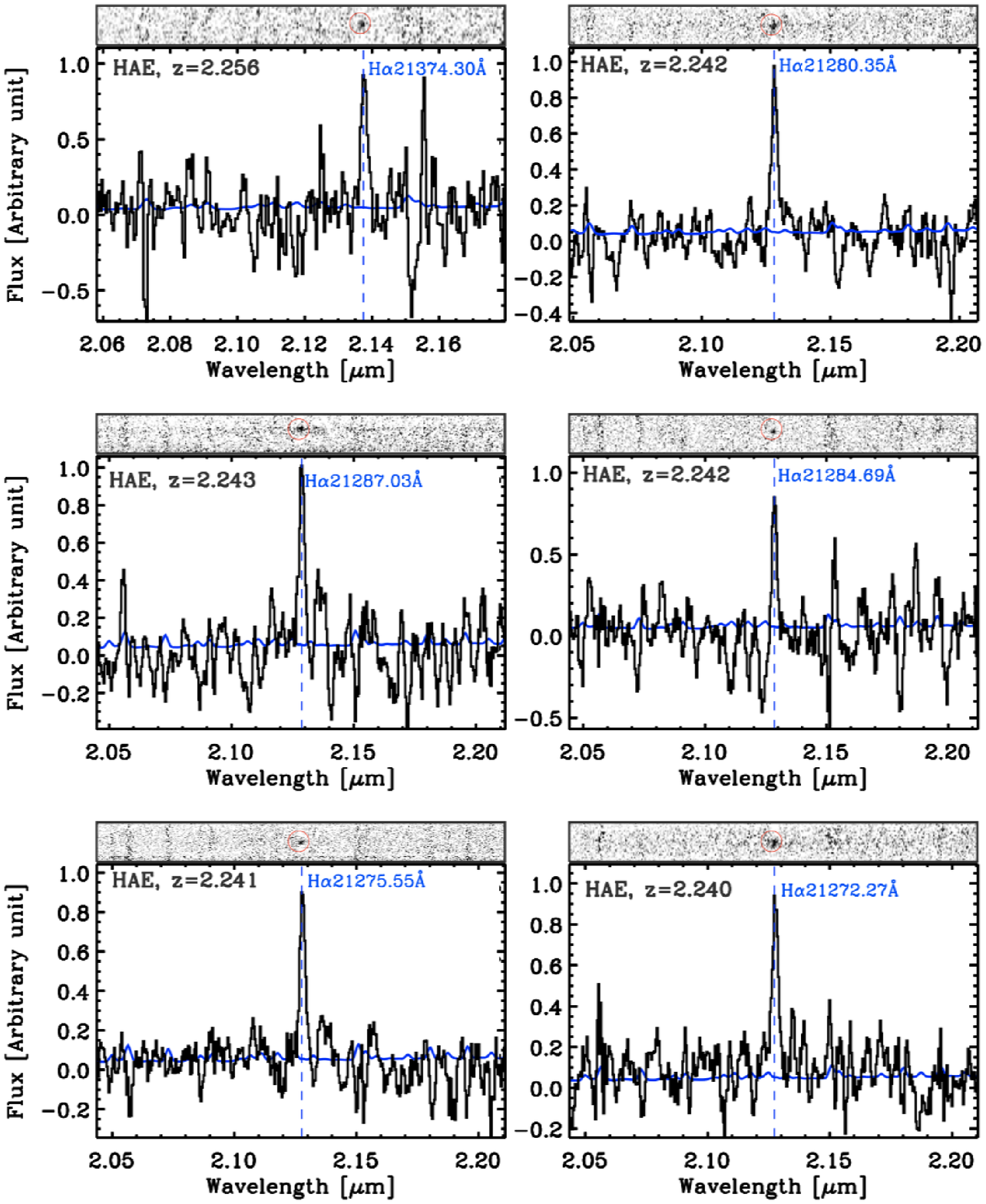}

\caption{ LBT/LUCI 1D and 2D spectra of six HAEs in BOSS1542. The blue line in each panel is the sky line and the vertical blue dashed line is the H$\alpha$ emission.} 
\label{fig:fig7}
\end{center}
\end{figure*}

\subsection{Redshift distributions of HAEs} \label{sec:reds}

\begin{figure*}[ht]
\setlength{\abovecaptionskip}{0pt}
\begin{center}
\includegraphics[trim=7mm 0mm 8mm 0mm,clip,height=0.38\textwidth]{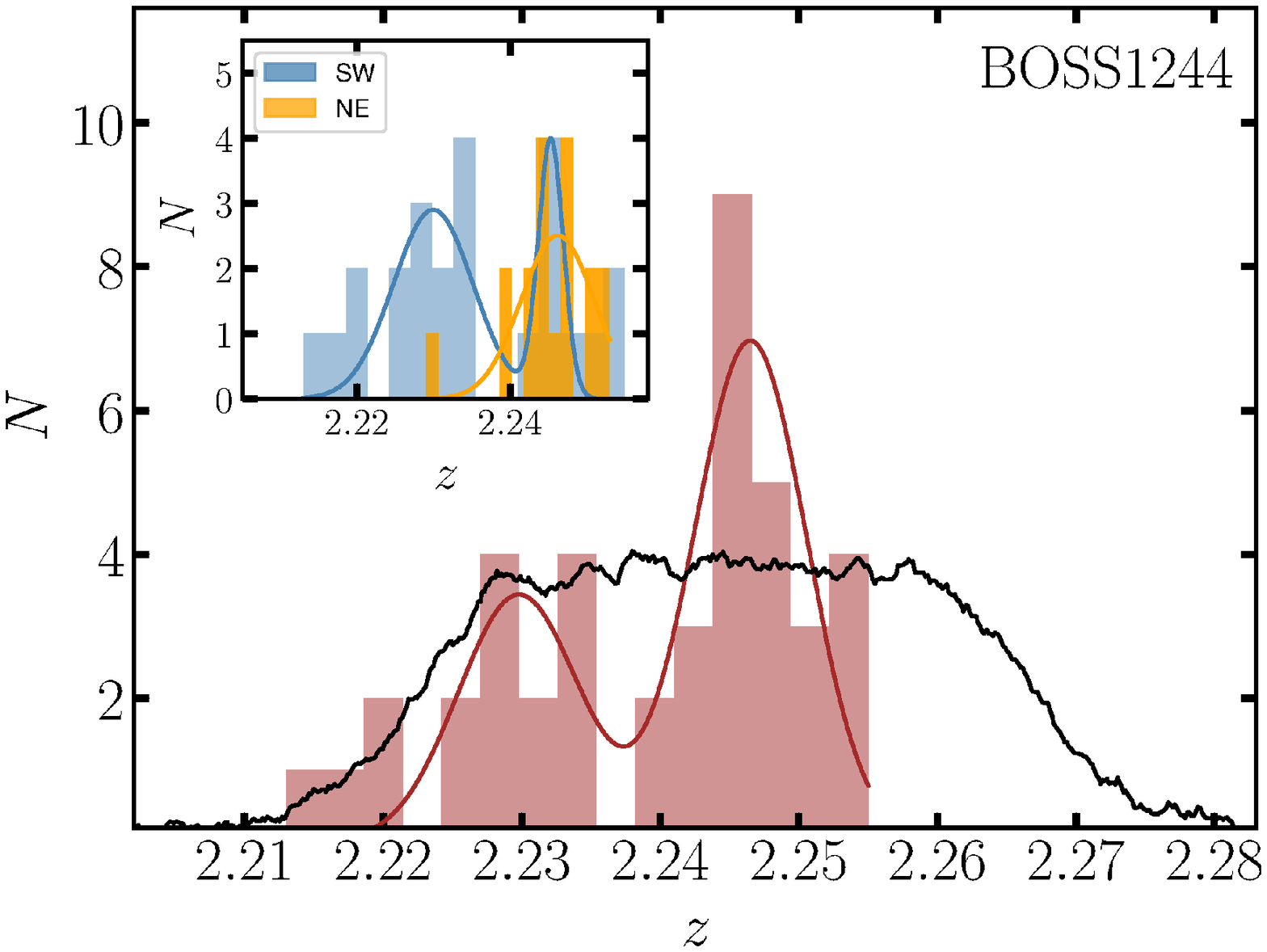}
\includegraphics[trim=7mm 0mm 8mm 0mm,clip,height=0.38\textwidth]{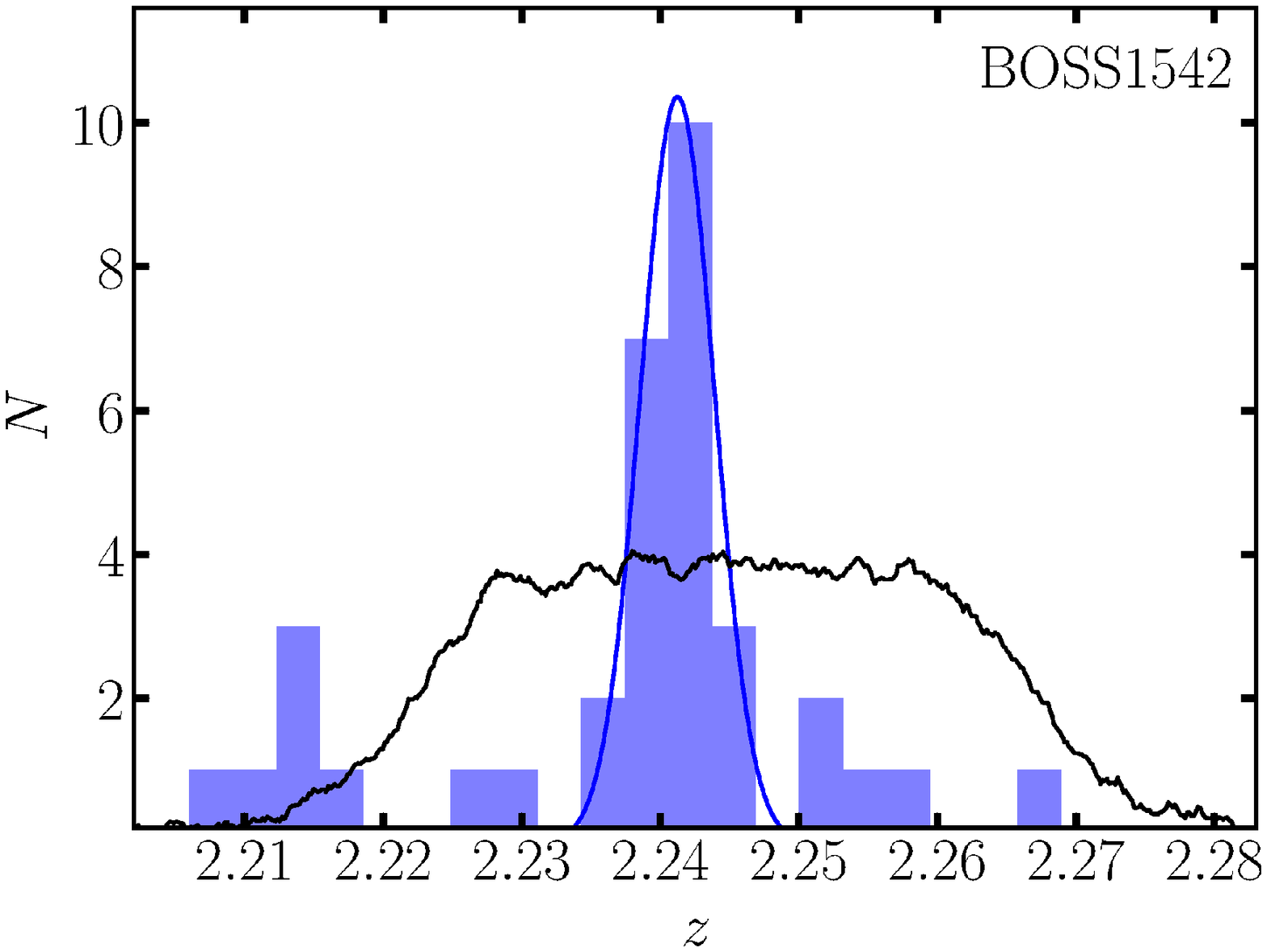}
\caption{The redshift distributions of the confirmed HAEs in BOSS1244 (left) and BOSS1542 (right).  The inner panel of left figure is the redshift distribution of HAEs in the BOSS1244 SW and NE regions.  The black solid line is the transmission curve of the NB filter. } 
\label{fig:fig8}
\end{center}
\end{figure*}

Figure~\ref{fig:fig8} shows the redshift distributions of HAEs of BOSS1244 and BOSS1542.  The two overdense systems exhibit different velocity (redshift) structures.  BOSS1244 presents two separated peaks in redshift distribution, indicating that there are two substructures in redshift space.  We use two Gaussian profiles to fit the redshift histogram, giving two redshift peaks at $z=2.230\pm 0.002$ and $z=2.246\pm 0.001$.  From the projected sky space shown in the left panel of Figure~\ref{fig:fig9}, there are two distinct components in sky space as well, i.e. South-West (SW) and  North-East (NE) regions. The SW region seems to be connected with the NE region (shown in the left panel of Figure~\ref{fig:fig111}), the projected separation between them is 13$\farcm$5 ($\sim 21.6\,$cMpc at $z=2.24$). The two components are covered by our MMT/MMIRS observations.  Further, we find that the SW region also shows the double peak in redshift distribution. One is at $z=2.230\pm 0.002$ and another is at $z=2.245\pm 0.001$, which is the consistent with  the redshift of NE region.  Moreover, from the projected sky space, the two substructures of SW region may be merging and forming a larger structure. Altogether, BOSS1244 has two distinct components both in sky and redshift space. In contrast, BOSS1542 shows a very extended filamentary structure over the scale of $\sim14\farcm6$, or 23.4\,cMpc from South to North region and may be forming a cosmic filament, and the redshift spike is $z=2.241\pm0.001$.

\subsection{Overdensity and Present-day Masses Estimate} \label{sec:overdensity}

Galaxy overdensity $\delta_{\rm g}$ is estimated from the galaxy surface density, where $\delta_{\rm g}$ is defined 
as $\delta_{\rm g}=\frac{\Sigma_{\rm group}}{\Sigma_{\rm field}}-1$, $\Sigma_{\rm group}$ is the HAE number per arcmin$^{2}$ 
within the overdensity, and $\Sigma_{\rm field}$ is the surface density of HAEs in the random fields. Our NIR spectroscopically-confirmed HAEs indicate that BOSS1244 and BOSS1542 are indeed extremely overdense.  The surface densities of HAE candidates at $z=2.24$ in the BOSS1244 and BOSS1542 fields are 0.585$\pm$0.037 and $0.559\pm0.037$\,HAE arcmin$^{-2}$, 
respectively. The surface density $\Sigma_{\rm field}$ of HAEs is calculated through some popular general fields with large HAE surveys. \cite{An2014} detected 28$\pm$5 HAE candidates at  $z=2.24$ with the same narrowband, detection depth and selection criteria over 383\,arcmin$^{2}$ area in ECDFS and the surface density $(7.31\pm1.38)\times10^{-2}$\,HAE arcmin$^{-2}$. \cite{Sobral2013} performed a large H$\alpha$ survey at $z=2.23, 1.47, 0.84 $ and 0.4 in the Cosmological Evolution Survey (COSMOS) and 
Ultra Deep Survey (UDS) fields. Using the same criteria of HAE candidates, the average surface densities in COSMOS and UDS are estimated to be $(8.25\pm0.36)\times10^{-2}$ and $(7.01\pm0.53)\times10^{-2}$\,HAE arcmin$^{-2}$, respectively.  
Note that there is an overdense region in COSMOS so that the surface density is higher than that in ECDFS and 
UDS, which is reported in \cite{Geach2012}. If the overdense region is masked, the surface density in COSMOS 
is $(7.42\pm0.35)\times10^{-2}$\,HAE arcmin$^{-2}$. In addition, 11 HAE candidates at $z=2.19\pm0.02$ over $\sim70\,$arcmin$^{2}$ are 
reported in the Great Observatories Origins Deep Survey North (GOODS-N) field \citep{Tadaki2011}, the surface 
density is $0.157\pm0.047$\,HAE arcmin$^{-2}$, which is about twice as high as other general fields (like ECDFS, COSMOS and UDS). This is mainly due to the limited NB survey area \citep{Tadaki2011}. For a comparison, we expect $31\pm6$ HAEs in our areas, based on integrating 
the H$\alpha$ luminosity function \citep{Sobral2013} and the completeness function of HAEs. The surface density 
in random fields is $(7.43\pm1.44)\times10^{-2}$\,HAE arcmin$^{-2}$. 
The method is described in \cite{Lee12014}.  Here we adopt as the average surface density ($7.25\pm0.51)\times10^{-2}$\,HAE arcmin$^{-2}$
of HAE candidates in the ECDFS, COSMOS and UDS fields for estimating HAE overdensities of the BOSS1244 and BOSS1542 fields.

 We do not map out all the HAE candidates in our NIR spectroscopic observations. However, considering the fore- and back-ground emitter contaminations, we assume that 80\% of our sample are true HAEs at $z=2.24\pm0.02$ based on the detect rate from our NIR spectroscopic analyses, and the galaxy overdensities $\delta_{\rm g}$ in BOSS1244 and BOSS1542 are computed to be 5.5$\pm$0.7 and 5.2$\pm$0.6, respectively.  The results are consistent with \cite{Zheng2020}.  These two overdensities are the most overdense fields currently known over the volumes of $\sim(40)^3$\,cMpc$^{3}$, and thus provide ideal laboratories to study galaxy properties in dense environments and the environmental dependence of galaxy mass assembly at the cosmic noon.

The characteristic size of a protocluster is $\sim15$\,cMpc (10.5\,$h^{-1}$\,cMpc), and the 
protocluster mass is typically calculated in a volume of (15\,cMpc)$^{3}$ \citep{Steidel1998,Steidel2005,Chiang2013,Stark2015,Muldrew2015}. 
To compare with our two overdensities (protoclusters),  we measure $\delta_{\rm g}$ in BOSS1244 and BOSS1542 over the 
typical protocluster scale (volume). Since the line-of-sight depth of 
our survey is 56.9\,cMpc, we measure the overdensity in a circular area 8.7\,cMpc in diameter, 
which corresponds to the volume of (15\,cMpc)$^{3}$.  In BOSS1244, two density peaks regions (NE and SW area) are separated in space and redshift obviously. 
The number of HAEs within a diameter of 8.7\,cMpc (5.4\,arcmin) centered on the position $[\alpha,\delta]=[190.892, 35.909]$
 in the SW dense region is $40\pm6$, corresponding to a surface density of 1.73$\pm$0.26, 
 and $\delta_{\rm g}=22.9\pm4.0$. The number of HAEs within a diameter of 8.7\,cMpc centered on the position $[\alpha,\delta]=[191.074, 36.079]$ in the NE dense region is $20\pm4$, corresponding to surface density of 0.87$\pm$0.17, and 
 $\delta_{\rm g}$ is 10.9$\pm$2.5, which is lower than that in the SW region.
 In BOSS1542, it shows a giant filament structure with $\sim14\farcm6$ ($\sim23.4$\,cMpc) along the 
 South to North region. We further calculate that the $\delta_{\rm g}$ through an elliptical area with 
 an axial ratio of $1/3$ centered on the position $[\alpha,\delta]=[235.682, 38.954]$. 
 The estimated $\delta_{\rm g}$ in the 
 (15\,cMpc)$^{3}$ volume is 20.5$\pm$3.9. 
 The uncertainties are estimated in the galaxy number counts within an overdense region by including 
 poisson shot noise and cosmic variance (clustering effect) \citep{Lee12014,Cai2017a}.

\begin{deluxetable*}{ccccc}
\tablenum{2}
\tablecaption{Galaxy overdensities and present-day masses in BOSS1244 and BOSS1542.   \label{tab:tab3}} 
\tablewidth{0pt}
\tablehead{
\colhead{Cluster} & \colhead{Redshift} & \colhead{Scale} & \colhead{$\delta_{\rm g}$}  & \colhead{Present-day Mass} \\
 \colhead{name} & \colhead{$<z>$} & \colhead{[cMpc]} & \colhead{}  & \colhead{[$10^{15}$\,$M_{\sun}$]}  
 }
\decimalcolnumbers
\startdata
    BOSS1244$^{\rm a}$ & 2.24 &  39.4 & 7.1$\pm$0.8 & 10.70$\pm$0.80 \\
    BOSS1244$^{\rm b}$ & 2.24 &  39.4 & 5.5$\pm$0.7 & 8.90$\pm$0.70 \\
    BOSS1244 NE & 2.246$\pm$0.001 & 15  & 10.9$\pm$2.5 & 0.83$\pm$0.11  \\
    BOSS1244 SW & 2.230$\pm$0.002 & 15 & 22.9$\pm$4.9 & 1.59$\pm$0.20  \\
    BOSS1542$^{\rm c}$ & 2.24 &  38.8 & 6.7$\pm$0.7 & 9.80$\pm$0.70 \\
    BOSS1542$^{\rm d}$ & 2.24 &  38.8 & 5.2$\pm$0.6 & 8.20$\pm$0.60 \\
    BOSS1542 Filament & 2.241$\pm$0.001 &  15 & 20.5$\pm$3.9 & 1.42$\pm$0.18 \\
\enddata
\footnotesize{$^{\rm a}$The emission line galaxy overdensity in BOSS1244. $^{\rm b}$The HAE overdensity in BOSS1244 assuming 80\% sample are true HAEs. $^{\rm c}$The emission line galaxy overdensity in BOSS1542. $^{\rm d}$The HAE overdensity in BOSS1542 assuming 80\% sample are true HAEs.}

\end{deluxetable*}

Galaxy formation models predict that galaxies inside large-scale overdensities should be older than those outside, 
because matter fluctuations inside overdensities are sitting on a large-scale pedestal and easier to collapse by 
crossing the threshold of $\delta_{\rm c}=1.69$. 
We estimate the total masses 
at $z\sim0$ in our two overdensities using the galaxy overdensity factor and the appropriate volume based on the  
approach outlined by \cite{Steidel1998,Steidel2005}, although there are some uncertainties such as systematic and random errors from the assumption of spherical collapse model \citep{Chiang2013, Overzier2016}.  The equation is given by:

 \begin{equation}
\centering 
\label{sec:equ2}
M_{z\rm =0}=\bar{\rho}V_{\rm true}(1+\delta_{\rm m})
\end{equation}
where $\bar{\rho}$ is the the mean comoving matter density of the universe, which is equal to $\frac{3 H_{\rm 0}^{2}}{8\pi G}\Omega _{\rm m}=4.1\times10^{10}$\,$M_{\sun}$\,cMpc$^{-3}$, and $V_{\rm true}$ is the volume in real space that encloses the observed galaxy overdensity after correcting the effects of redshift-space distortions. So Equation~\ref{sec:equ2} is equivalent to $M_{z=0}=[4.1\times10^{10}\,M_{\sun}](1+\delta_{\rm m})[V_{\rm true}/$\,cMpc$^{3}$], 
and  $V_{\rm true}=V_{\rm apparent}/C$. $V_{\rm apparent}$ is the observed comoving volume. The observed volume $V_{\rm apparent}$ is 60,964\,cMpc$^{3}$ 
and 58,332\,cMpc$^{3}$ in BOSS1244 and BOSS1542, respectively. Namely, $\sim20\farcm4\times20\farcm4$ region in BOSS1244 and 
$20\farcm0\times20\farcm0$ region in BOSS1542 are covered on the plane of sky between $z=2.223-2.267$ (the sight of comoving distance 
is about 56.9\,cMpc). The matter overdensity $\delta_{\rm m}$ is related to the galaxy overdensity by $1+b\delta_{\rm m}=C(1+\delta _{\rm g})$, 
where $b$ is the HAE bias factor. 
We take $b_{\rm HAE}=2.4_{-0.2}^{+0.1}$ from \cite{Geach2012} to be the HAE bias at $z=2.24$.
$C$ is the correction factor which is an estimation of the effects of redshift-space distortions caused by peculiar 
velocities \citep{Steidel1998}, which is a function of matter overdensity $\delta_{\rm m}$ and redshift $z$. In the case of spherical 
collapse, the correction factor $C$ can be estimated using the expression of $C=1+f-f(1-\delta _{\rm m})^{\frac{1}{3}}$, where 
$f=\Omega_{\rm m}(z)^{4/7}$, which we take to be $f=0.96$ at $z=2.24$.

We obtain the correction factor $C=0.62\pm0.03$ and matter overdensity $\delta_{\rm m}=1.69\pm0.15$ in BOSS1244, while $C=0.64\pm0.03$ and matter overdensity $\delta_{\rm m}=1.62\pm0.14$ in BOSS1542.  
According to the theory of density perturbation, we use the 
approximation for spherical collapse from the equation~18 in \cite{Mo1996}
to the linear matter density $\delta_{\rm L}$.  
The linear overdensity $\delta_{\rm L}$ in BOSS1244 and BOSS1542 is 0.74 and 0.73.  
If it evolves to the redshift of $z=0$, 
the linear overdensities $\delta_{\rm L}$ in BOSS1244 and BOSS1542  are 2.17 and 2.14, respectively, which is exceeding 
the collapse threshold of $\delta_{\rm c}=1.69$. We thus expect the entire BOSS1244 and 
BOSS1542 overdensities to be virialized by $z=0$. 
Using the Equation~\ref{sec:equ2}, the 
total masses at $z=0$ in the overall BOSS1244 and BOSS1542 fields are $(0.89\pm0.07)\times10^{16}$\,$M_{\sun}$ 
and $(0.82\pm0.06)\times10^{16}$\,$M_{\sun}$, respectively. 
We find that the same volume without an overdensity $\delta_{\rm m}$ results in a mass of $(3.70\pm0.18)\times10^{15}$\,$M_{\sun}$ 
in BOSS1244 and $(3.48\pm0.16)\times10^{15}$\,$M_{\sun}$ in BOSS1542. 
\cite{Kurk2004b} explained that the masses may be mostly intergalactic gas which will disappear out of the cluster with the Hubble flow, and eventually evolve into the mass of the bound system, although they gave an error in computing the value of $\bar{\rho}V_{\rm true}$, being pointed out in \cite{Steidel2005}.

\begin{deluxetable*}{cccccc}
\tablenum{3}
\tablecaption{Dynamical properties in this work and other protoclusters at $z=2-3$.   \label{tab:tab4}} 
\tablewidth{0pt}
\tablehead{
\colhead{Cluster} & \colhead{Redshift} & \colhead{$\sigma _{\rm los}$} & \colhead{$r_{200}$}  & \colhead{$M_{200}$} & \colhead{Reference} \\
 \colhead{name} & \colhead{$z$} & \colhead{[km\,s$^{-1}$]} & \colhead{[Mpc]}  & \colhead{[$10^{13}$\,$M_{\sun}$]}  
 }
\decimalcolnumbers
\startdata
    SSA~22 Blue & 3.069 & 350$\pm$53 &  $0.19\pm0.03$ & $1.6\pm0.7$ & \cite{Topping2016} \\
    SSA~22 Red & 3.095 & 540$\pm$40 &  $0.29\pm0.02$ & $5.9\pm1.3$ & \cite{Topping2016} \\
    MRC~0943–242 & 2.92 &  715$\pm$105 & $0.41\pm0.06 $ & $15\pm6$ & \cite{Venemans2007} \\
    MRC~0052–241 & 2.86 &  980 $\pm$120 & $0.57\pm0.07 $ & $38\pm14$ & \cite{Venemans2007} \\
    USS~1558-003 C1 & 2.53 & 284 & 0.19 & 1.00 & \cite{Shimakawa2014} \\
    USS~1558-003 C2 & 2.53 & 574 & 0.38 & 8.70 & \cite{Shimakawa2014} \\
    CL~J1001 & 2.506 & 530$\pm$120 & 0.36$\pm$0.08 & 7.94$\pm$3.80 & \cite{Wang2016} \\
    PCL1002 & 2.47 & 426 & $0.29$ & $3.7$ & \cite{Casey2015} \\
    PKS~1138-262 & 2.16 & 683 & 0.53 & 17.10 & \cite{Shimakawa2014} \\
    BOSS1441 & 2.32 & 943$\pm$500  & 0.7$\pm$0.4 & 40.0$_{-39}^{+70}$  & \cite{Cai2017a}  \\
    CC2.2 & 2.23 & 645$\pm$69 & 0.49$\pm$0.05 & 14.0$\pm$5.0 & \cite{Darvish2020} \\
    BOSS1244 NE & 2.246 &  377$\pm$99 & 0.28$\pm$0.07 & 2.80$\pm$2.20 & This work \\
    BOSS1244 SW & 2.230 &  405$\pm$202 & 0.30$\pm$0.15 & 3.00$\pm$5.00 & This work \\
    BOSS1542 Filament & 2.241 &  247$\pm$32 & 0.19$\pm$0.02 & 0.79$\pm$0.31 & This work \\
\enddata
\end{deluxetable*}

 The progenitors of galaxy clusters at $z\sim0$ have a characteristic size of 15\,cMpc, we thus estimate the total present-day masses of the density peak structures in the volume of (15\,cMpc)$^{3}$. The density peak regions in BOSS1244 and BOSS1542 are expected 
 to be virialized at $z\sim0$. As described in Section~\ref{sec:reds}, BOSS1244 has two different components: the NE region with $\delta_{\rm g}=10.9\pm2.5$ and SW region 
 with $\delta_{\rm g}=22.9\pm4.0$ over the 15\,cMpc scale. Using the Equation~\ref{sec:equ2}, 
 the total masses at $z=0$ in NE and SW are expected to be $(0.83\pm0.11)\times10^{15}$\,$M_{\sun}$ and 
 $(1.59\pm0.20)\times10^{15}$\,$M_{\sun}$, respectively. The mass in SW is about twice the mass in NE, which is related to the HAE overdensity. 
 In contrast, BOSS1542 displays a huge filamentary structure and $\delta_{\rm g}$ is $20.5\pm3.9$ over the 15\,cMpc scale 
 at $z=2.241$, and the present-day mass is $(1.42\pm0.18)\times10^{15}$\,$M_{\sun}$.  We summarize $\delta_{\rm g}$ and 
 present-day masses of the BOSS1244 and BOSS1542 fields in Table~\ref{tab:tab3}.

 Furthermore, we estimate the total enclosed mass from the scaling relation based on cosmological simulation. 
\citet{Chiang2013} presented the correlation between mass 
overdensity $\delta_{\rm m}$ at different redshifts and the descendant cluster mass $M_{z=0}$ in the 
(13.1\,cMpc)$^{3}$ and (24.1\,cMpc)$^{3}$ tophat box windows. We thus estimate $M_{z=0}$ based on
 our calculated matter overdensity in the same volume. Using $b=2.4$ as above, in the volume of 
 (13.1\,cMpc)$^{3}$, the matter overdensity $\delta_{\rm m}$ in BOSS1244 SW and NE regions is $12.6\pm2.5$ 
 and $5.6\pm1.6$, respectively, and $\delta_{\rm m}$ in BOSS1542 is $9.3\pm2.1$. 
 According to the correlation between overdensity and present-day total mass,
 we estimate the descendant cluster masses $M_{z=0}$ are $>2\times10^{15}$\,$M_{\sun}$ in BOSS1244 SW region and in $(1.24\pm0.24)\times10^{15}$\,$M_{\sun}$ 
 NE region, and $M_{z=0}$ in BOSS1542 is $>2\times10^{15}$\,$M_{\sun}$. 
 By comparison of the method of \cite{Steidel1998}, we find that mass 
 estimation presented by \cite{Chiang2013} is about twice as high as the method in \cite{Steidel1998}, but both 
 mass estimates suggest that BOSS1244 and BOSS1542 protoclusters over the 15\,cMpc scale will evolve into 
 Coma-type ($\geq10^{15}$\,$M_{\sun}$) clusters in the present epoch.

\subsection{Velocity Dispersions of Two Overdensities} \label{sec:velocity}

As described in \ref{sec:reds}, BOSS1244 shows two distinct spikes at $z=2.230$ in the SW region and $z=2.246$ 
in the NE region, and the SW region appears to be much denser than that the NE region. The line-of-sight depth between 
$z=2.230$ and $z=2.246$ is 21\,cMpc. 
In both regions, we estimate the line-of-sight 
velocity dispersion from our measured spectroscopic redshifts using two Gaussian functions. The velocity dispersion 
in the SW region is $405\pm202$\,km\,s$^{-1}$, similar to the velocity dispersion in the NE region ($377\pm99$\,km\,s$^{-1}$). 
 SW region in BOSS1244 also presents two substructures in redshift distribution
(the left inner panel of Figure~\ref{fig:fig8}) and the velocity dispersions of these two substructures 
 are $484\pm181$\,km\,s$^{-1}$ at $z=2.230$ and $152\pm58$\,km\,s$^{-1}$ at $z=2.245$. 
The lower velocity dispersion infers that the structure might perpendicular to the line-of-sight \citep{Venemans2007}, or might be due to the smaller number HAEs that we estimate the velocity dispersion.
For BOSS1542, our NIR spectroscopic observations present an extended and narrow filamentary structure on the scale of 23.4\,cMpc 
(the right panel of Figure~\ref{fig:fig8}).
The estimated velocity dispersion is $247\pm32$\,km\,s$^{-1}$ at $z=2.241$, 
which is much lower than that in BOSS1244. The lower velocity dispersion may indicate that BOSS1542 is a dynamically 
young protocluster \citep{Dey2016}.  

We also use the bi-weight method to check the line-of-sight velocity dispersion from the measured 
spectroscopic redshifts considering that this method is shown to be robust against a few outliers 
and for non-Gaussian underlying distributions \citep{Beers1990}. The velocity dispersions of NE 
and SW regions are $320\pm60$\,km\,s$^{-1}$ at $z=2.247$ and $431\pm99$\,km\,s$^{-1}$ at 
$z=2.231$, respectively.  The velocity dispersions of two components in the SW region are 
$304\pm99$\,km\,s$^{-1}$ at $z=2.246$ and $430\pm119$\,km\,s$^{-1}$ at $z=2.230$. The 
velocity dispersion in the BOSS1542 filament within 2$\sigma$ is $255\pm48$\,km\,s$^{-1}$ 
at $z=2.241$. All these estimates are consistent with the results obtained with the Gaussian methods. We adopt 
the measured results using the Gaussian method hereafter. 

\subsection{Dynamical Mass} \label{sec:dynamical}

Protocluster systems are not virialized, and the velocity dispersion of member galaxies traces the dynamical 
state of system rather than the halo mass \citep{Wang2016, Darvish2020}. Still,  dynamical mass estimates provide upper limits for the actual masses, given that the galaxies most likely populate multiple halos within the protocluster system rather
than one virialized system \citep{Lemaux2014, Dey2016, Overzier2016}.  
We apply the method for virialized systems to our protoclusters to draw an upper limit of dynamical mass and examine their dynamical state.

We first assume that BOSS1244 and BOSS1542 protoclusters at $z=2.24$ are virialized 
and the halos of the two protoclusters are spherical region within which the average density is $200\, \rho _{\rm c}(z)$.  
The virial mass $M_{200}=\frac{4\pi }{3}r_{200}^{3}200\rho _{\rm c}(z)$, where $r_{\rm 200}$ is the virial radius and $\rho _{\rm c}(z)=3H^{2}(z)/(8\pi G)$
is the critical density of the universe at redshift of $z$. Using a spherical symmetry combined with the virial theorem and the line-of-sight velocity 
dispersion $\sigma_{\rm los}$, the virial radius is $r_{200}= GM_{200}/(3\sigma _{\rm los}^{2})$, where $G$ is the gravitational constant.  Using 
above three formulae, we can derive $r_{200}$ and $M_{200}$ as a function of $\sigma_{\rm los}$ and $H(z)$: $M_{200}=(\sqrt{3}\sigma_{\rm los})^{3}/(10GH(z))$ and $r_{200}=\sqrt{3}\sigma_{\rm los}/(10H(z))$. 
For BOSS1542 protocluster, we estimate virial radius $r_{200}=0.19\pm0.02$\,Mpc and virial mass $M_{200}=(0.79\pm0.31)\times10^{13}$\,$M_{\sun}$.  
For BOSS1244 protocluster, we estimate $r_{200}=0.28\pm0.07$\,Mpc and $M_{200}=(2.80\pm2.20)\times10^{13}$\,$M_{\sun}$ in the NE region, and $r_{200}=0.30\pm0.15$\,Mpc and $M_{200}=(3.00\pm5.00)\times10^{13}$\,$M_{\sun}$ are computed in the SW region. The large errors in mass are due to the relatively large velocity dispersion errors.

We also use the scaling relation between velocity dispersion and total mass presented in \cite{Evrard2008} to estimate the dynamical total masses. 
\begin{equation}
\centering 
\label{sec:equ3}
\sigma _{\rm DM}(M,z)=\sigma _{\rm DM,15} \left [\frac{h(z)M_{200}}{10^{15} M_{\odot }}  \right ]^{\alpha } 
\end{equation}
where $\sigma _{\rm DM,15}$ is normalization at mass $10^{15}\,M_{\odot }$ and $\alpha$ is the logarithmic slope. 
Using $\sigma _{\rm DM,15}=1082.9\pm4.0$\,km\,s$^{-1}$ and $\alpha=0.3361\pm0.0026$, we can derive the 
halo masses in our protoclusters \citep[see also][]{Munari2013}. 
The virial masses $M_{200}$ in the BOSS1244 NE region, BOSS1244 SW region and BOSS1542 filament are $(1.90\pm1.50)\times10^{13}$\,$M_{\sun}$, $(2.30\pm3.50)\times10^{13}$\,$M_{\sun}$  and $(5.30\pm2.10)\times10^{12}$\,$M_{\sun}$, respectively.
The derived $M_{200}$ using the scaling relation is consistent with our estimate based on the virial theorem. 

We find that the dynamical masses of two substructures in BOSS1244 are $\sim2-3$ times that of BOSS1542,  indicating that they may be in different dynamic evolution states.
\cite{Wang2016} discovered an X-ray detected galaxy cluster with a  halo of $M_{200}\sim10^{13.7\pm0.2}$\,$M_{\sun}$ based on the velocity 
dispersion of $530\pm120\,$km\,s$^{-1}$ at $z=2.506$.  \cite{Shimakawa2014} presented two protoclusters PKS~1138-262 
at $z=2.16$ and USS~1558-003 at $z=2.53$, and the dynamical mass of the core is estimated to be 
$M_{200}=1.71\times10^{14}$\,$M_{\sun}$with the velocity dispersion of $683\,$km\,s$^{-1}$ and 
$M_{200}=0.87\times10^{14}$\,$M_{\sun}$  with the velocity dispersion of $574\,$km\,s$^{-1}$, respectively.  
Recently, \cite{Darvish2020} showed a new protocluster CC2.2 in COSMOS at $z=2.23$. The redshift and 
selection technique of HAEs are the same as ours used for BOSS1244 and BOSS1542. The dynamical mass of 
CC2.2 is $M_{200}=(1.40\pm0.5)\times10^{14}$\,$M_{\sun}$ and the velocity dispersion is $645\pm69\,$km\,s$^{-1}$.  They are summarized in Table~\ref{tab:tab4}.
These protocluster systems will evolve into fully collapsed and virialized Coma-type structures with a total mass of $M_{200}\sim10^{15}$\,$M_{\sun}$ at $z=0$, so they are likely to be in transition phase between protoclusters and mature clusters.
However, the dynamical masses in the BOSS1244 
and BOSS1542 protoclusters are $\sim1-2$ order-of-magnitude lower than that in above mentioned protoclusters, 
suggesting that BOSS1244 and BOSS1542 protoclusters are pre-virialized, younger systems. This provides 
evidence on earlier phase during which the clusters and their members are actually forming.

\section{Discussions} \label{sec:discuss}

\subsection{CoSLAs and Quasar Pairs in the Two Overdensities} \label{sec:cmd}

Protoclusters are not only traced by overdensities of galaxies, but also by intergalactic hydrogen gas that can produce Ly$\alpha$ absorption in the spectra of background quasars. 
Four out of five background quasars at $2.3<z<3.2$ in BOSS1244/BOSS1542 are used to measure the Ly$\alpha$ absorption. 
These background quasars $2-30$\,cMpc away from the center of two fields in sky projection,  with the effective optical depths of IGM absorption of $3-5\times$ the average optical depth ($\sim$0.25) at $z=2.24$.  
We find that no CoSLAs are in the dense region in BOSS1244, and all of them are distributed in the periphery of the BOSS1244 field. In contrast, two CoSLAs are located in the BOSS1542 filamentary structure and others are in the outskirts of the BOSS1542 field.

In BOSS1244, we note that the background quasar marked ``1" is located between SW and NE regions and is $\sim 6.85\,$cMpc away from the field center. The line-of-sight distance between SW and NE region and the CoSLAs at $z=2.24$ detected by background quasar is $7.8\,$cMpc and  $13.0\,$cMpc, respectively.  From the upper left of the figure~1 of \cite{Zheng2020}, we find that this CoSLA  shows a double peak in the line-of-sight distance with $> 3\times$ the average optical depth, corresponding to the comoving distance of $0-10\,$cMpc and $10-20\,$cMpc. It confirms that BOSS1244 should consists of two separated components along the line-of-sight direction, which is consistent with our spectroscopic observations. The background quasar marked ``2" is around the SW region with the line-of-sight distance of $10-20\,$cMpc, the background quasars marked ``3" and ``4" are around the NE region with the line-of-sight distance of $0-10\,$cMpc.  Similarly,  BOSS1542 also shows the consistent results from the effective optical depth profile described in the figure~2 of \cite{Zheng2020}. In short, these reflect the importance of NIR spectroscopic observations and indicate that the redshift we measured is reliable.

Furthermore, there are five BOSS quasars at $z=2.23-2.26$ in BOSS1244 and three BOSS quasars at $z=2.23-2.25$ in BOSS1542, as shown in Table~\ref{tab:tab5}.  
We find that the measured redshifts of the HAEs are consistent with the redshifts of these quasars around them, and most of these BOSS quasars are in the overdense regions of BOSS1244 and BOSS1542,  indicating that these quasars are likely to be associated with the two overdensities. 
In BOSS1244, we find that  QSO2/QSO5 with projected separation $\sim1.2$\,pMpc and velocity offset $\sim1570$\,km s$^{-1}$ meets the definition of quasar pairs with a projected separation $<2 (=1.4 h^{-1}$)\,pMpc and velocity offset $<2300$\,km\,s$^{-1}$ \citep{Onoue2018}.   In BOSS1542, QSO2/QSO3 with projected separation $\sim1.9$\,pMpc and velocity offset $\sim740$\,km s$^{-1}$ is a quasar pair. Interestingly, the quasar pairs QSO2/QSO5 in BOSS1244 and QSO2/QSO3 in BOSS1542 are located in the overdense region, suggesting that the overdensities of galaxy are associated with quasar pairs.   Some works show that quasar pair environment has a systematic larger overdensity of galaxies around quasars in pairs with respect to that of isolated quasars at $z<1.0$ \citep{Farina2011,Hennawi2015,Sandrinelli2018,Onoue2018}.
However, some works suggest no enhancement in the galaxy density around the quasar pair at low-$z$ or high-$z$ \citep{Green2011,Sandrinelli2014, Fukugita2004}. 
 \cite{Onoue2018} pointed out that pairs of luminous quasars at $z>3$ and $z\sim1.0$ are better tracers of protoclusters than single quasars, but are not tracing the most overdense protoclusters.  In \cite{Cai2017a}, we find that the first MAMMOTH overdensity BOSS1441 has two quasar pairs with the projected separations are $\sim1.8$\,pMpc/$2.0$\,pMpc and the velocity offsets are $\sim 900$\,km s$^{-1}$/$100$\,km s$^{-1}$, which also reside in the density peak of the overdensity.  Therefore, these observations provide further evidence that overdensity of galaxy at $z=2-3$ could also be traced by the quasar pairs.

\begin{deluxetable}{c|cccc}
\tablenum{4}
\tablecaption{BOSS QSOs (including background QSOs) in the BOSS1244 and BOSS1542 fields.   \label{tab:tab5}} 
\tablewidth{0pt}
\tablehead{
\colhead{Field} & ID & \colhead{R.A.}  & \colhead{Decl.} & \colhead{Redshift}  \\
 \colhead{} & \colhead{} & \colhead{(J2000.0)} & \colhead{(J2000.0)}   & \colhead{$z$}  
}

\startdata
    & BG QSO1 & 191.11745 & 36.122048 & 3.212 \\
    & BG QSO2 & 190.70921 & 35.966921 & 3.143  \\
    & BG QSO3 & 190.81858 & 36.145731 & 2.888 \\
    & BG QSO4 & 191.08365 & 36.151788 & 2.863 \\
BOSS1244    & BG QSO5 & 190.90702 & 36.031874 & 2.437   \\
    & QSO1 (HAE) & 191.10603 & 36.132581 & 2.260  \\
    & QSO2 (HAE) & 190.91308 & 35.901003 & 2.245  \\
    & QSO3 (HAE) & 191.01176 & 35.935079 & 2.235  \\
    & QSO4 (HAE) & 191.11257 & 36.152239 & 2.233   \\
    & QSO5 (HAE) & 190.90429 & 35.941498 & 2.228   \\
\hline
    & BG QSO1 & 235.47301 & 38.774797 & 2.781 \\
    & BG QSO2 & 235.43314 & 39.149935 & 2.717 \\
    & BG QSO3 & 235.72489 & 38.834426 & 2.477 \\
    & BG QSO4 & 235.63317 & 39.148571 & 2.356   \\
 BOSS1542   & BG QSO5  & 235.66231 & 38.913783 & 2.267  \\
    & QSO1 (HAE) & 235.62312 & 39.021790 & 2.250 \\
    & QSO2 (HAE) & 235.70368 & 38.960585 & 2.242  \\
    & QSO3 (HAE) & 235.69590 & 38.895094 & 2.234   \\
\enddata
\end{deluxetable}

\subsection{Internal Structures} \label{sec:ks}

Kinematical structures of distant protoclusters provide essential information on the mass assembly history of galaxy clusters. 
Figure~\ref{fig:fig9} and Figure~\ref{fig:fig10} show the three-dimensional spatial distribution of HAEs in BOSS1244 and BOSS1542, 
the color-coded filled points are the confirmed HAEs with NIR spectroscopy. 
In terms of the large-scale geometry, the BOSS1244 and BOSS1542 protoclusters are very different, 
suggesting that they are in different stages of evolution.

\subsubsection{BOSS1244 with multiple components } \label{sec:boss1244}

\begin{figure*}[ht]
\setlength{\abovecaptionskip}{-50pt}
\begin{center}
\includegraphics[trim=0mm 55mm 0mm 0mm,clip,height=0.65\textwidth]{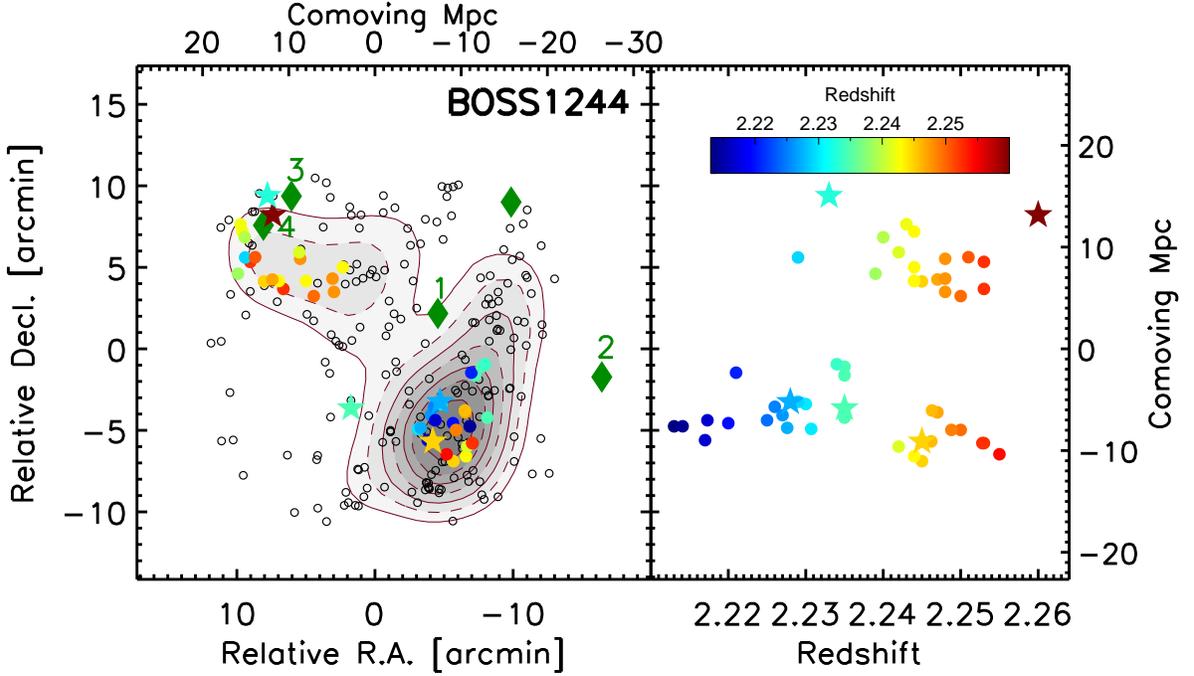}
\caption{Three-dimensional distribution of the protocluster galaxies in the BOSS1244 field. Sky coordinates are listed relative to the centre of the BOSS1244 field. The color-coded points and stars are the confirmed HAEs and quasars at $z=2.213-2.260$. {\bf Left:} HAEs plotted as Decl. vs. R.A. The black circles are the HAE candidates in BOSS1244. {\bf Right:}  Confirmed HAEs plotted Decl. vs. $z$. There is a separation in the distributions of galaxies both in Decl. and $z$.} 

\label{fig:fig9}
\end{center}
\end{figure*}

BOSS1244 shows two distinct components (NE and SW) both in space and in redshift distribution, and two substructures in the SW region. The redshift difference
 between SW and NE regions is $\Delta z=0.016$ (the line-of-sight depth is 20.8\,cMpc), and the projected separation between them is 13\farcm5 ($\sim21.6$\,cMpc). 
 The right panel in Figure~\ref{fig:fig9} is the spectroscopic confirmed HAEs, we find that serveral substructures are in the SW and NE regions.  
  In the SW region, there are about five subgroups with $\ge 3-5$ galaxies at $2.213<z<2.221$, $2.224<z<2.229$, $2.234<z<2.236$, $2.242<z<2.247$ and $2.248<z<2.257$. There are two subgroups with $\ge 5$ galaxies in the NE region at $2.238<z<2.245$ and $2.246<z<2.255$. These subgroups are clustered together along the line-of-sight direction.  The hierarchical structure formation model predicts that the larger structures are formed from smaller substructures through the continuous merging \cite[e.g.,][]{Kauffmann1993, Kauffmann1999}. The subgroups in SW and NE regions might continue to grow into larger structures.

We want to know whether the SW and NE protoclusters can collapse to a single rich cluster, or evolve into two independent massive halos as parts of a supercluster. A typical size of the effective radius is $\sim$3.2\,cMpc at $z\sim2.2$ for the progenitors of $(1-3)\times10^{14}$\,$M_{\sun}$ halos at $z=0$ and the size is about 6.5\,cMpc for  those of $>10^{15}$\,$M_{\sun}$\citep{Chiang2013}.  According to theoretical simulations, \cite{Muldrew2015} predicted that the average radius of protoclusters is $\sim9-13\,$cMpc at $z\sim2$ for the descendants of $(4-10)\times10^{14}$\,$M_{\sun}$ halos.  The main protoclusters and their surrounding groups/clusters could merge into a single galaxy cluster by $z = 0$ only if a descendant halo mass at $z = 0$ is $>10^{15}$\,$M_{\sun}$ \citep{Muldrew2015}.   The present-day total mass in SW component is $(1.59\pm0.20)\times10^{15}$\,$M_{\sun}$, which  is twice that of the NE component. It is possible that the NE component falls to the larger SW component through the gravitational potential well, and merge to a larger structure eventually.

For what concerns merger events, \cite{Lee2016} reconstructed a 3D tomographic map of the foreground Ly$\alpha$ forest absorption at $z=2-3$ using the background LBGs and quasars in COSMOS. Using mock tomographic maps, they found that very few of protoclusters with an elongated shape will collapse to one single cluster at $z\sim0$. The proto-supercluster was identified with seven density peaks subsequently \citep{Cucciati2018}.  \cite{Topping2018} presented the Small MultiDark Planck Simulation on searching for the $z\sim3$ protoclusters with a double peak in redshift distribution and the two peaks separated by 2000\,km\,s$^{-1}$, like SSA22 protocluster \citep{Topping2016}. They found that such double-peaked overdensities are not going to merge into a single cluster at $z = 0$.  The redshift separation between SW and NE protoclusters in BOSS1244 is 0.016, and the velocity offset is $\sim1486\,$km\,s$^{-1}$.  Their projected separation is $\sim$21.6\,cMpc, which is much larger than the typical size of protoclusters. The BOSS1244 structure is similar to the SSA22 protocluster, so it may evolve into two separated clusters at $z\sim0$.

In the local universe ($z<0.5$),  superclusters are typically a few cMpc to $\sim$100\,cMpc in size \citep{Rosati1999, Kim2016}, with wide ranges of mass from  a few $10^{14}$\,$M_{\sun}$ to $10^{16}$\,$M_{\sun}$ \citep{Swinbank2007,Bagchi2017}, and most of superclusters are composed of two or three galaxy clusters, and a few of them include nearly 10 clusters \citep{Lubin2000,Lemaux2012}.   Figure~\ref{fig:fig9} displays the three-dimensional distribution (sky and redshift positions) of our spectroscopically-identified  protocluster galaxies in BOSS1244.  We find that there are many smaller components and they may be forming large-scale structures, and the present-day total mass in the whole system is $\sim10^{16}$\,$M_{\sun}$.  It suggests that BOSS1244 may also be evolved into a supercluster with two massive galaxy clusters at $z\sim0$.

\subsubsection{BOSS1542 with an extended filamentary structure} \label{sec:boss1542}

\begin{figure*}[ht]
\setlength{\abovecaptionskip}{-50pt}
\begin{center}
\includegraphics[trim=0mm 55mm 0mm 0mm,clip,height=0.65\textwidth]{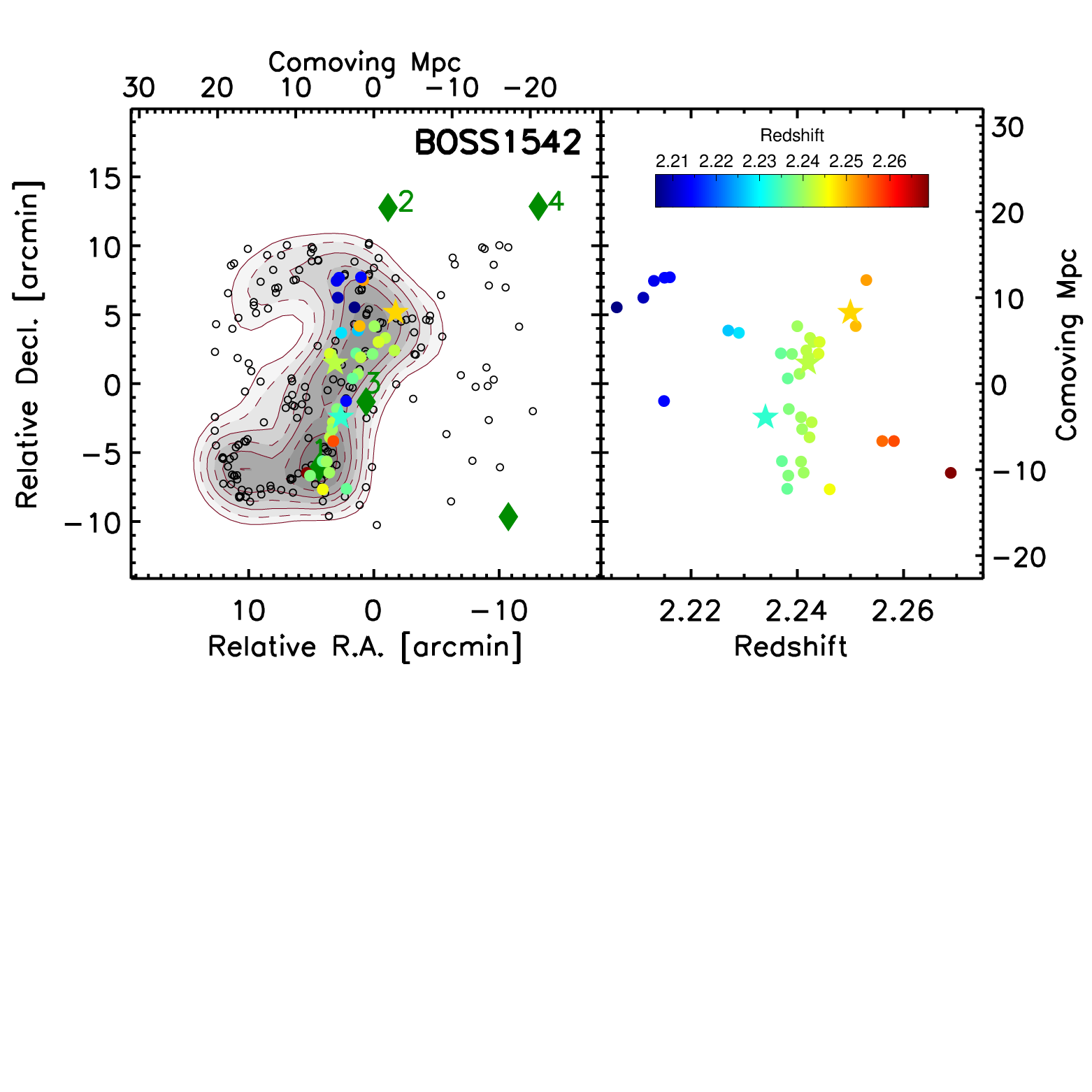}
\caption{Three-dimensional distribution of the protocluster galaxies in the BOSS1542 field. Sky coordinates are listed relative to the centre of the BOSS1542 field. The color-coded points and stars are the confirmed HAEs and quasars at $z=2.206-2.269$. {\bf Left:} HAEs plotted as Decl. vs. R.A. The black circles are the HAE candidates in BOSS1244. {\bf Right:}  Confirmed HAEs plotted Decl. vs. $z$. There is no significant difference in the distributions of galaxies in Decl.}
\label{fig:fig10}
\end{center}
\end{figure*}

Filaments are ubiquitous in the universe and and account for $\sim40-60\%$ of the matter in the universe, but only $\sim6-16\%$ of the volume \citep{Tempel2014,Martizzi2019}. Cosmic filaments are elongated relatively high density structures of matter, tens of Mpc in length, and intersect at the location of galaxy clusters. They form through a gravitational collapse of matter driven by gravity \citep{Codis2012,Laigle2015,Kraljic2018,Kuchner2020}. BOSS1542 shows an enormous H$\alpha$ filamentary structure with a projected length of 23.4\,cMpc running in the North-South direction, shown in  Figure~\ref{fig:fig10}.  The galaxy overdensities $\delta_{\rm g}$ in the filament region over a typical protocluster scale (15\,cMpc) is $20.5\pm3.9$, corresponding to the present-day halo mass of $(1.42\pm0.18)\times10^{15}$\,$M_{\sun}$. It indicates that the H$\alpha$ filament structure in BOSS1542 will evolve into a Coma-type galaxy cluster at $z\sim0$.  

Recently, \cite{Umehata2019} presented a cold-gas filament of the cosmic web in Ly$\alpha$ emission in the core of the SSA22 protocluster at $z=3.09$.  The network of filaments in SSA22 is found to connect individual galaxies across a large volume, allowing it to power star formation and black hole growth in active galaxy populations at $z\sim3$. It is suggested that similar structures may be a general feature of protoclusters in the early universe \citep{Martin2014,Kikuta2019}. In BOSS1542, we find that star-forming galaxies (HAEs), quasars and  CoSLAs detected by background quasars reside in this filamentary, the velocity dispersion of this structure is $247\pm32\,$km\,s$^{-1}$, suggesting that it is a dynamically young structure. Cosmological simulations of structure formation predict that the majority of gas in the intergalactic medium (IGM) is distributed in a cosmic web of sheets and filaments by gravitational collapse \citep{Bond1996}.  The cold gas is falling along the filaments driven by gravity and the filaments are able to provide most of the gas required for the growth of galaxies and SMBHs. Using absorption spectroscopy of background sources to trace neutral hydrogen in the IGM, which will provide insights into the nature of the cosmic web.  The filament in BOSS1542 is providing a direct detection of the cosmic web in the early universe.  In the future, we will search for LAEs, LABs, X-ray luminous AGNs and SMGs in the BOSS1542 filamentary structure to understand how the universe evolved through time, how galaxies grow and mature, and how the changing environments affect galaxy properties at $z=2-3$.

From the right panel of Figure~\ref{fig:fig10}, our NIR spectroscopic observations show that BOSS1542 is dominated by one component with an extended filament at $2.235<z<2.245$. However, at the intersection of this filamentary structure, several substructures without NIR spectroscopy confirmation are gathering together. For these regions, we need much more NIR spectroscopy to reveal their dynamical nature. Similarly, given the total mass of $\sim 10^{16}$\,$M_{\sun}$ at $z\sim0$, BOSS1542 may also evolve into a supercluster in the late time.

\subsubsection{Comparison with other protoclusters at $z>2$} \label{sec:comparison} 

\begin{figure*}[ht]
\setlength{\abovecaptionskip}{0pt}
\begin{center}
\includegraphics[trim=2mm 0mm 0mm 0mm,height=0.38\textwidth]{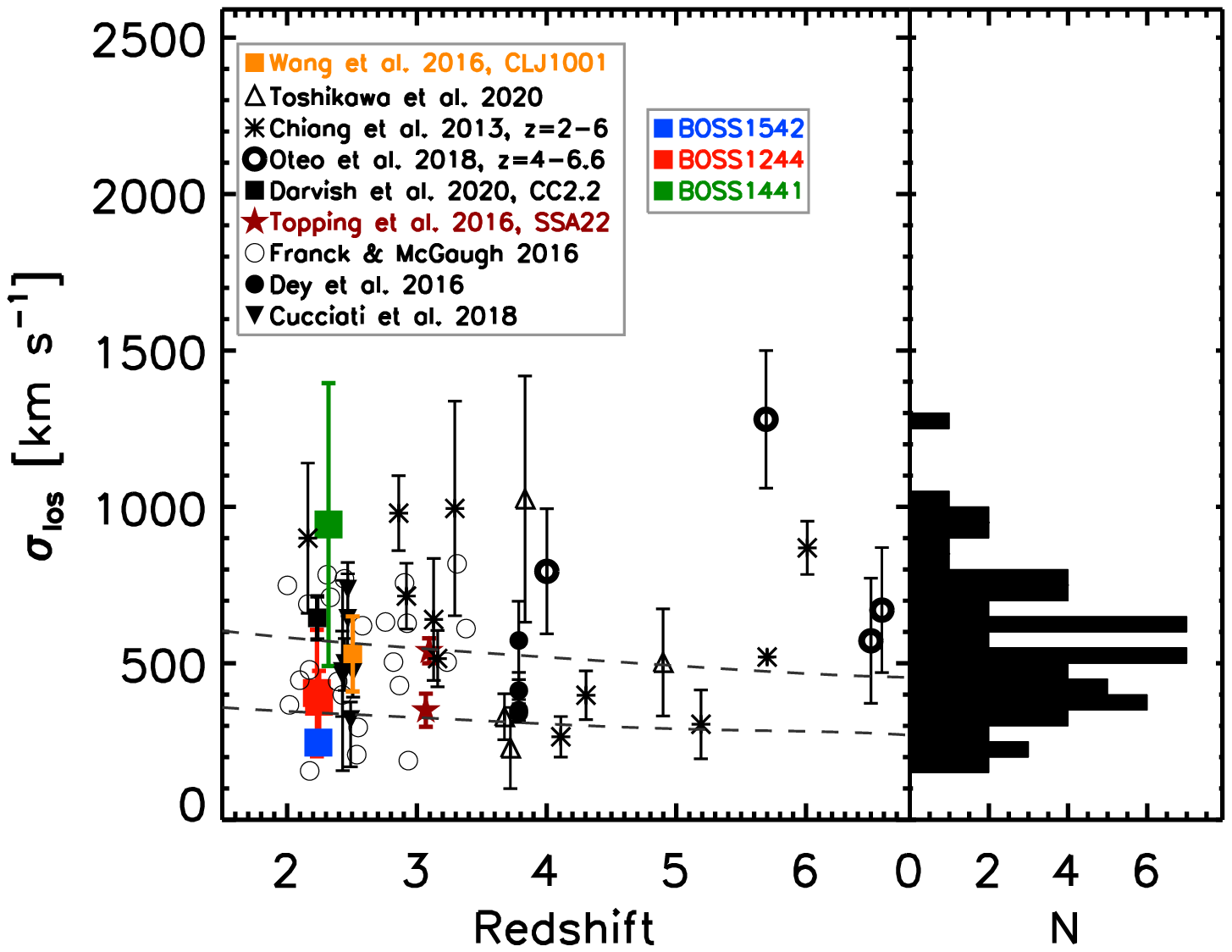}
\includegraphics[trim=2mm 0mm 0mm 0mm,height=0.38\textwidth]{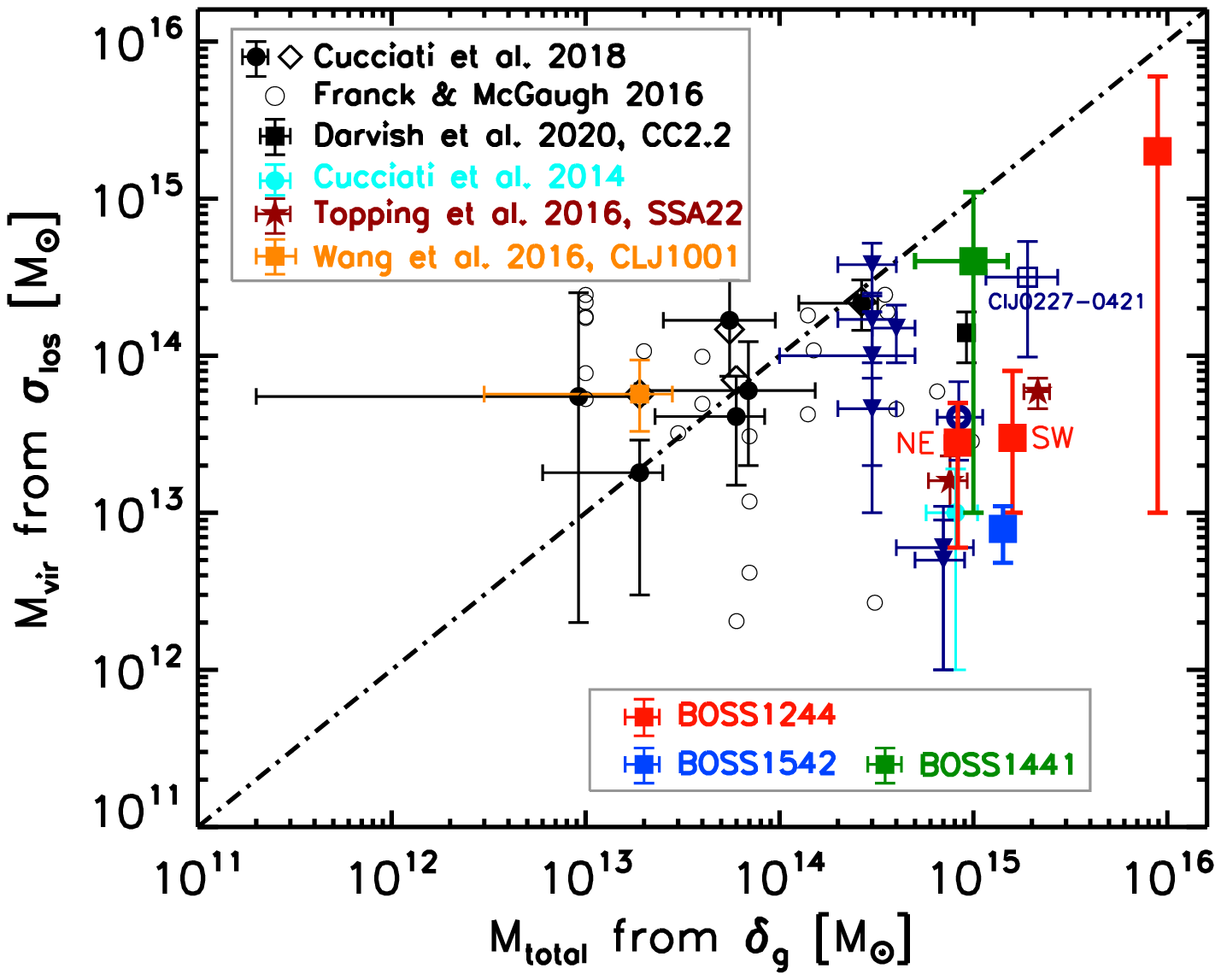}
\caption{{\bf Left panel:} velocity dispersion of protoclusters as a function of redshift and the histogram of the velocity dispersion of protoclusters. The red squares are the SW and NE protoclusters in BOSS1244, the blue square is the protocluster in BOSS1542, and other black symbols are the other protoclusters from the literature \citep{Chiang2013, Lemaux2014, Dey2016, Wang2016, Franck2016, Topping2016, Cai2017a, Cucciati2014, Cucciati2018, Oteo2018, Toshikawa2020, Darvish2020, Hill2020}. The dashed lines show the redshift evolution of dark matter velocity dispersion of $1\times10^{14}$\,$M_{\sun}$ and $5\times10^{14}$\,$M_{\sun}$ halos at $z=0$, which is from \cite{Toshikawa2020}.   {\bf Right panel:} virial mass $M_{\rm vir}$ derived from velocity dispersion, vs. the present-day total mass $M_{\rm tot}$ estimated from galaxy overdensity. The red squares are the SW and NE protoclusters in BOSS1244 and the blue square is the protocluster in BOSS1542. The rightmost red square is our estimated velocity dispersion of the entire BOSS1244 system. Other symbols are the other protoclusters from the literature \citep{Cucciati2014, Lemaux2014, Franck2016, Cai2017a, Topping2016, Topping2018, Cucciati2018, Darvish2020}. The dot-dashed line shows the 1:1 relation between $M_{\rm vir}$ and $M_{\rm tot}$.  } 
\label{fig:fig11}
\end{center}
\end{figure*}

We compare other protoclusters at $z=2-3$ from literature with our protoclusters. The protoclusters reported in the literature has a median present-day halo masses of log$(M/M_{\sun})_{z=0}=14.6$ \citep{Overzier2016}. Not only will our entire overdensities ($\sim10^{16}$\,$M_{\sun}$) be among the most massive clusters known today, but the density peaks ($\delta_{\rm g}>20$) will be evolved into massive galaxy clusters of $>10^{15}$\,$M_{\sun}$ at $z=0$. 

\cite{Lemaux2014} discovered a massive protocluster (Cl~J0227-0421) at $z=3.29$ with a galaxy overdensity of $10.5\pm2.8$. They estimated the dynamical mass of $\sim3\times10^{14}$\,$M_{\sun}$, and expected a halo mass of $\sim(3.67-8.69)\times10^{15}$\,$M_{\sun}$ for the $z = 0$ cluster.  \cite{Cucciati2018} presented a proto-supercluster at $z=2.45$ by identifying seven galaxy groups within a volume of $\sim60\times60\times150\,$cMpc$^{3}$. The total mass in each galaxy group at $z=0$ is about $9.2\times10^{12}-2.6\times10^{14}$\,$M_{\sun}$, and the estimated present-day mass in the effective volume of $9.5\times10^{4}\,$cMpc$^{3}$ is $\sim4.8\times10^{15}$\,$M_{\sun}$. The effective volume of the proto-supercluster is slightly larger than that in our protoclusters ($6.2\times10^{4}\,$cMpc$^{3}$), but estimated present-day mass is $2-3$ times less than ours. The main reason is that the estimation of mass depends not only on volume, but also on the galaxy overdensity.  \cite{Topping2018} investigated the nature of large-scale structure in the SSA22 protocluster region at $z = 3.09$ with a galaxy overdensity of $7.6\pm1.4$. The expected total halo mass at $z=0$ is  $(3.19\pm0.40)\times10^{15}$\,$M_{\sun}$  over a volume of $\sim12\times14\times43\,$cMpc$^{3}$. They revealed two separated overdensities at $z=3.065$ and $z=3.095$, corresponding to the present-day masses are  $(0.76\pm0.17)\times10^{15}$\,$M_{\sun}$ and $(2.15\pm0.32)\times10^{15}$\,$M_{\sun}$, respectively.  Our findings of the large-scale structures in BOSS1244 seem to exhibit a consistent size and  number of components with similar present-day masses.  Recently, \cite{Darvish2020} reported a protocluster (CC2.2) in the COSMOS field at $z=2.23$ traced by the spectroscopic confirmation of HAEs. The galaxy overdensity is $\sim6.6$ over the volume of $\sim5500$\,cMpc$^{3}$, corresponding to the halo mass of $\sim9.20\times10^{14}$\,$M_{\sun}$ at $z=0$.  The dynamical mass is $\sim(1-2)\times10^{14}$\,$M_{\sun}$, suggesting the CC2.2 protocluster is not fully virialized at $z=2.23$.  Similarly, our protoclusters are also identified by the NIR spectroscopy of HAEs at the same redshift, suggesting that our protocluster system are dynamically younger and in the process of galaxy accretion and merging, which are far from virialization, especially for the BOSS1542 protocluster. The NE protocluster in BOSS1244 is  similar to the CC2.2 protocluster given their consistent galaxy overdensities and expected halo masses at $z=0$. 

\cite{Cai2017a} reported the first MAMMOTH protocluster BOSS1441 at $z=2.32$. This protocluster contains a dominant dark matter halo that would likely collapse into a virialized cluster with mass $M_{z=0}=10^{15}$\,$M_{\sun}$.  For BOSS1441, BOSS1244 and BOSS1542, the three overdensities selected from the same technique display dramatically different  in morphology, representing different stages of galaxy cluster assembly. We suspect BOSS1441 is a nearly-virialized protocluster, BOSS1244 is forming one or two protoclusters, and BOSS1542 is forming a filamentary structure. These processes are also expected in the cosmological simulations \citep{Shandarin1989,vandeWeygaer2008,Shandarin2009,Libeskind2018}. Namely, considering a non-spherical collapse of an ellipsoildal overdensity for the real perturbations, the expansion turns into collapse along shortest axis first to form a ``pancake" sheet structure, then along the intermediate axis to form filaments, and only at the end to fully collapse along the longest axis to form clusters \citep[e.g.,][]{Libeskind2018}. 

For the rarity of these two structures, \cite{Cai2017a} found 11 fields at $z=2-3$ that contain groups of strong IGM Ly$\alpha$ absorption over the volume of $10^{8}\,$cMpc$^{3}$, but 30\% of the 11 absorptions have $>10^{15}$\,$M_{\sun}$, where BOSS1441, BOSS1244 and BOSS1542 are included and identified.  Namely, the volume density of high-mass of $>10^{15}$\,$M_{\sun}$ is $\sim3\times10^{-8}\,$cMpc$^{-3}$.  However, BOSS1244 and BOSS1542 have currently the highest galaxy overdensities over the volume of $\sim(39\,\rm cMpc)^{3}$, corresponding to the present-day total masses of $\sim10^{16}$\,$M_{\sun}$, so the occurrence rate of such structures is expected to be $2\times10^{-8}\,$cMpc$^{-3}$ at $z=2.2$.  
\cite{Darvish2020} detected one protocluster (CC2.2) at $z=2.23$ in the COSMOS field over the volume of $\sim5.48\times10^{5}\,$cMpc$^{-3}$ and the corresponding volume density is $1.8\times10^{-6}\,$cMpc$^{-3}$, which is higher than our result. This reason is due to their limited effective volume.  According to simulations, \cite{Topping2018} found that the presence of two massive halos separated by 2000\,km\,s$^{-1}$ around the SSA22 protocluster is 7.4\,Gpc$^{-3}$ at $z\sim3$.  The BOSS1244 with two massive protoclusters separated by 1500\,km\,s$^{-1}$ is similar to the SSA22 protocluster, and their occurrence rate of such a structure is consistent.

\subsection{Velocity Structure} \label{sec:vd}
BOSS1244 and BOSS1542 are the most massive structures consisting of multiple components. BOSS1244 has two distinct protoclusters in the SW and NE regions, the velocity dispersions are $405\pm202\,$km\,s$^{-1}$ and $377\pm99\,$km\,s$^{-1}$, respectively. Furthermore, the SW protocluster in BOSS1244 appears to be two components with the velocity dispersions of $484\pm181$\,km\,s$^{-1}$ and $152\pm58$\,km\,s$^{-1}$.  In contrast, the estimated velocity dispersion in BOSS1542 filament is $247\pm32$\,km\,s$^{-1}$. Note that we do not map all the HAEs in BOSS1244 and BOSS1542 due to the limited NIR spectroscopic observations.

The line-of-sight velocity dispersions of density peaks in BOSS1244 and BOSS1542 are relatively small compared with those of studied protoclusters in the literature. For example, The velocity dispersions of the proto-supercluster with seven density peaks at $z=2.45$ are  $320-737\,$km\,s$^{-1}$ \citep{Cucciati2018}. 
\cite{Venemans2007} discovered six protoclusters around eight radio galaxies at $z=2.0-5.2$, the velocity dispersions for these clusters were measured to be ranging from $\sim250$\,km\,s$^{-1}$ to 1000\,km\,s$^{-1}$. In that sample, MRC~1138-262 (PKS~1138-262) at $z=2.16$ and MRC~0052-241 at $z=2.86$ protoclusters showed bimodal redshift distributions:  MRC~1138-262 had double peaks with velocity dispersions of 280\,km\,s$^{-1}$ and 520\,km\,s$^{-1}$, and MRC~0052-241 had double peaks with velocity dispersions of 185\,km\,s$^{-1}$ and 230\,km\,s$^{-1}$. The best-studied system SSA22 protocluster (similar to the BOSS1244 protocluster) at $z=3.09$ showed two substructures with velocity dispersions of 350\,km\,s$^{-1}$ and 540\,km\,s$^{-1}$ \citep{Topping2016}.  
Using NIR spectroscopy of HAE sample, \cite{Darvish2020} estimated that the velocity dispersion of protocluster CC2.2 at $z=2.23$ is 645\,km\,s$^{-1}$, which is higher than the velocity dispersions of our protoclusters.
We calculate the velocity dispersion of BOSS1441 protocluster at $z=2.32$ from 18 Ly$\alpha$ spectroscopy is 943\,km\,s$^{-1}$ \citep{Cai2017a}.  It indicates that all these protoclusters may be in different dynamical state. The velocity dispersions of other known protocluster candidates at $z>2$ have been compiled \citep{Kuiper2011, Cucciati2014, Toshikawa2014,Wang2016, Lemaux2014, Lemaux2018, Miller2018, Chanchaiworawit2019} in Figure~\ref{fig:fig11}. The left panel of Figure~\ref{fig:fig11} presents a relationship between velocity dispersion and redshift of protoclusters. No significant correlation is seen between velocity dispersion and redshift, although the velocity dispersion is expected to increase with protocluster growth. 
Most protoclusters have velocity dispersions of $<600$\,km\,s$^{-1}$, but some protoclusters show higher velocity dispersions, even $>1000$\,km\,s$^{-1}$.  We estimate the velocity dispersion for the whole BOSS1244 field is $1670\pm963$\,km\,s$^{-1}$ using the Gaussian fitting.  We also apply the biweight method and estimate the velocity dispersion to be $1058\pm142$\,km\,s$^{-1}$, consistent  with previous mentioned. Previous works explain that the protoclusters with higher velocity dispersions are in merging processes and forming more massive structures, and their dynamical state may be far from virialization \citep[e.g.,][]{Dey2016, Toshikawa2020}.

The right panel of Figure~\ref{fig:fig11} presents dynamical masses based on the estimation of velocity dispersion as a function of present-day masses based on the Equation~\ref{sec:equ2}.
The dot-dashed line lies on the 1:1 relation. \cite{Cucciati2018} presented  a ``Hyperion"  proto-supercluster with seven density peaks, the estimated two sets of masses are surprisingly consistent with the agreement of $<2\sigma$. Moreover, one of the seven peaks (orange square in Figure~\ref{fig:fig11}) has already been identified as a virialized structure \citep{Wang2016}, but they suggested $M_{\rm total}$ may be underestimated given the most distant in reconstruction from the 1:1 relation between $M_{\rm total}$ and $M_{\rm vir}$.  
The evolution of a density fluctuation from the beginning of collapse to virialization can take a few gigayears. Furthermore, galaxies outside the peaks' volumes may be included in the velocity distribution along the line-of-sight. Thus, the estimated dynamical masses based on the velocity dispersion of protoclusters have relatively small changes, while $M_{\rm total}$ varies by changing the overdensity threshold to define the density peaks \citep{Cucciati2018}.

We may infer the level of virialization of a density peak based on the comparison between $M_{\rm total}$ and $M_{\rm vir}$, although there are some uncertainties. We have compiled some known protoclusters in the literature, as shown in the right panel of Figure~\ref{fig:fig11}.  We find that some protoclusters locate on the relation between $M_{\rm total}$ and $M_{\rm vir}$, while some protoclusters show that $M_{\rm vir}$ is $2-3$ orders of magnitude lower than $M_{\rm total}$. We argue that some protoclusters are near-virialized \citep[e.g.,][]{Wang2016, Cai2017a, Cucciati2018, Darvish2020}, representing an important transition phase between protoclusters and mature clusters, and some protocluster systems are far from virialization \citep[e.g.,][]{Cucciati2014, Lemaux2014, Topping2016}, providing direct evidence on the earlier formation phase. BOSS1441 protocluster might be a near-virialized structure, while the protoclusters in BOSS1244 and BOSS1542 show much smaller $M_{\rm total}$/$M_{\rm vir}$ ratios, indicating that these structures are far from virialization, they may be collapsing and forming much more massive large-scale structures.  Therefore, we conclude that the protoclusters in BOSS1441, BOSS1244 and BOSS1542 are likely to be at different stages of their evolution, and will become virialized structures later.

Recently, \cite{Shimakawa2018a} proposed a speculation of formation and evolution histories of galaxy clusters. They divided the formation of galaxy clusters into three stages: growing phase at $z\ge 3$, maturing phase at $z=2-3$ and declining phase at $z\le2$. In the growing stage, cold gas stream in the hot gas is able to support the active star formation in massive galaxies \citep[e.g.,][]{Keres2005, Dekel2006, Ocvirk2008, Keres2009, vandeVoort2011}. The protocluster USS~1558 at $z=2.5$ is suggested to be a growing protocluster \citep{Shimakawa2018a, Shimakawa2018b}.  In the maturing stage, the members of protoclusters may be undergoing a rapid transition from dusty starbursts to quenching populations, red sequence and high AGN fraction can be seen \citep{Williams2009, Whitaker2011}. The protocluster PKS~1138 at $z=2.2$ is considering to be a maturing protocluster \citep{Shimakawa2018a}.  Due to the insufficient gas accretion, cluster members in the hot inter-cluster medium enriched by superheated plasma would no longer maintain their star formation in the declining phase \citep[e.g.,][]{Keres2005,Hughes2013,Hayashi2017}, including environmental quenching \citep[e.g.,][]{Bamford2009, Peng2010, Gobat2015, Ji2018}. Based on the cold flows and shock-heated medium as a function of total mass and redshift described in \cite{Dekel2006} and the aforementioned scenario, we predict that BOSS1542 is in the cold in the evolutionary stage of cool gas filament, may still be in the growing phase like USS~1558, BOSS1441 may be similar to PKS~1138, it is probably a maturing protocluster, and BOSS1244 could be the transitional stage between BOSS1542 and BOSS1441. The discovery of these three MAMMOTH protoclusters provides us with new insights into the formation and evolution of galaxy cluster at the present epoch.

\section{Summary} \label{sec:conclusion}

BOSS1244 and BOSS1542 are two extreme overdensities traced by Ly$\alpha$ absorbers within $\sim20$\,cMpc at $z=2.24\pm0.02$. They have been confirmed with HAEs identified using the NIR NB imaging technique. Using the NIR MMT/MMIRS and LBT/LUCI instruments, 46 and 36 HAEs are spectroscopically identified in BOSS1244 and BOSS1542, respectively.  We analyze the properties of the two overdensities taking advantage of NIR spectroscopy. The results are summarized as follows.

\begin{itemize}

\item[(1)]  We identify 46/36 HAEs at $z\sim2.24$ in the BOSS1244/BOSS1542 field through NIR spectroscopic observations. The detection rate in BOSS1244 is $80\%$, while the success rate is $57\%$  in BOSS1542. This is due to the shorter exposure time in the slit-masks in BOSS1542, and $\sim4\%$ contaminations in HAE sample given an [O\,{\small III}] emitter galaxy at $z=3.302$ is detected. These confirmed HAEs suggest that the BOSS1244 and BOSS1542 fields are indeed extremely overdense.

\item[(2)]  In BOSS1244, there are two distinct peaks in the SW and NE regions  at $z=2.230$ and $z=2.246$ segregated on the sky and redshift distribution, the projected separation is about $13\farcm5$ (21.6\,cMpc). The estimated  line-of-sight velocity dispersions are $405\pm202$\,km\,s$^{-1}$ and  $377\pm99$\,km\,s$^{-1}$, respectively.  Moreover, two substructures in the SW region of the BOSS1244 are found.  Comparatively, BOSS1542 presents an enormous filamentary structure at $z=2.241$ with a very small velocity dispersion of $247\pm32$\,km\,s$^{-1}$, suggesting that it might be a dynamically younger system and providing a direct detection of  cosmic web in the early universe.

\item[(3)]  We recompute the HAE overdensities $\delta_{\rm g}$ in BOSS1244 and BOSS1542.  
Assuming 80\% HAE candidates are true HAEs at $z=2.24$, $\delta_{g}$ in the BOSS1244 and BOSS1542 fields are $5.5\pm0.7$ and $5.2\pm0.6$, respectively.  For the protocluster scale of 15\,cMpc, $\delta_{\rm g}$ in the BOSS1244 SW, NE regions and BOSS1542 filament are $22.9\pm4.9$ and $10.9\pm2.5$,$20.5\pm3.9$, respectively. Therefore, BOSS1244 and BOSS1542 are the most overdense galaxy protoclusters ($\delta_{\rm g}>20.0$) discovered to date at $z>2$. 

 \item[(4)]  The BOSS1244 and BOSS1542 overdensities are expected to be virialized at $z\sim0$, so we can calculate their present-day total masses $M_{\rm total}$ based on the galaxy overdensity $\delta_{\rm g}$.  
BOSS1244 and BOSS1542 are expected to evolve into a cluster with halo masses of $(0.89\pm0.07)\times10^{16}$\,$M_{\sun}$ and $(0.82\pm0.06)\times10^{16}$\,$M_{\sun}$.  On the scale of 15\,cMpc, the present-day masses in BOSS1244 SW and NE density peaks are $(1.59\pm0.20)\times10^{15}$\,$M_{\sun}$ and $(0.83\pm0.11)\times10^{15}$\,$M_{\sun}$, and the expected total mass in the BOSS1542 filament is $(1.42\pm0.18)\times10^{15}$\,$M_{\sun}$. The masses without galaxy overdensities in BOSS1244 and BOSS1542 are $(3.70\pm0.18 )\times10^{15}$\,$M_{\sun}$ and $(3.48\pm0.16 )\times10^{15}$\,$M_{\sun}$ in BOSS1542.  
For the density peaks in BOSS1244 and BOSS1542, they will evolve into Coma-type galaxy clusters at $z=0$.  

\item[(5)] The dynamical masses $M_{\rm vir}$ in BOSS1244 and BOSS1542 are estimated using the line-of-sight velocity dispersion assuming that these systems are virialized. The dynamical masses for the BOSS1244 SW, NE region, and BOSS1542 filament are ($2.30\pm3.50)\times10^{13}$\,$M_{\sun}$ , $(1.90\pm1.50)\times10^{13}$\,$M_{\sun}$, and $M_{\rm vir}=(5.30\pm2.10)\times10^{12}$\,$M_{\sun}$ , respectively. The log($M_{\rm total}$/$M_{\rm vir}$) ratios are $1.64-2.43$, indicating that our protoclusters are far from virialization, especially for the BOSS1542 structures.  We caveat that it may be not very accurate to use log($M_{\rm total}$/$M_{\rm vir}$) to judge whether a system is virialized beacuse the estimated velocity dispersions include the galaxies outside peaks' volume and $M_{\rm total}$ varies by changing the overdensity threshold to define the density peaks, but it is very helpful to understand the protocluster systems quantitatively.

\item[(6)] We stress that BOSS1441, BOSS1244 and  BOSS1542 protoclusters display dramatically different morphologies, representing different stages of galaxy cluster assembly. Namely, BOSS1441 may be a near-virialized protocluster, BOSS1244 is forming one or two protoclusters, and BOSS1542 is forming a filament.  Besides, two quasar pairs in BOSS1441, one quasar pair in BOSS1244 and BOSS1542, these quasar pairs may work with CoSLAs to trace the most massive large-scale structures of universe.

\end{itemize}

Taken together, our results imply that BOSS1244 and BOSS1542 are dynamically young and pre-virialization. Using the obtained high-quality data, we will investigate the properties of galaxies in the overdense environemnts at $z=2-3$, including SFR, gas-phase metallicity, morphology and AGN fraction relative to galaxies in the general fields in forthcoming works. Moreover, much more follow-up spectroscopy are needed to further explore BOSS1244 and BOSS1542.

\acknowledgments
We thank anonymous referee for the valuable and helpful comments. We thank Mengtao Tang, Jinyi Yang, Chun Ly, Igor Chilingarian, Yongming Liang and FangXia An for their useful discussions on this work. 
This work is supported by the National Science Foundation of China (11773076, 12073078 and 12073014),  the National Key R\&D Program of China (grant No. 2017YFA0402703 and No. 2018YFA0404503), the Major Science and Technology Project of Qinghai Province (2019-ZJ-A10),  and the Chinese Academy of Sciences (CAS) through a China-Chile Joint Research Fund (CCJRF \#1809) administered by the CAS South America Centre for Astronomy (CASSACA). DDS gratefully acknowledges the supports from the China Scholarship Council (201806340059) in $2018-2020$.

\vspace{5mm}
\facilities{MMT (MMIRS),  LBT (LUCI)}

\clearpage
\appendix

\begin{figure}[ht]
\setlength{\abovecaptionskip}{0pt}
\begin{center}
\includegraphics[trim=1mm 2mm 0mm 2mm,clip,height=0.5\textwidth]{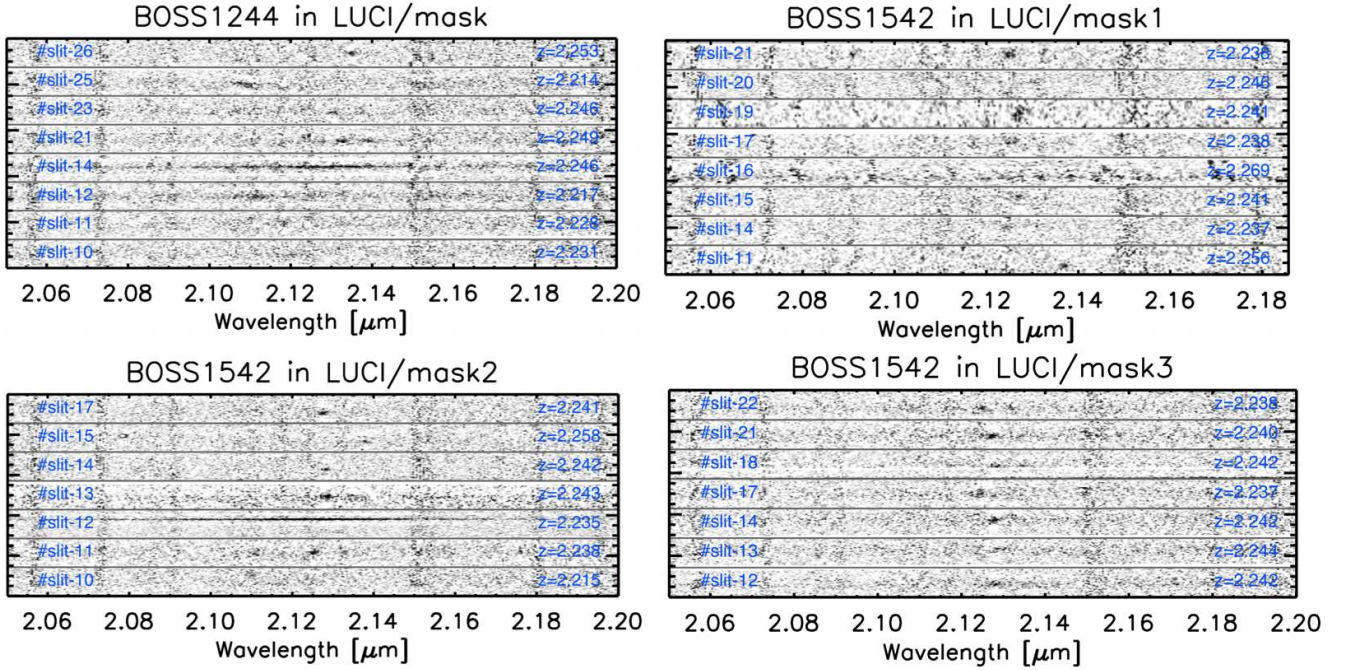}
\caption{2D spectra of HAEs in BOSS1244 and BOSS1542 observed via LBT/LUCI slit masks. Skyline residuals are visible as vertical lines with a higher noise level. The slit and spectrum ID in each mask are indicated. The upper left panel is the mask in the BOSS1244 field. The rest are the three masks in the BOSS1542 field. } 
\label{fig:fig13}
\end{center}
\end{figure}

We show all the 2D spectra of HAEs in BOSS1244 and BOSS1542 based on the LBT/ LUCI observations, as shown in Figure~\ref{fig:fig13}.

Table~\ref{tab:tabA1} and Table~\ref{tab:tabA2} list the catalog of spectroscopically-confirmed HAEs and an [O\,{\small III}] emitter in BOSS1244 and BOSS1542.  

\begin{deluxetable}{c|ccccc}
\tablenum{A1}
\tablecaption{The sky positions, redshifts and total magnitudes at $K_{\rm s}$-band for the spectroscopically-confirmed HAEs in BOSS1244.   \label{tab:tabA1}} 
\tablewidth{0pt}
\tablehead{
\colhead{SlitMask} &  \colhead{ID} & \colhead{R.A.}  & \colhead{Decl.}  & \colhead{Redshift} &  \colhead{$K_{\rm s}$}  \\
 \colhead{} & \colhead{} &  \colhead{(J2000.0)} & \colhead{(J2000.0)}  & \colhead{$z$}  & \colhead{(mag)}  
}

\startdata
    & slit-2 & 12:43:24.01 & $+$35:58:47.98  & 2.234 & 23.30$\pm$0.14\\
    & slit-3 & 12:43:24.95 & $+$35:58:38.59  & 2.235 & 22.79$\pm$0.08\\
    & slit-4 & 12:43:27.98 & $+$35:58:16.75  & 2.221 & 20.56$\pm$0.01\\
    & slit-5 & 12:43:26.88 & $+$35:58:06.74  & 2.235 & 22.45$\pm$0.08\\
    & slit-13$^{\rm a}$ & 12:43:37.03 & $+$35:56:29.58  & 2.229 & 20.39$\pm$0.01\\
    & slit-14 & 12:43:36.69 & $+$35:56:21.21  & 2.230 & 21.71$\pm$0.04\\
    & slit-15 & 12:43:38.80 & $+$35:56:11.17  & 2.226 & 21.80$\pm$0.05\\    
    & slit-17 & 12:43:29.71 & $+$35:55:50.79  & 2.247 & 23.09$\pm$0.11\\
    & slit-18 & 12:43:39.68 & $+$35:55:40.21  & 2.227 & 22.08$\pm$0.04\\    
 MMIRS-mask1   & slit-19 & 12:43:23.35 & $+$35:55:31.22  & 2.235 & 20.56$\pm$0.01\\
    & slit-20 & 12:43:38.43 & $+$35:55:21.58  & 2.225 & 21.69$\pm$0.04\\
    & slit-21 & 12:43:33.30 & $+$35:55:10.79  & 2.220 & 22.52$\pm$0.07\\
    & slit-22$^{\rm b}$ & 12:43:28.42 & $+$35:54:59.30  & 2.213 & 21.54$\pm$0.04\\    
    & slit-23$^{\rm b}$ & 12:43:32.44 & $+$35:54:46.00  & 2.250 & 21.77$\pm$0.03\\
    & slit-26$^{\rm b}$ & 12:43:27.69 & $+$35:53:57.44  & 2.253 & 22.08$\pm$0.06\\
    & slit-27 & 12:43:29.22 & $+$35:53:44.75  & 2.242 & 22.55$\pm$0.07\\
    & slit-29 & 12:43:35.15 & $+$35:53:16.44  & 2.255 & 21.74$\pm$0.04\\
    & slit-30 & 12:43:29.54 & $+$35:53:08.90  & 2.244 & 22.21$\pm$0.04\\
    & slit-31 & 12:43:33.17 & $+$35:52:50.99  & 2.245 & 22.18$\pm$0.05\\
\hline
   & slit-1 & 12:44:13.61 & $+$36:02:58.60  & 2.243 & 22.72$\pm$0.08\\
   & slit-2 & 12:44:34.29 & $+$36:06:55.98  & 2.244 & 22.42$\pm$0.09\\
   & slit-3 & 12:44:33.60 & $+$36:06:35.31  & 2.240 & 22.96$\pm$0.13\\
   & slit-4 & 12:44:35.55 & $+$36:04:20.98  & 2.239 & 22.41$\pm$0.07\\
   & slit-5 & 12:44:33.48 & $+$36:05:20.73  & 2.229 & 22.66$\pm$0.06\\
   & slit-6 & 12:44:32.00 & $+$36:05:04.60  & 2.253 & 21.95$\pm$0.05\\
   & slit-7 & 12:44:30.59 & $+$36:05:21.79  & 2.251 & 21.71$\pm$0.03\\
   & slit-8 & 12:44:28.04 & $+$36:03:52.50  & 2.245 & 21.54$\pm$0.03\\
MMIRS-mask2   & slit-9 & 12:44:25.62 & $+$36:03:59.77  & 2.247 & 21.87$\pm$0.04\\
   & slit-10 & 12:44:23.73 & $+$36:03:54.10  & 2.244 & 23.27$\pm$0.09\\  
   & slit-11 & 12:44:22.41 & $+$36:03:24.96  & 2.253 & 22.58$\pm$0.09\\
   & slit-12 & 12:44:17.83 & $+$36:05:39.97  & 2.242 & 23.06$\pm$0.10\\
   & slit-13 & 12:44:17.54 & $+$36:05:15.89  & 2.248 & 22.67$\pm$0.10\\
   & slit-16 & 12:44:15.86 & $+$36:03:54.11  & 2.244 & 22.91$\pm$0.07\\
   & slit-17 & 12:44:13.61 & $+$36:02:58.60  & 2.250 & 22.44$\pm$0.08\\
   & slit-20 & 12:44:08.17 & $+$36:04:02.92  & 2.248 & 22.27$\pm$0.09\\   
   & slit-21 & 12:44:07.75 & $+$36:03:13.92  & 2.248 & 21.14$\pm$0.04\\
   & slit-22 & 12:44:05.18 & $+$36:04:44.78  & 2.244 & 23.26$\pm$0.10\\
\hline
   & slit-10 & 12:43:43.41 & $+$35:54:49.69  & 2.231 & 22.06$\pm$0.06\\
   & slit-11 & 12:43:42.66 & $+$35:54:53.55  & 2.228 & 23.12$\pm$0.09\\
   & slit-12 & 12:43:40.61 & $+$35:54:08.32  & 2.217 & 20.50$\pm$0.02\\
LUCI-mask   & slit-14 & 12:43:39.09 & $+$35:54:03.71  & 2.246 & 20.54$\pm$0.01\\
   & slit-21$^{\rm c}$ & 12:43:32.43 & $+$35:54:46.07  & 2.249 & 21.77$\pm$0.03\\
   & slit-23 & 12:43:29.86 & $+$35:55:57.53  & 2.246 & 22.27$\pm$0.04\\
   & slit-25$^{\rm c}$ & 12:43:28.44 & $+$35:54:59.41  & 2.214 & 21.54$\pm$0.04\\
   & slit-26$^{\rm c}$ & 12:43:27.68 & $+$35:53:57.40  & 2.253 & 22.08$\pm$0.06\\   
\enddata
\tablecomments{$^{\rm a}$slit-13 in MMIRS-mask1 is the QSO5 listed in Table~\ref{tab:tab5}. $^{\rm b}$slit-22, slit-23 and slit-26 in MMIRS-mask1 are the same with $^{\rm c}$slit-25, slit-21 and slit-26 in LUCI-mask. Their measured redshifts are consistent.}
\end{deluxetable}

\begin{deluxetable}{c|ccccc}
\tablenum{A2}
\tablecaption{The sky positions, redshifts  and total magnitudes at $K_{\rm s}$-band for the spectroscopically-confirmed HAEs in BOSS1542.   \label{tab:tabA2}} 
\tablewidth{0pt}
\tablehead{
\colhead{SlitMask} &  \colhead{ID} & \colhead{R.A.}  & \colhead{Decl.}  & \colhead{Redshift} & \colhead{$K_{\rm s}$}   \\
 \colhead{} & \colhead{} &  \colhead{(J2000.0)} & \colhead{(J2000.0)}  & \colhead{$z$}  & \colhead{(mag)}
}

\startdata
   & slit-1 & 15:42:50.52 & $+$38:58:18.48  & 2.244 & 22.59$\pm$0.10\\
   & slit-4 & 15:42:36.90 & $+$38:58:17.55  & 2.239 & 22.54$\pm$0.06\\
   & slit-6$^{\rm a}$ & 15:42:34.96 & $+$38:59:09.63  & 2.243 & 22.37$\pm$0.09\\
   & slit-7 & 15:42:46.90 & $+$38:59:49.53  & 2.229 &22.33$\pm$0.07\\
   & slit-8 & 15:42:41.55 & $+$38:59:59.62  & 2.227 &23.31$\pm$0.14\\
   & slit-10 & 15:42:41.04 & $+$39:00:19.09  & 2.251 & 22.09$\pm$0.05\\   
   & slit-11 & 15:42:36.31 & $+$39:00:18.43  & 2.240 & 22.69$\pm$0.10\\
MMIRS-mask2   & slit-15 & 15:42:42.66 & $+$39:01:40.51  & 2.206 & 22.17$\pm$0.06\\
   & slit-17 & 15:42:47.98 & $+$39:02:22.66  & 2.211 & 22.05$\pm$0.06\\
   & slit-21 & 15:42:48.39 & $+$39:03:35.89  & 2.213 & 23.48$\pm$0.17\\   
   & slit-23 & 15:42:47.58 & $+$39:03:49.58  & 2.215 & 22.82$\pm$0.11\\
   & slit-24 & 15:42:39.88 & $+$39:03:39.26  & 2.253 & 23.71 $\pm$0.22\\
   & slit-25 & 15:42:40.60 & $+$39:03:51.47  & 2.216 &22.78$\pm$0.10\\
   & slit-27$^{\rm b}$ & 15:42:29.55 & $+$39:03:47.10  & 3.302 & 22.72$\pm$0.07\\
\hline
   & slit-11 & 15:42:49.44 & $+$38:51:58.53  & 2.256 & 22.10$\pm$0.06\\
   & slit-14 & 15:42:52.83 & $+$38:50:31.18  & 2.237 & 22.44$\pm$0.07\\
   & slit-15 & 15:42:51.55 & $+$38:50:29.29  & 2.241 & 21.63$\pm$0.06\\
LUCI-mask1   & slit-16 & 15:42:57.86 & $+$38:49:40.00  & 2.269 & 18.94 $\pm$0.02\\
   & slit-17 & 15:42:56.87 & $+$38:49:28.27  & 2.238 & 22.62$\pm$0.11\\
   & slit-19 & 15:42:50.72 & $+$38:49:41.08  & 2.241 & 22.27$\pm$0.08\\
   & slit-20 & 15:42:52.78 & $+$38:48:28.48  & 2.246 & 22.77$\pm$0.11\\
   & slit-21 & 15:42:45.24 & $+$38:48:30.69  & 2.238 & 22.06$\pm$0.05\\
\hline   
   & slit-10 & 15:42:45.35 & $+$38:54:52.78  & 2.215 & 23.06$\pm$0.15\\
   & slit-11 & 15:42:48.24 & $+$38:54:18.03  & 2.238 & 22.12$\pm$0.08\\
   & slit-12$^{\rm c}$ & 15:42:47.01 & $+$38:53:42.34  & 2.241 & 19.42$\pm$0.01\\
LUCI-mask2   & slit-13 & 15:42:49.56 & $+$38:53:20.48  & 2.243 & 20.77$\pm$0.07\\
   & slit-14 & 15:42:50.51 & $+$38:52:14.80  & 2.242 & 22.78$\pm$0.07\\
   & slit-15 & 15:42:49.44 & $+$38:51:58.44  & 2.258 & 22.10$\pm$0.06\\
   & slit-17 & 15:42:49.77 & $+$38:52:50.46  & 2.241 & 22.94$\pm$0.11\\
\hline
   & slit-12 & 15:42:32.86 & $+$38:59:27.41  & 2.242 & 22.12$\pm$0.05\\   
   & slit-13$^{\rm d}$ & 15:42:34.97 & $+$38:59:09.56  & 2.244 & 22.37$\pm$0.09\\
   & slit-14 & 15:42:29.92 & $+$38:58:33.08  & 2.242 & 22.88$\pm$0.12\\ 
LUCI-mask3   & slit-17 & 15:42:42.12 & $+$38:58:20.20  & 2.237 & 21.64$\pm$0.06\\ 
   & slit-18 & 15:42:40.69 & $+$38:58:03.00  & 2.242 & 20.24 $\pm$0.05 \\ 
   & slit-21 & 15:42:41.55 & $+$38:56:50.88  & 2.240 & 21.75$\pm$0.05\\ 
   & slit-22 & 15:42:43.27 & $+$38:56:31.21  & 2.238 & 21.64$\pm$0.05\\        
\enddata
\tablecomments{$^{\rm a}$slit-6 in MMIRS-mask2 is the same with $^{\rm d}$slit-13 in LUCI-mask3, and the measured redshift is consistent. $^{\rm b}$slit-27 in MMIRS-mask2 is an [O\,{\small III}] emitter. $^{\rm c}$slit-12 in LUCI-mask2 is the QSO3 listed in Table~\ref{tab:tab5}.}
\end{deluxetable}

\clearpage
\bibliography{ms.bib}{}

\begin{thebibliography}{}
\expandafter\ifx\csname natexlab\endcsname\relax\def\natexlab#1{#1}\fi
\providecommand{\url}[1]{\href{#1}{#1}}
\providecommand{\dodoi}[1]{doi:~\href{http://doi.org/#1}{\nolinkurl{#1}}}
\providecommand{\doeprint}[1]{\href{http://ascl.net/#1}{\nolinkurl{http://ascl.net/#1}}}
\providecommand{\doarXiv}[1]{\href{https://arxiv.org/abs/#1}{\nolinkurl{https://arxiv.org/abs/#1}}}

\bibitem[{{An} {et~al.}(2014){An}, {Zheng}, {Wang}, {Huang}, {Kong}, {Wang},
  {Fang}, {Zhu}, {Gu}, {Wu}, {Hao}, \& {Xia}}]{An2014}
{An}, F.~X., {Zheng}, X.~Z., {Wang}, W.-H., {et~al.} 2014, \apj, 784, 152,
  \dodoi{10.1088/0004-637X/784/2/152}

\bibitem[{{Arrigoni Battaia} {et~al.}(2015){Arrigoni Battaia}, {Hennawi},
  {Prochaska}, \& {Cantalupo}}]{Arrigoni2015}
{Arrigoni Battaia}, F., {Hennawi}, J.~F., {Prochaska}, J.~X., \& {Cantalupo},
  S. 2015, \apj, 809, 163, \dodoi{10.1088/0004-637X/809/2/163}

\bibitem[{{Bagchi} {et~al.}(2017){Bagchi}, {Sankhyayan}, {Sarkar},
  {Raychaudhury}, {Jacob}, \& {Dabhade}}]{Bagchi2017}
{Bagchi}, J., {Sankhyayan}, S., {Sarkar}, P., {et~al.} 2017, \apj, 844, 25,
  \dodoi{10.3847/1538-4357/aa7949}

\bibitem[{{Bamford} {et~al.}(2009){Bamford}, {Nichol}, {Baldry}, {Land},
  {Lintott}, {Schawinski}, {Slosar}, {Szalay}, {Thomas}, {Torki}, {Andreescu},
  {Edmondson}, {Miller}, {Murray}, {Raddick}, \& {Vandenberg}}]{Bamford2009}
{Bamford}, S.~P., {Nichol}, R.~C., {Baldry}, I.~K., {et~al.} 2009, \mnras, 393,
  1324, \dodoi{10.1111/j.1365-2966.2008.14252.x}

\bibitem[{{Beers} {et~al.}(1990){Beers}, {Flynn}, \& {Gebhardt}}]{Beers1990}
{Beers}, T.~C., {Flynn}, K., \& {Gebhardt}, K. 1990, \aj, 100, 32,
  \dodoi{10.1086/115487}

\bibitem[{{Belli} {et~al.}(2018){Belli}, {Contursi}, \& {Davies}}]{Belli2018}
{Belli}, S., {Contursi}, A., \& {Davies}, R.~I. 2018, \mnras, 478, 2097,
  \dodoi{10.1093/mnras/sty1236}

\bibitem[{{Bond} {et~al.}(1996){Bond}, {Kofman}, \& {Pogosyan}}]{Bond1996}
{Bond}, J.~R., {Kofman}, L., \& {Pogosyan}, D. 1996, \nat, 380, 603,
  \dodoi{10.1038/380603a0}

\bibitem[{{Borisova} {et~al.}(2016){Borisova}, {Cantalupo}, {Lilly}, {Marino},
  {Gallego}, {Bacon}, {Blaizot}, {Bouch{\'e}}, {Brinchmann}, {Carollo},
  {Caruana}, {Finley}, {Herenz}, {Richard}, {Schaye}, {Straka}, {Turner},
  {Urrutia}, {Verhamme}, \& {Wisotzki}}]{Borisova2016}
{Borisova}, E., {Cantalupo}, S., {Lilly}, S.~J., {et~al.} 2016, \apj, 831, 39,
  \dodoi{10.3847/0004-637X/831/1/39}

\bibitem[{{Bower} {et~al.}(1992){Bower}, {Lucey}, \& {Ellis}}]{Bower1992}
{Bower}, R.~G., {Lucey}, J.~R., \& {Ellis}, R.~S. 1992, \mnras, 254, 601,
  \dodoi{10.1093/mnras/254.4.601}

\bibitem[{{Cai} {et~al.}(2016){Cai}, {Fan}, {Peirani}, {Bian}, {Frye},
  {McGreer}, {Prochaska}, {Lau}, {Tejos}, {Ho}, \& {Schneider}}]{Cai2016}
{Cai}, Z., {Fan}, X., {Peirani}, S., {et~al.} 2016, \apj, 833, 135,
  \dodoi{10.3847/1538-4357/833/2/135}

\bibitem[{{Cai} {et~al.}(2017a){Cai}, {Fan}, {Bian}, {Zabludoff}, {Yang},
  {Prochaska}, {McGreer}, {Zheng}, {Kashikawa}, {Wang}, {Frye}, {Green}, \&
  {Jiang}}]{Cai2017a}
{Cai}, Z., {Fan}, X., {Bian}, F., {et~al.} 2017a, \apj, 839, 131,
  \dodoi{10.3847/1538-4357/aa6a1a}

\bibitem[{{Cai} {et~al.}(2017b){Cai}, {Fan}, {Yang}, {Bian}, {Prochaska},
  {Zabludoff}, {McGreer}, {Zheng}, {Green}, {Cantalupo}, {Frye}, {Hamden},
  {Jiang}, {Kashikawa}, \& {Wang}}]{Cai2017b}
{Cai}, Z., {Fan}, X., {Yang}, Y., {et~al.} 2017b, \apj, 837, 71,
  \dodoi{10.3847/1538-4357/aa5d14}

\bibitem[{{Cai} {et~al.}(2018){Cai}, {Hamden}, {Matuszewski}, {Prochaska},
  {Li}, {Cantalupo}, {Arrigoni Battaia}, {Martin}, {Neill}, {O'Sullivan},
  {Wang}, {Moore}, \& {Morrissey}}]{Cai2018}
{Cai}, Z., {Hamden}, E., {Matuszewski}, M., {et~al.} 2018, \apjl, 861, L3,
  \dodoi{10.3847/2041-8213/aacce6}

\bibitem[{{Cai} {et~al.}(2019){Cai}, {Cantalupo}, {Prochaska}, {Arrigoni
  Battaia}, {Burchett}, {Li}, {Chisholm}, {Bundy}, \& {Hennawi}}]{Cai2019}
{Cai}, Z., {Cantalupo}, S., {Prochaska}, J.~X., {et~al.} 2019, \apjs, 245, 23,
  \dodoi{10.3847/1538-4365/ab4796}

\bibitem[{{Cantalupo} {et~al.}(2014){Cantalupo}, {Arrigoni-Battaia},
  {Prochaska}, {Hennawi}, \& {Madau}}]{Cantalupo2014}
{Cantalupo}, S., {Arrigoni-Battaia}, F., {Prochaska}, J.~X., {Hennawi}, J.~F.,
  \& {Madau}, P. 2014, \nat, 506, 63, \dodoi{10.1038/nature12898}

\bibitem[{{Casasola} {et~al.}(2018){Casasola}, {Magrini}, {Combes}, {Sani},
  {Fritz}, {Rodighiero}, {Poggianti}, {Bianchi}, \& {Liuzzo}}]{Casasola2018}
{Casasola}, V., {Magrini}, L., {Combes}, F., {et~al.} 2018, \aap, 618, A128,
  \dodoi{10.1051/0004-6361/201833052}

\bibitem[{{Casey} {et~al.}(2015){Casey}, {Cooray}, {Capak}, {Fu}, {Kovac},
  {Lilly}, {Sanders}, {Scoville}, \& {Treister}}]{Casey2015}
{Casey}, C.~M., {Cooray}, A., {Capak}, P., {et~al.} 2015, \apjl, 808, L33,
  \dodoi{10.1088/2041-8205/808/2/L33}

\bibitem[{{Cen} \& {Ostriker}(2000)}]{Cen2000}
{Cen}, R., \& {Ostriker}, J.~P. 2000, \apj, 538, 83, \dodoi{10.1086/309090}

\bibitem[{{Chanchaiworawit} {et~al.}(2019){Chanchaiworawit}, {Guzm{\'a}n},
  {Salvador-Sol{\'e}}, {Rodr{\'\i}guez Espinosa}, {Calvi}, {Manrique},
  {Gallego}, {Herrero}, {Mar{\'\i}n-Franch}, \&
  {Mas-Hesse}}]{Chanchaiworawit2019}
{Chanchaiworawit}, K., {Guzm{\'a}n}, R., {Salvador-Sol{\'e}}, E., {et~al.}
  2019, \apj, 877, 51, \dodoi{10.3847/1538-4357/ab1a34}

\bibitem[{{Chiang} {et~al.}(2013){Chiang}, {Overzier}, \&
  {Gebhardt}}]{Chiang2013}
{Chiang}, Y.-K., {Overzier}, R., \& {Gebhardt}, K. 2013, \apj, 779, 127,
  \dodoi{10.1088/0004-637X/779/2/127}

\bibitem[{{Chiang} {et~al.}(2014){Chiang}, {Overzier}, \&
  {Gebhardt}}]{Chiang2014}
---. 2014, \apjl, 782, L3, \dodoi{10.1088/2041-8205/782/1/L3}

\bibitem[{{Chilingarian} {et~al.}(2015){Chilingarian}, {Beletsky}, {Moran},
  {Brown}, {McLeod}, \& {Fabricant}}]{Chilingarian2015}
{Chilingarian}, I., {Beletsky}, Y., {Moran}, S., {et~al.} 2015, \pasp, 127,
  406, \dodoi{10.1086/680598}

\bibitem[{{Codis} {et~al.}(2012){Codis}, {Pichon}, {Devriendt}, {Slyz},
  {Pogosyan}, {Dubois}, \& {Sousbie}}]{Codis2012}
{Codis}, S., {Pichon}, C., {Devriendt}, J., {et~al.} 2012, \mnras, 427, 3320,
  \dodoi{10.1111/j.1365-2966.2012.21636.x}

\bibitem[{{Collins} {et~al.}(2009){Collins}, {Stott}, {Hilton}, {Kay},
  {Stanford}, {Davidson}, {Hosmer}, {Hoyle}, {Liddle}, {Lloyd-Davies}, {Mann},
  {Mehrtens}, {Miller}, {Nichol}, {Romer}, {Sahl{\'e}n}, {Viana}, \&
  {West}}]{Collins2009}
{Collins}, C.~A., {Stott}, J.~P., {Hilton}, M., {et~al.} 2009, \nat, 458, 603,
  \dodoi{10.1038/nature07865}

\bibitem[{{Cooke} {et~al.}(2014){Cooke}, {Hatch}, {Muldrew}, {Rigby}, \&
  {Kurk}}]{Cooke2014}
{Cooke}, E.~A., {Hatch}, N.~A., {Muldrew}, S.~I., {Rigby}, E.~E., \& {Kurk},
  J.~D. 2014, \mnras, 440, 3262, \dodoi{10.1093/mnras/stu522}

\bibitem[{{Cooke} {et~al.}(2016){Cooke}, {Hatch}, {Stern}, {Rettura},
  {Brodwin}, {Galametz}, {Wylezalek}, {Bridge}, {Conselice}, {De Breuck},
  {Gonzalez}, \& {Jarvis}}]{Cooke2016}
{Cooke}, E.~A., {Hatch}, N.~A., {Stern}, D., {et~al.} 2016, \apj, 816, 83,
  \dodoi{10.3847/0004-637X/816/2/83}

\bibitem[{{Cucciati} {et~al.}(2014){Cucciati}, {Zamorani}, {Lemaux},
  {Bardelli}, {Cimatti}, {Le F{\`e}vre}, {Cassata}, {Garilli}, {Le Brun},
  {Maccagni}, {Pentericci}, {Tasca}, {Thomas}, {Vanzella}, {Zucca}, {Amorin},
  {Capak}, {Cassar{\`a}}, {Castellano}, {Cuby}, {de la Torre}, {Durkalec},
  {Fontana}, {Giavalisco}, {Grazian}, {Hathi}, {Ilbert}, {Moreau}, {Paltani},
  {Ribeiro}, {Salvato}, {Schaerer}, {Scodeggio}, {Sommariva}, {Talia},
  {Taniguchi}, {Tresse}, {Vergani}, {Wang}, {Charlot}, {Contini}, {Fotopoulou},
  {L{\'o}pez-Sanjuan}, {Mellier}, \& {Scoville}}]{Cucciati2014}
{Cucciati}, O., {Zamorani}, G., {Lemaux}, B.~C., {et~al.} 2014, \aap, 570, A16,
  \dodoi{10.1051/0004-6361/201423811}

\bibitem[{{Cucciati} {et~al.}(2018){Cucciati}, {Lemaux}, {Zamorani}, {Le
  F{\`e}vre}, {Tasca}, {Hathi}, {Lee}, {Bardelli}, {Cassata}, {Garilli}, {Le
  Brun}, {Maccagni}, {Pentericci}, {Thomas}, {Vanzella}, {Zucca}, {Lubin},
  {Amorin}, {Cassar{\`a}}, {Cimatti}, {Talia}, {Vergani}, {Koekemoer}, {Pforr},
  \& {Salvato}}]{Cucciati2018}
{Cucciati}, O., {Lemaux}, B.~C., {Zamorani}, G., {et~al.} 2018, \aap, 619, A49,
  \dodoi{10.1051/0004-6361/201833655}

\bibitem[{{Daddi} {et~al.}(2004){Daddi}, {Cimatti}, {Renzini}, {Fontana},
  {Mignoli}, {Pozzetti}, {Tozzi}, \& {Zamorani}}]{Daddi2004}
{Daddi}, E., {Cimatti}, A., {Renzini}, A., {et~al.} 2004, \apj, 617, 746,
  \dodoi{10.1086/425569}

\bibitem[{{Daddi} {et~al.}(2009){Daddi}, {Dannerbauer}, {Stern}, {Dickinson},
  {Morrison}, {Elbaz}, {Giavalisco}, {Mancini}, {Pope}, \&
  {Spinrad}}]{Daddi2009}
{Daddi}, E., {Dannerbauer}, H., {Stern}, D., {et~al.} 2009, \apj, 694, 1517,
  \dodoi{10.1088/0004-637X/694/2/1517}

\bibitem[{{Darvish} {et~al.}(2020){Darvish}, {Scoville}, {Martin}, {Sobral},
  {Mobasher}, {Rettura}, {Matthee}, {Capak}, {Chartab}, {Hemmati}, {Masters},
  {Nayyeri}, {O'Sullivan}, {Paulino-Afonso}, {Sattari}, {Shahidi}, {Salvato},
  {Lemaux}, {F{\`e}vre}, \& {Cucciati}}]{Darvish2020}
{Darvish}, B., {Scoville}, N.~Z., {Martin}, C., {et~al.} 2020, \apj, 892, 8,
  \dodoi{10.3847/1538-4357/ab75c3}

\bibitem[{{Dekel} \& {Birnboim}(2006)}]{Dekel2006}
{Dekel}, A., \& {Birnboim}, Y. 2006, \mnras, 368, 2,
  \dodoi{10.1111/j.1365-2966.2006.10145.x}

\bibitem[{{Dey} {et~al.}(2016){Dey}, {Lee}, {Reddy}, {Cooper}, {Inami}, {Hong},
  {Gonzalez}, \& {Jannuzi}}]{Dey2016}
{Dey}, A., {Lee}, K.-S., {Reddy}, N., {et~al.} 2016, \apj, 823, 11,
  \dodoi{10.3847/0004-637X/823/1/11}

\bibitem[{{Dressler}(1980)}]{Dressler1980}
{Dressler}, A. 1980, \apj, 236, 351, \dodoi{10.1086/157753}

\bibitem[{{Evrard} {et~al.}(2008){Evrard}, {Bialek}, {Busha}, {White}, {Habib},
  {Heitmann}, {Warren}, {Rasia}, {Tormen}, {Moscardini}, {Power}, {Jenkins},
  {Gao}, {Frenk}, {Springel}, {White}, \& {Diemand}}]{Evrard2008}
{Evrard}, A.~E., {Bialek}, J., {Busha}, M., {et~al.} 2008, \apj, 672, 122,
  \dodoi{10.1086/521616}

\bibitem[{{Farina} {et~al.}(2011){Farina}, {Falomo}, \& {Treves}}]{Farina2011}
{Farina}, E.~P., {Falomo}, R., \& {Treves}, A. 2011, \mnras, 415, 3163,
  \dodoi{10.1111/j.1365-2966.2011.18931.x}

\bibitem[{{Franck} \& {McGaugh}(2016)}]{Franck2016}
{Franck}, J.~R., \& {McGaugh}, S.~S. 2016, \apj, 833, 15,
  \dodoi{10.3847/0004-637X/833/1/15}

\bibitem[{{Fukugita} {et~al.}(2004){Fukugita}, {Nakamura}, {Schneider}, {Doi},
  \& {Kashikawa}}]{Fukugita2004}
{Fukugita}, M., {Nakamura}, O., {Schneider}, D.~P., {Doi}, M., \& {Kashikawa},
  N. 2004, \apjl, 603, L65, \dodoi{10.1086/383222}

\bibitem[{{Geach} {et~al.}(2012){Geach}, {}, {Hickox}, {Wake}, {Smail}, {Best},
  {Baugh}, \& {Stott}}]{Geach2012}
{Geach}, J.~E., {}, D., {Hickox}, R.~C., {et~al.} 2012, \mnras, 426, 679,
  \dodoi{10.1111/j.1365-2966.2012.21725.x}

\bibitem[{{Geach} {et~al.}(2008){Geach}, {Smail}, {Best}, {Kurk}, {Casali},
  {Ivison}, \& {Coppin}}]{Geach2008}
{Geach}, J.~E., {Smail}, I., {Best}, P.~N., {et~al.} 2008, \mnras, 388, 1473,
  \dodoi{10.1111/j.1365-2966.2008.13481.x}

\bibitem[{{Gobat} {et~al.}(2015){Gobat}, {Daddi}, {B{\'e}thermin}, {Pannella},
  {Finoguenov}, {Gozaliasl}, {Le Floc'h}, {Schreiber}, {Strazzullo}, {Sargent},
  {Wang}, {Hwang}, {Valentino}, {Cappelluti}, {Li}, \& {Hasinger}}]{Gobat2015}
{Gobat}, R., {Daddi}, E., {B{\'e}thermin}, M., {et~al.} 2015, \aap, 581, A56,
  \dodoi{10.1051/0004-6361/201526274}

\bibitem[{{Goto} {et~al.}(2003){Goto}, {Yamauchi}, {Fujita}, {Okamura},
  {Sekiguchi}, {Smail}, {Bernardi}, \& {Gomez}}]{Goto2003}
{Goto}, T., {Yamauchi}, C., {Fujita}, Y., {et~al.} 2003, \mnras, 346, 601,
  \dodoi{10.1046/j.1365-2966.2003.07114.x}

\bibitem[{{Green} {et~al.}(2011){Green}, {Myers}, {Barkhouse}, {Aldcroft},
  {Trichas}, {Richards}, {Ruiz}, \& {Hopkins}}]{Green2011}
{Green}, P.~J., {Myers}, A.~D., {Barkhouse}, W.~A., {et~al.} 2011, \apj, 743,
  81, \dodoi{10.1088/0004-637X/743/1/81}

\bibitem[{{Hatch} {et~al.}(2011){Hatch}, {De Breuck}, {Galametz}, {Miley},
  {Overzier}, {R{\"o}ttgering}, {Doherty}, {Kodama}, {Kurk}, {Seymour},
  {Venemans}, {Vernet}, \& {Zirm}}]{Hatch2011}
{Hatch}, N.~A., {De Breuck}, C., {Galametz}, A., {et~al.} 2011, \mnras, 410,
  1537, \dodoi{10.1111/j.1365-2966.2010.17538.x}

\bibitem[{{Hatch} {et~al.}(2014){Hatch}, {Wylezalek}, {Kurk}, {Stern}, {De
  Breuck}, {Jarvis}, {Galametz}, {Gonzalez}, {Hartley}, {Mortlock}, {Seymour},
  \& {Stevens}}]{Hatch2014}
{Hatch}, N.~A., {Wylezalek}, D., {Kurk}, J.~D., {et~al.} 2014, \mnras, 445,
  280, \dodoi{10.1093/mnras/stu1725}

\bibitem[{{Hayashi} {et~al.}(2012){Hayashi}, {Kodama}, {Tadaki}, {Koyama}, \&
  {Tanaka}}]{Hayashi2012}
{Hayashi}, M., {Kodama}, T., {Tadaki}, K.-i., {Koyama}, Y., \& {Tanaka}, I.
  2012, \apj, 757, 15, \dodoi{10.1088/0004-637X/757/1/15}

\bibitem[{{Hayashi} {et~al.}(2017){Hayashi}, {Kodama}, {Kohno}, {Yamaguchi},
  {Tadaki}, {Hatsukade}, {Koyama}, {Shimakawa}, {Tamura}, \&
  {Suzuki}}]{Hayashi2017}
{Hayashi}, M., {Kodama}, T., {Kohno}, K., {et~al.} 2017, \apjl, 841, L21,
  \dodoi{10.3847/2041-8213/aa71ad}

\bibitem[{{Hennawi} {et~al.}(2015){Hennawi}, {Prochaska}, {Cantalupo}, \&
  {Arrigoni-Battaia}}]{Hennawi2015}
{Hennawi}, J.~F., {Prochaska}, J.~X., {Cantalupo}, S., \& {Arrigoni-Battaia},
  F. 2015, Science, 348, 779, \dodoi{10.1126/science.aaa5397}

\bibitem[{{Hill} {et~al.}(2020){Hill}, {Chapman}, {Scott}, {Apostolovski},
  {Aravena}, {B{\'e}thermin}, {Bradford}, {Canning}, {De Breuck}, {Dong},
  {Gonzalez}, {Greve}, {Hayward}, {Hezaveh}, {Litke}, {Malkan}, {Marrone},
  {Phadke}, {Reuter}, {Rotermund}, {Spilker}, {Vieira}, \&
  {Wei{\ss}}}]{Hill2020}
{Hill}, R., {Chapman}, S., {Scott}, D., {et~al.} 2020, \mnras,
  \dodoi{10.1093/mnras/staa1275}

\bibitem[{{Hughes} {et~al.}(2013){Hughes}, {Cortese}, {Boselli}, {Gavazzi}, \&
  {Davies}}]{Hughes2013}
{Hughes}, T.~M., {Cortese}, L., {Boselli}, A., {Gavazzi}, G., \& {Davies},
  J.~I. 2013, \aap, 550, A115, \dodoi{10.1051/0004-6361/201218822}

\bibitem[{{Husband} {et~al.}(2016){Husband}, {Bremer}, {Stott}, \&
  {Murphy}}]{Husband2016}
{Husband}, K., {Bremer}, M.~N., {Stott}, J.~P., \& {Murphy}, D.~N.~A. 2016,
  \mnras, 462, 421, \dodoi{10.1093/mnras/stw1520}

\bibitem[{{Ji} {et~al.}(2018){Ji}, {Giavalisco}, {Williams}, {Faber},
  {Ferguson}, {Guo}, {Liu}, \& {Lee}}]{Ji2018}
{Ji}, Z., {Giavalisco}, M., {Williams}, C.~C., {et~al.} 2018, \apj, 862, 135,
  \dodoi{10.3847/1538-4357/aacc2c}

\bibitem[{{Jiang} {et~al.}(2018){Jiang}, {Wu}, {Bian}, {Chiang}, {Ho}, {Shen},
  {Zheng}, {Bailey}, {Blanc}, {Crane}, {Fan}, {Mateo}, {Olszewski},
  {Oyarz{\'u}n}, {Wang}, \& {Wu}}]{Jiang2018}
{Jiang}, L., {Wu}, J., {Bian}, F., {et~al.} 2018, Nature Astronomy, 2, 962,
  \dodoi{10.1038/s41550-018-0587-9}

\bibitem[{{Kartaltepe} {et~al.}(2019){Kartaltepe}, {Casey}, {Dickinson},
  {Hathi}, {Koekemoer}, {Lemaux}, {Postman}, \& {Rudnick}}]{Kartaltepe2019}
{Kartaltepe}, J., {Casey}, C., {Dickinson}, M., {et~al.} 2019, \baas, 51, 395.
\newblock \doarXiv{1903.05026}

\bibitem[{{Kauffmann} {et~al.}(1999){Kauffmann}, {Colberg}, {Diaferio}, \&
  {White}}]{Kauffmann1999}
{Kauffmann}, G., {Colberg}, J.~M., {Diaferio}, A., \& {White}, S. D.~M. 1999,
  \mnras, 303, 188, \dodoi{10.1046/j.1365-8711.1999.02202.x}

\bibitem[{{Kauffmann} \& {White}(1993)}]{Kauffmann1993}
{Kauffmann}, G., \& {White}, S.~D.~M. 1993, \mnras, 261, 921,
  \dodoi{10.1093/mnras/261.4.921}

\bibitem[{{Kelson}(2003)}]{Kelson2003}
{Kelson}, D.~D. 2003, \pasp, 115, 688, \dodoi{10.1086/375502}

\bibitem[{{Kere{\v{s}}} {et~al.}(2009){Kere{\v{s}}}, {Katz}, {Fardal},
  {Dav{\'e}}, \& {Weinberg}}]{Keres2009}
{Kere{\v{s}}}, D., {Katz}, N., {Fardal}, M., {Dav{\'e}}, R., \& {Weinberg},
  D.~H. 2009, \mnras, 395, 160, \dodoi{10.1111/j.1365-2966.2009.14541.x}

\bibitem[{{Kere{\v{s}}} {et~al.}(2005){Kere{\v{s}}}, {Katz}, {Weinberg}, \&
  {Dav{\'e}}}]{Keres2005}
{Kere{\v{s}}}, D., {Katz}, N., {Weinberg}, D.~H., \& {Dav{\'e}}, R. 2005,
  \mnras, 363, 2, \dodoi{10.1111/j.1365-2966.2005.09451.x}

\bibitem[{{Kikuta} {et~al.}(2019){Kikuta}, {Matsuda}, {Cen}, {Steidel}, {Yagi},
  {Hayashino}, {Imanishi}, {Komiyama}, {Momose}, \& {Saito}}]{Kikuta2019}
{Kikuta}, S., {Matsuda}, Y., {Cen}, R., {et~al.} 2019, \pasj, 71, L2,
  \dodoi{10.1093/pasj/psz055}

\bibitem[{{Kim} {et~al.}(2016){Kim}, {Im}, {Lee}, {Edge}, {Hyun}, {Kim},
  {Choi}, {Hong}, {Jeon}, {Jun}, {Karouzos}, {Kim}, {Kim}, {Kim}, {Park},
  {Taak}, \& {Yoon}}]{Kim2016}
{Kim}, J.-W., {Im}, M., {Lee}, S.-K., {et~al.} 2016, \apjl, 821, L10,
  \dodoi{10.3847/2041-8205/821/1/L10}

\bibitem[{{Koyama} {et~al.}(2021){Koyama}, {Polletta}, {Tanaka}, {Kodama},
  {Dole}, {Soucail}, {Frye}, {Lehnert}, \& {Scodeggio}}]{Koyama2021}
{Koyama}, Y., {Polletta}, M. d.~C., {Tanaka}, I., {et~al.} 2021, \mnras, 503,
  L1, \dodoi{10.1093/mnrasl/slab013}

\bibitem[{{Kraljic} {et~al.}(2018){Kraljic}, {Arnouts}, {Pichon}, {Laigle}, {de
  la Torre}, {Vibert}, {Cadiou}, {Dubois}, {Treyer}, {Schimd}, {Codis}, {de
  Lapparent}, {Devriendt}, {Hwang}, {Le Borgne}, {Malavasi}, {Milliard},
  {Musso}, {Pogosyan}, {Alpaslan}, {Bland-Hawthorn}, \& {Wright}}]{Kraljic2018}
{Kraljic}, K., {Arnouts}, S., {Pichon}, C., {et~al.} 2018, \mnras, 474, 547,
  \dodoi{10.1093/mnras/stx2638}

\bibitem[{{Kuchner} {et~al.}(2020){Kuchner}, {Arag{\'o}n-Salamanca}, {Pearce},
  {Gray}, {Rost}, {Mu}, {Welker}, {Cui}, {Haggar}, {Laigle}, {Knebe},
  {Kraljic}, {Sarron}, \& {Yepes}}]{Kuchner2020}
{Kuchner}, U., {Arag{\'o}n-Salamanca}, A., {Pearce}, F.~R., {et~al.} 2020,
  \mnras, \dodoi{10.1093/mnras/staa1083}

\bibitem[{{Kuiper} {et~al.}(2011){Kuiper}, {Hatch}, {Venemans}, {Miley},
  {R{\"o}ttgering}, {Kurk}, {Overzier}, {Pentericci}, {Bland-Hawthorn}, \&
  {Cepa}}]{Kuiper2011}
{Kuiper}, E., {Hatch}, N.~A., {Venemans}, B.~P., {et~al.} 2011, \mnras, 417,
  1088, \dodoi{10.1111/j.1365-2966.2011.19324.x}

\bibitem[{{Kurk} {et~al.}(2004{\natexlab{a}}){Kurk}, {Pentericci}, {Overzier},
  {R{\"o}ttgering}, \& {Miley}}]{Kurk2004a}
{Kurk}, J.~D., {Pentericci}, L., {Overzier}, R.~A., {R{\"o}ttgering}, H.~J.~A.,
  \& {Miley}, G.~K. 2004{\natexlab{a}}, \aap, 428, 817,
  \dodoi{10.1051/0004-6361:20041819}

\bibitem[{{Kurk} {et~al.}(2004{\natexlab{b}}){Kurk}, {Pentericci},
  {R{\"o}ttgering}, \& {Miley}}]{Kurk2004b}
{Kurk}, J.~D., {Pentericci}, L., {R{\"o}ttgering}, H.~J.~A., \& {Miley}, G.~K.
  2004{\natexlab{b}}, \aap, 428, 793, \dodoi{10.1051/0004-6361:20040075}

\bibitem[{{Kurk} {et~al.}(2000){Kurk}, {R{\"o}ttgering}, {Pentericci}, {Miley},
  {van Breugel}, {Carilli}, {Ford}, {Heckman}, {McCarthy}, \&
  {Moorwood}}]{Kurk2000}
{Kurk}, J.~D., {R{\"o}ttgering}, H.~J.~A., {Pentericci}, L., {et~al.} 2000,
  \aap, 358, L1.
\newblock \doarXiv{astro-ph/0005058}

\bibitem[{{Laigle} {et~al.}(2015){Laigle}, {Pichon}, {Codis}, {Dubois}, {Le
  Borgne}, {Pogosyan}, {Devriendt}, {Peirani}, {Prunet}, {Rouberol}, {Slyz}, \&
  {Sousbie}}]{Laigle2015}
{Laigle}, C., {Pichon}, C., {Codis}, S., {et~al.} 2015, \mnras, 446, 2744,
  \dodoi{10.1093/mnras/stu2289}

\bibitem[{{Le F{\`e}vre} {et~al.}(2015){Le F{\`e}vre}, {Tasca}, {Cassata},
  {Garilli}, {Le Brun}, {Maccagni}, {Pentericci}, {Thomas}, {Vanzella},
  {Zamorani}, {Zucca}, {Amorin}, {Bardelli}, {Capak}, {Cassar{\`a}},
  {Castellano}, {Cimatti}, {Cuby}, {Cucciati}, {de la Torre}, {Durkalec},
  {Fontana}, {Giavalisco}, {Grazian}, {Hathi}, {Ilbert}, {Lemaux}, {Moreau},
  {Paltani}, {Ribeiro}, {Salvato}, {Schaerer}, {Scodeggio}, {Sommariva},
  {Talia}, {Taniguchi}, {Tresse}, {Vergani}, {Wang}, {Charlot}, {Contini},
  {Fotopoulou}, {L{\'o}pez-Sanjuan}, {Mellier}, \& {Scoville}}]{Fevre2015}
{Le F{\`e}vre}, O., {Tasca}, L.~A.~M., {Cassata}, P., {et~al.} 2015, \aap, 576,
  A79, \dodoi{10.1051/0004-6361/201423829}

\bibitem[{{Lee} {et~al.}(2014{\natexlab{a}}){Lee}, {Hennawi}, {White}, {Croft},
  \& {Ozbek}}]{Lee2014}
{Lee}, K.-G., {Hennawi}, J.~F., {White}, M., {Croft}, R. A.~C., \& {Ozbek}, M.
  2014{\natexlab{a}}, \apj, 788, 49, \dodoi{10.1088/0004-637X/788/1/49}

\bibitem[{{Lee} {et~al.}(2016){Lee}, {Hennawi}, {White}, {Prochaska},
  {Font-Ribera}, {Schlegel}, {Rich}, {Suzuki}, {Stark}, {Le F{\`e}vre},
  {Nugent}, {Salvato}, \& {Zamorani}}]{Lee2016}
{Lee}, K.-G., {Hennawi}, J.~F., {White}, M., {et~al.} 2016, \apj, 817, 160,
  \dodoi{10.3847/0004-637X/817/2/160}

\bibitem[{{Lee} {et~al.}(2018){Lee}, {Krolewski}, {White}, {Schlegel},
  {Nugent}, {Hennawi}, {M{\"u}ller}, {Pan}, {Prochaska}, {Font-Ribera},
  {Suzuki}, {Glazebrook}, {Kacprzak}, {Kartaltepe}, {Koekemoer}, {Le
  F{\`e}vre}, {Lemaux}, {Maier}, {Nanayakkara}, {Rich}, {Sanders}, {Salvato},
  {Tasca}, \& {Tran}}]{Lee2018}
{Lee}, K.-G., {Krolewski}, A., {White}, M., {et~al.} 2018, \apjs, 237, 31,
  \dodoi{10.3847/1538-4365/aace58}

\bibitem[{{Lee} {et~al.}(2014{\natexlab{b}}){Lee}, {Dey}, {Hong}, {Reddy},
  {Wilson}, {Jannuzi}, {Inami}, \& {Gonzalez}}]{Lee12014}
{Lee}, K.-S., {Dey}, A., {Hong}, S., {et~al.} 2014{\natexlab{b}}, \apj, 796,
  126, \dodoi{10.1088/0004-637X/796/2/126}

\bibitem[{{Lemaux} {et~al.}(2012){Lemaux}, {Gal}, {Lubin}, {Kocevski},
  {Fassnacht}, {McGrath}, {Squires}, {Surace}, \& {Lacy}}]{Lemaux2012}
{Lemaux}, B.~C., {Gal}, R.~R., {Lubin}, L.~M., {et~al.} 2012, \apj, 745, 106,
  \dodoi{10.1088/0004-637X/745/2/106}

\bibitem[{{Lemaux} {et~al.}(2014){Lemaux}, {Cucciati}, {Tasca}, {Le F{\`e}vre},
  {Zamorani}, {Cassata}, {Garilli}, {Le Brun}, {Maccagni}, {Pentericci},
  {Thomas}, {Vanzella}, {Zucca}, {Amor{\'\i}n}, {Bardelli}, {Capak},
  {Cassar{\`a}}, {Castellano}, {Cimatti}, {Cuby}, {de la Torre}, {Durkalec},
  {Fontana}, {Giavalisco}, {Grazian}, {Hathi}, {Ilbert}, {Moreau}, {Paltani},
  {Ribeiro}, {Salvato}, {Schaerer}, {Scodeggio}, {Sommariva}, {Talia},
  {Taniguchi}, {Tresse}, {Vergani}, {Wang}, {Charlot}, {Contini}, {Fotopoulou},
  {Gal}, {Kocevski}, {L{\'o}pez-Sanjuan}, {Lubin}, {Mellier}, {Sadibekova}, \&
  {Scoville}}]{Lemaux2014}
{Lemaux}, B.~C., {Cucciati}, O., {Tasca}, L.~A.~M., {et~al.} 2014, \aap, 572,
  A41, \dodoi{10.1051/0004-6361/201423828}

\bibitem[{{Lemaux} {et~al.}(2018){Lemaux}, {Le F{\`e}vre}, {Cucciati},
  {Ribeiro}, {Tasca}, {Zamorani}, {Ilbert}, {Thomas}, {Bardelli}, {Cassata},
  {Hathi}, {Pforr}, {Smol{\v{c}}i{\'c}}, {Delvecchio}, {Novak}, {Berta},
  {McCracken}, {Koekemoer}, {Amor{\'\i}n}, {Garilli}, {Maccagni}, {Schaerer},
  \& {Zucca}}]{Lemaux2018}
{Lemaux}, B.~C., {Le F{\`e}vre}, O., {Cucciati}, O., {et~al.} 2018, \aap, 615,
  A77, \dodoi{10.1051/0004-6361/201730870}

\bibitem[{{Liang} {et~al.}(2021){Liang}, {Kashikawa}, {Cai}, {Fan},
  {Prochaska}, {Shimasaku}, {Tanaka}, {Uchiyama}, {Ito}, {Shimakawa},
  {Nagamine}, {Shimizu}, {Onoue}, \& {Toshikawa}}]{Liang2021}
{Liang}, Y., {Kashikawa}, N., {Cai}, Z., {et~al.} 2021, \apj, 907, 3,
  \dodoi{10.3847/1538-4357/abcd93}

\bibitem[{{Libeskind} {et~al.}(2018){Libeskind}, {van de Weygaert}, {Cautun},
  {Falck}, {Tempel}, {Abel}, {Alpaslan}, {Arag{\'o}n-Calvo}, {Forero-Romero},
  {Gonzalez}, {Gottl{\"o}ber}, {Hahn}, {Hellwing}, {Hoffman}, {Jones},
  {Kitaura}, {Knebe}, {Manti}, {Neyrinck}, {Nuza}, {Padilla}, {Platen},
  {Ramachandra}, {Robotham}, {Saar}, {Shand arin}, {Steinmetz}, {Stoica},
  {Sousbie}, \& {Yepes}}]{Libeskind2018}
{Libeskind}, N.~I., {van de Weygaert}, R., {Cautun}, M., {et~al.} 2018, \mnras,
  473, 1195, \dodoi{10.1093/mnras/stx1976}

\bibitem[{{Lubin} {et~al.}(2000){Lubin}, {Brunner}, {Metzger}, {Postman}, \&
  {Oke}}]{Lubin2000}
{Lubin}, L.~M., {Brunner}, R., {Metzger}, M.~R., {Postman}, M., \& {Oke}, J.~B.
  2000, \apjl, 531, L5, \dodoi{10.1086/312518}

\bibitem[{{Martin} {et~al.}(2014){Martin}, {Chang}, {Matuszewski}, {Morrissey},
  {Rahman}, {Moore}, {Steidel}, \& {Matsuda}}]{Martin2014}
{Martin}, D.~C., {Chang}, D., {Matuszewski}, M., {et~al.} 2014, \apj, 786, 107,
  \dodoi{10.1088/0004-637X/786/2/107}

\bibitem[{{Martizzi} {et~al.}(2019){Martizzi}, {Vogelsberger}, {Artale},
  {Haider}, {Torrey}, {Marinacci}, {Nelson}, {Pillepich}, {Weinberger},
  {Hernquist}, {Naiman}, \& {Springel}}]{Martizzi2019}
{Martizzi}, D., {Vogelsberger}, M., {Artale}, M.~C., {et~al.} 2019, \mnras,
  486, 3766, \dodoi{10.1093/mnras/stz1106}

\bibitem[{{McLeod} {et~al.}(2012){McLeod}, {Fabricant}, {Nystrom}, {McCracken},
  {Amato}, {Bergner}, {Brown}, {Burke}, {Chilingarian}, {Conroy}, {Curley},
  {Furesz}, {Geary}, {Hertz}, {Holwell}, {Matthews}, {Norton}, {Park}, {Roll},
  {Zajac}, {Epps}, \& {Martini}}]{McLeod2012}
{McLeod}, B., {Fabricant}, D., {Nystrom}, G., {et~al.} 2012, \pasp, 124, 1318,
  \dodoi{10.1086/669044}

\bibitem[{{Miller} {et~al.}(2019){Miller}, {Bolton}, \& {Hatch}}]{Miller2019}
{Miller}, J. S.~A., {Bolton}, J.~S., \& {Hatch}, N. 2019, \mnras, 489, 5381,
  \dodoi{10.1093/mnras/stz2504}

\bibitem[{{Miller} {et~al.}(2018){Miller}, {Chapman}, {Aravena}, {Ashby},
  {Hayward}, {Vieira}, {Wei{\ss}}, {Babul}, {B{\'e}thermin}, {Bradford},
  {Brodwin}, {Carlstrom}, {Chen}, {Cunningham}, {De Breuck}, {Gonzalez},
  {Greve}, {Harnett}, {Hezaveh}, {Lacaille}, {Litke}, {Ma}, {Malkan},
  {Marrone}, {Morningstar}, {Murphy}, {Narayanan}, {Pass}, {Perry}, {Phadke},
  {Rennehan}, {Rotermund}, {Simpson}, {Spilker}, {Sreevani}, {Stark},
  {Strandet}, \& {Strom}}]{Miller2018}
{Miller}, T.~B., {Chapman}, S.~C., {Aravena}, M., {et~al.} 2018, \nat, 556,
  469, \dodoi{10.1038/s41586-018-0025-2}

\bibitem[{{Mo} \& {White}(1996)}]{Mo1996}
{Mo}, H.~J., \& {White}, S.~D.~M. 1996, \mnras, 282, 347,
  \dodoi{10.1093/mnras/282.2.347}

\bibitem[{{Mukae} {et~al.}(2020){Mukae}, {Ouchi}, {Hill}, {Gebhardt}, {Cooper},
  {Jeong}, {Saito}, {Fabricius}, {Gawiser}, {Ciardullo}, {Farrow}, {Davis},
  {Zeimann}, {Finkelstein}, {Gronwall}, {Liu}, {Zhang}, {Byrohl}, {Ono},
  {Schneider}, {Jarvis}, {Casey}, \& {Mawatari}}]{Mukae2020}
{Mukae}, S., {Ouchi}, M., {Hill}, G.~J., {et~al.} 2020, \apj, 903, 24,
  \dodoi{10.3847/1538-4357/abb81b}

\bibitem[{{Muldrew} {et~al.}(2015){Muldrew}, {Hatch}, \& {Cooke}}]{Muldrew2015}
{Muldrew}, S.~I., {Hatch}, N.~A., \& {Cooke}, E.~A. 2015, \mnras, 452, 2528,
  \dodoi{10.1093/mnras/stv1449}

\bibitem[{{Munari} {et~al.}(2013){Munari}, {Biviano}, {Borgani}, {Murante}, \&
  {Fabjan}}]{Munari2013}
{Munari}, E., {Biviano}, A., {Borgani}, S., {Murante}, G., \& {Fabjan}, D.
  2013, \mnras, 430, 2638, \dodoi{10.1093/mnras/stt049}

\bibitem[{{Ocvirk} {et~al.}(2008){Ocvirk}, {Pichon}, \&
  {Teyssier}}]{Ocvirk2008}
{Ocvirk}, P., {Pichon}, C., \& {Teyssier}, R. 2008, \mnras, 390, 1326,
  \dodoi{10.1111/j.1365-2966.2008.13763.x}

\bibitem[{{Onoue} {et~al.}(2018){Onoue}, {Kashikawa}, {Uchiyama}, {Akiyama},
  {Harikane}, {Imanishi}, {Komiyama}, {Matsuoka}, {Nagao}, {Nishizawa},
  {Oguri}, {Ouchi}, {Tanaka}, {Toba}, \& {Toshikawa}}]{Onoue2018}
{Onoue}, M., {Kashikawa}, N., {Uchiyama}, H., {et~al.} 2018, \pasj, 70, S31,
  \dodoi{10.1093/pasj/psx092}

\bibitem[{{Oteo} {et~al.}(2018){Oteo}, {Ivison}, {Dunne}, {Manilla-Robles},
  {Maddox}, {Lewis}, {de Zotti}, {Bremer}, {Clements}, {Cooray}, {Dannerbauer},
  {Eales}, {Greenslade}, {Omont}, {Perez─Fourn{\'o}n}, {Riechers}, {Scott},
  {van der Werf}, {Weiss}, \& {Zhang}}]{Oteo2018}
{Oteo}, I., {Ivison}, R.~J., {Dunne}, L., {et~al.} 2018, \apj, 856, 72,
  \dodoi{10.3847/1538-4357/aaa1f1}

\bibitem[{{Ouchi} {et~al.}(2005){Ouchi}, {Shimasaku}, {Akiyama}, {Sekiguchi},
  {Furusawa}, {Okamura}, {Kashikawa}, {Iye}, {Kodama}, {Saito}, {Sasaki},
  {Simpson}, {Takata}, {Yamada}, {Yamanoi}, {Yoshida}, \&
  {Yoshida}}]{Ouchi2005}
{Ouchi}, M., {Shimasaku}, K., {Akiyama}, M., {et~al.} 2005, \apjl, 620, L1,
  \dodoi{10.1086/428499}

\bibitem[{{Overzier}(2016)}]{Overzier2016}
{Overzier}, R.~A. 2016, \aapr, 24, 14, \dodoi{10.1007/s00159-016-0100-3}

\bibitem[{{Pascarelle} {et~al.}(1996){Pascarelle}, {Windhorst}, {Keel}, \&
  {Odewahn}}]{Pascarelle1996}
{Pascarelle}, S.~M., {Windhorst}, R.~A., {Keel}, W.~C., \& {Odewahn}, S.~C.
  1996, \nat, 383, 45, \dodoi{10.1038/383045a0}

\bibitem[{{Peebles}(1970)}]{Peebles1970}
{Peebles}, P.~J.~E. 1970, \aj, 75, 13, \dodoi{10.1086/110933}

\bibitem[{{Peng} {et~al.}(2010){Peng}, {Lilly}, {Kova{\v{c}}}, {Bolzonella},
  {Pozzetti}, {Renzini}, {Zamorani}, {Ilbert}, {Knobel}, {Iovino}, {Maier},
  {Cucciati}, {Tasca}, {Carollo}, {Silverman}, {Kampczyk}, {de Ravel},
  {Sanders}, {Scoville}, {Contini}, {Mainieri}, {Scodeggio}, {Kneib}, {Le
  F{\`e}vre}, {Bardelli}, {Bongiorno}, {Caputi}, {Coppa}, {de la Torre},
  {Franzetti}, {Garilli}, {Lamareille}, {Le Borgne}, {Le Brun}, {Mignoli},
  {Perez Montero}, {Pello}, {Ricciardelli}, {Tanaka}, {Tresse}, {Vergani},
  {Welikala}, {Zucca}, {Oesch}, {Abbas}, {Barnes}, {Bordoloi}, {Bottini},
  {Cappi}, {Cassata}, {Cimatti}, {Fumana}, {Hasinger}, {Koekemoer},
  {Leauthaud}, {Maccagni}, {Marinoni}, {McCracken}, {Memeo}, {Meneux}, {Nair},
  {Porciani}, {Presotto}, \& {Scaramella}}]{Peng2010}
{Peng}, Y.-j., {Lilly}, S.~J., {Kova{\v{c}}}, K., {et~al.} 2010, \apj, 721,
  193, \dodoi{10.1088/0004-637X/721/1/193}

\bibitem[{{Pentericci} {et~al.}(2000){Pentericci}, {Kurk}, {R{\"o}ttgering},
  {Miley}, {van Breugel}, {Carilli}, {Ford}, {Heckman}, {McCarthy}, \&
  {Moorwood}}]{Pentericci2000}
{Pentericci}, L., {Kurk}, J.~D., {R{\"o}ttgering}, H.~J.~A., {et~al.} 2000,
  \aap, 361, L25.
\newblock \doarXiv{astro-ph/0008143}

\bibitem[{{Planck Collaboration} {et~al.}(2015){Planck Collaboration},
  {Aghanim}, {Altieri}, {Arnaud}, {Ashdown}, {Aumont}, {Baccigalupi}, {Banday},
  {Barreiro}, {Bartolo}, {Battaner}, {Beelen}, {Benabed}, {Benoit-L{\'e}vy},
  {Bernard}, {Bersanelli}, {Bethermin}, {Bielewicz}, {Bonavera}, {Bond},
  {Borrill}, {Bouchet}, {Boulanger}, {Burigana}, {Calabrese}, {Canameras},
  {Cardoso}, {Catalano}, {Chamballu}, {Chary}, {Chiang}, {Christensen},
  {Clements}, {Colombi}, {Couchot}, {Crill}, {Curto}, {Danese}, {Dassas},
  {Davies}, {Davis}, {de Bernardis}, {de Rosa}, {de Zotti}, {Delabrouille},
  {Diego}, {Dole}, {Donzelli}, {Dor{\'e}}, {Douspis}, {Ducout}, {Dupac},
  {Efstathiou}, {Elsner}, {En{\ss}lin}, {Falgarone}, {Flores-Cacho}, {Forni},
  {Frailis}, {Fraisse}, {Franceschi}, {Frejsel}, {Frye}, {Galeotta}, {Galli},
  {Ganga}, {Giard}, {Gjerl{\o}w}, {Gonz{\'a}lez-Nuevo}, {G{\'o}rski},
  {Gregorio}, {Gruppuso}, {Gu{\'e}ry}, {Hansen}, {Hanson}, {Harrison}, {Helou},
  {Hern{\'a}ndez-Monteagudo}, {Hildebrandt}, {Hivon}, {Hobson}, {Holmes},
  {Hovest}, {Huffenberger}, {Hurier}, {Jaffe}, {Jaffe}, {Keih{\"a}nen},
  {Keskitalo}, {Kisner}, {Kneissl}, {Knoche}, {Kunz}, {Kurki-Suonio},
  {Lagache}, {Lamarre}, {Lasenby}, {Lattanzi}, {Lawrence}, {Le Floc'h},
  {Leonardi}, {Levrier}, {Liguori}, {Lilje}, {Linden-V{\o}rnle},
  {L{\'o}pez-Caniego}, {Lubin}, {Mac{\'\i}as-P{\'e}rez}, {MacKenzie}, {Maffei},
  {Mandolesi}, {Maris}, {Martin}, {Martinache}, {Mart{\'\i}nez-Gonz{\'a}lez},
  {Masi}, {Matarrese}, {Mazzotta}, {Melchiorri}, {Mennella}, {Migliaccio},
  {Moneti}, {Montier}, {Morgante}, {Mortlock}, {Munshi}, {Murphy}, {Natoli},
  {Negrello}, {Nesvadba}, {Novikov}, {Novikov}, {Omont}, {Pagano}, {Pajot},
  {Pasian}, {Perdereau}, {Perotto}, {Perrotta}, {Pettorino}, {Piacentini},
  {Piat}, {Plaszczynski}, {Pointecouteau}, {Polenta}, {Popa}, {Pratt},
  {Prunet}, {Puget}, {Rachen}, {Reach}, {Reinecke}, {Remazeilles}, {Renault},
  {Ristorcelli}, {Rocha}, {Roudier}, {Rusholme}, {Sandri}, {Santos}, {Savini},
  {Scott}, {Spencer}, {Stolyarov}, {Sunyaev}, {Sutton}, {Sygnet}, {Tauber},
  {Terenzi}, {Toffolatti}, {Tomasi}, {Tristram}, {Tucci}, {Umana},
  {Valenziano}, {Valiviita}, {Valtchanov}, {Van Tent}, {Vieira}, {Vielva},
  {Wade}, {Wandelt}, {Wehus}, {Welikala}, {Zacchei}, \& {Zonca}}]{Planck2015}
{Planck Collaboration}, {Aghanim}, N., {Altieri}, B., {et~al.} 2015, \aap, 582,
  A30, \dodoi{10.1051/0004-6361/201424790}

\bibitem[{{Rosati} {et~al.}(1999){Rosati}, {Stanford}, {Eisenhardt}, {Elston},
  {Spinrad}, {Stern}, \& {Dey}}]{Rosati1999}
{Rosati}, P., {Stanford}, S.~A., {Eisenhardt}, P.~R., {et~al.} 1999, \aj, 118,
  76, \dodoi{10.1086/300934}

\bibitem[{{Sandrinelli} {et~al.}(2014){Sandrinelli}, {Falomo}, {Treves},
  {Farina}, \& {Uslenghi}}]{Sandrinelli2014}
{Sandrinelli}, A., {Falomo}, R., {Treves}, A., {Farina}, E.~P., \& {Uslenghi},
  M. 2014, \mnras, 444, 1835, \dodoi{10.1093/mnras/stu1526}

\bibitem[{{Sandrinelli} {et~al.}(2018){Sandrinelli}, {Falomo}, {Treves},
  {Scarpa}, \& {Uslenghi}}]{Sandrinelli2018}
{Sandrinelli}, A., {Falomo}, R., {Treves}, A., {Scarpa}, R., \& {Uslenghi}, M.
  2018, \mnras, 474, 4925, \dodoi{10.1093/mnras/stx2822}

\bibitem[{{Santos} {et~al.}(2014){Santos}, {Altieri}, {Tanaka}, {Valtchanov},
  {Saintonge}, {Dickinson}, {Foucaud}, {Kodama}, {Rawle}, \&
  {Tadaki}}]{Santos2014}
{Santos}, J.~S., {Altieri}, B., {Tanaka}, M., {et~al.} 2014, \mnras, 438, 2565,
  \dodoi{10.1093/mnras/stt2376}

\bibitem[{{Santos} {et~al.}(2015){Santos}, {Altieri}, {Valtchanov}, {Nastasi},
  {Bohringer}, {Cresci}, {Elbaz}, {Fassbender}, {Rosati}, {Tozzi}, \&
  {Verdugo}}]{Santos2015}
{Santos}, J.~S., {Altieri}, B., {Valtchanov}, I., {et~al.} 2015, \mnras, 447,
  L65, \dodoi{10.1093/mnrasl/slu180}

\bibitem[{{Shandarin} \& {Sunyaev}(2009)}]{Shandarin2009}
{Shandarin}, S.~F., \& {Sunyaev}, R.~A. 2009, \aap, 500, 19,
  \dodoi{10.1051/0004-6361/200912144}

\bibitem[{{Shandarin} \& {Zeldovich}(1989)}]{Shandarin1989}
{Shandarin}, S.~F., \& {Zeldovich}, Y.~B. 1989, Reviews of Modern Physics, 61,
  185, \dodoi{10.1103/RevModPhys.61.185}

\bibitem[{{Shi} {et~al.}(2020){Shi}, {Toshikawa}, {Cai}, {Lee}, \&
  {Fang}}]{Shi2020}
{Shi}, K., {Toshikawa}, J., {Cai}, Z., {Lee}, K.-S., \& {Fang}, T. 2020, \apj,
  899, 79, \dodoi{10.3847/1538-4357/aba626}

\bibitem[{{Shimakawa} {et~al.}(2014){Shimakawa}, {Kodama}, {Tadaki}, {Tanaka},
  {Hayashi}, \& {Koyama}}]{Shimakawa2014}
{Shimakawa}, R., {Kodama}, T., {Tadaki}, K.~I., {et~al.} 2014, \mnras, 441, L1,
  \dodoi{10.1093/mnrasl/slu029}

\bibitem[{{Shimakawa} {et~al.}(2018{\natexlab{a}}){Shimakawa}, {Koyama},
  {R{\"o}ttgering}, {Kodama}, {Hayashi}, {Hatch}, {Dannerbauer}, {Tanaka},
  {Tadaki}, {Suzuki}, {Fukagawa}, {Cai}, \& {Kurk}}]{Shimakawa2018a}
{Shimakawa}, R., {Koyama}, Y., {R{\"o}ttgering}, H. J.~A., {et~al.}
  2018{\natexlab{a}}, \mnras, 481, 5630, \dodoi{10.1093/mnras/sty2618}

\bibitem[{{Shimakawa} {et~al.}(2018{\natexlab{b}}){Shimakawa}, {Kodama},
  {Hayashi}, {Prochaska}, {Tanaka}, {Cai}, {Suzuki}, {Tadaki}, \&
  {Koyama}}]{Shimakawa2018b}
{Shimakawa}, R., {Kodama}, T., {Hayashi}, M., {et~al.} 2018{\natexlab{b}},
  \mnras, 473, 1977, \dodoi{10.1093/mnras/stx2494}

\bibitem[{{Shimasaku} {et~al.}(2003){Shimasaku}, {Ouchi}, {Okamura},
  {Kashikawa}, {Doi}, {Furusawa}, {Hamabe}, {Hayashino}, {Kawabata}, {Kimura},
  {Kodaira}, {Komiyama}, {Matsuda}, {Miyazaki}, {Miyazaki}, {Nakata}, {Ohta},
  {Ohyama}, {Sekiguchi}, {Shioya}, {Tamura}, {Taniguchi}, {Yagi}, {Yamada}, \&
  {Yasuda}}]{Shimasaku2003}
{Shimasaku}, K., {Ouchi}, M., {Okamura}, S., {et~al.} 2003, \apjl, 586, L111,
  \dodoi{10.1086/374880}

\bibitem[{{Skibba} {et~al.}(2009){Skibba}, {Bamford}, {Nichol}, {Lintott},
  {Andreescu}, {Edmondson}, {Murray}, {Raddick}, {Schawinski}, {Slosar},
  {Szalay}, {Thomas}, \& {Vandenberg}}]{Skibba2009}
{Skibba}, R.~A., {Bamford}, S.~P., {Nichol}, R.~C., {et~al.} 2009, \mnras, 399,
  966, \dodoi{10.1111/j.1365-2966.2009.15334.x}

\bibitem[{{Sobral} {et~al.}(2012){Sobral}, {Best}, {Matsuda}, {Smail}, {Geach},
  \& {Cirasuolo}}]{Sobral2012}
{Sobral}, D., {Best}, P.~N., {Matsuda}, Y., {et~al.} 2012, \mnras, 420, 1926,
  \dodoi{10.1111/j.1365-2966.2011.19977.x}

\bibitem[{{Sobral} {et~al.}(2013){Sobral}, {Smail}, {Best}, {Geach}, {Matsuda},
  {Stott}, {Cirasuolo}, \& {Kurk}}]{Sobral2013}
{Sobral}, D., {Smail}, I., {Best}, P.~N., {et~al.} 2013, \mnras, 428, 1128,
  \dodoi{10.1093/mnras/sts096}

\bibitem[{{Springel} {et~al.}(2005){Springel}, {White}, {Jenkins}, {Frenk},
  {Yoshida}, {Gao}, {Navarro}, {Thacker}, {Croton}, {Helly}, {Peacock}, {Cole},
  {Thomas}, {Couchman}, {Evrard}, {Colberg}, \& {Pearce}}]{Springel2005}
{Springel}, V., {White}, S. D.~M., {Jenkins}, A., {et~al.} 2005, \nat, 435,
  629, \dodoi{10.1038/nature03597}

\bibitem[{{Stark} {et~al.}(2015){Stark}, {White}, {Lee}, \&
  {Hennawi}}]{Stark2015}
{Stark}, C.~W., {White}, M., {Lee}, K.-G., \& {Hennawi}, J.~F. 2015, \mnras,
  453, 311, \dodoi{10.1093/mnras/stv1620}

\bibitem[{{Steidel} {et~al.}(1998){Steidel}, {Adelberger}, {Dickinson},
  {Giavalisco}, {Pettini}, \& {Kellogg}}]{Steidel1998}
{Steidel}, C.~C., {Adelberger}, K.~L., {Dickinson}, M., {et~al.} 1998, \apj,
  492, 428, \dodoi{10.1086/305073}

\bibitem[{{Steidel} {et~al.}(2005){Steidel}, {Adelberger}, {Shapley}, {Erb},
  {Reddy}, \& {Pettini}}]{Steidel2005}
{Steidel}, C.~C., {Adelberger}, K.~L., {Shapley}, A.~E., {et~al.} 2005, \apj,
  626, 44, \dodoi{10.1086/429989}

\bibitem[{{Steidel} {et~al.}(2000){Steidel}, {Adelberger}, {Shapley},
  {Pettini}, {Dickinson}, \& {Giavalisco}}]{Steidel2000}
---. 2000, \apj, 532, 170, \dodoi{10.1086/308568}

\bibitem[{{Steidel} {et~al.}(2003){Steidel}, {Adelberger}, {Shapley},
  {Pettini}, {Dickinson}, \& {Giavalisco}}]{Steidel2003}
---. 2003, \apj, 592, 728, \dodoi{10.1086/375772}

\bibitem[{{Sunyaev} \& {Zeldovich}(1972)}]{Sunyaev1972}
{Sunyaev}, R.~A., \& {Zeldovich}, Y.~B. 1972, \aap, 20, 189

\bibitem[{{Swinbank} {et~al.}(2007){Swinbank}, {Edge}, {Smail}, {Stott},
  {Bremer}, {Sato}, {Van Breukelen}, {Jarvis}, {Waddington}, {Clewley},
  {Bergeron}, {Cotter}, {Dye}, {Geach}, {Gonzalez-Solares}, {Hirst}, {Ivison},
  {Rawlings}, {Simpson}, {Smith}, {Verma}, \& {Yamada}}]{Swinbank2007}
{Swinbank}, A.~M., {Edge}, A.~C., {Smail}, I., {et~al.} 2007, \mnras, 379,
  1343, \dodoi{10.1111/j.1365-2966.2007.12037.x}

\bibitem[{{Tadaki} {et~al.}(2011){Tadaki}, {Kodama}, {Koyama}, {Hayashi},
  {Tanaka}, \& {Tokoku}}]{Tadaki2011}
{Tadaki}, K.-I., {Kodama}, T., {Koyama}, Y., {et~al.} 2011, \pasj, 63, 437,
  \dodoi{10.1093/pasj/63.sp2.S437}

\bibitem[{{Tanaka} {et~al.}(2011){Tanaka}, {De Breuck}, {Kurk}, {Taniguchi},
  {Kodama}, {Matsuda}, {Packham}, {Zirm}, {Kajisawa}, {Ichikawa}, {Seymour},
  {Stern}, {Stockton}, {Venemans}, \& {Vernet}}]{Tanaka2011}
{Tanaka}, I., {De Breuck}, C., {Kurk}, J.~D., {et~al.} 2011, \pasj, 63, 415,
  \dodoi{10.1093/pasj/63.sp2.S415}

\bibitem[{{Tempel} {et~al.}(2014){Tempel}, {Stoica}, {Mart{\'\i}nez},
  {Liivam{\"a}gi}, {Castellan}, \& {Saar}}]{Tempel2014}
{Tempel}, E., {Stoica}, R.~S., {Mart{\'\i}nez}, V.~J., {et~al.} 2014, \mnras,
  438, 3465, \dodoi{10.1093/mnras/stt2454}

\bibitem[{{Topping} {et~al.}(2016){Topping}, {Shapley}, \&
  {Steidel}}]{Topping2016}
{Topping}, M.~W., {Shapley}, A.~E., \& {Steidel}, C.~C. 2016, \apjl, 824, L11,
  \dodoi{10.3847/2041-8205/824/1/L11}

\bibitem[{{Topping} {et~al.}(2018){Topping}, {Shapley}, {Steidel}, {Naoz}, \&
  {Primack}}]{Topping2018}
{Topping}, M.~W., {Shapley}, A.~E., {Steidel}, C.~C., {Naoz}, S., \& {Primack},
  J.~R. 2018, \apj, 852, 134, \dodoi{10.3847/1538-4357/aa9f0f}

\bibitem[{{Toshikawa} {et~al.}(2020){Toshikawa}, {Malkan}, {Kashikawa},
  {Overzier}, {Uchiyama}, {Ota}, {Ishikawa}, \& {Ito}}]{Toshikawa2020}
{Toshikawa}, J., {Malkan}, M.~A., {Kashikawa}, N., {et~al.} 2020, \apj, 888,
  89, \dodoi{10.3847/1538-4357/ab5e85}

\bibitem[{{Toshikawa} {et~al.}(2012){Toshikawa}, {Kashikawa}, {Ota},
  {Morokuma}, {Shibuya}, {Hayashi}, {Nagao}, {Jiang}, {Malkan}, {Egami},
  {Shimasaku}, {Motohara}, \& {Ishizaki}}]{Toshikawa2012}
{Toshikawa}, J., {Kashikawa}, N., {Ota}, K., {et~al.} 2012, \apj, 750, 137,
  \dodoi{10.1088/0004-637X/750/2/137}

\bibitem[{{Toshikawa} {et~al.}(2014){Toshikawa}, {Kashikawa}, {Overzier},
  {Shibuya}, {Ishikawa}, {Ota}, {Shimasaku}, {Tanaka}, {Hayashi}, {Niino}, \&
  {Onoue}}]{Toshikawa2014}
{Toshikawa}, J., {Kashikawa}, N., {Overzier}, R., {et~al.} 2014, \apj, 792, 15,
  \dodoi{10.1088/0004-637X/792/1/15}

\bibitem[{{Toshikawa} {et~al.}(2016){Toshikawa}, {Kashikawa}, {Overzier},
  {Malkan}, {Furusawa}, {Ishikawa}, {Onoue}, {Ota}, {Tanaka}, {Niino}, \&
  {Uchiyama}}]{Toshikawa2016}
---. 2016, \apj, 826, 114, \dodoi{10.3847/0004-637X/826/2/114}

\bibitem[{{Umehata} {et~al.}(2019){Umehata}, {Fumagalli}, {Smail}, {Matsuda},
  {Swinbank}, {Cantalupo}, {Sykes}, {Ivison}, {Steidel}, {Shapley}, {Vernet},
  {Yamada}, {Tamura}, {Kubo}, {Nakanishi}, {Kajisawa}, {Hatsukade}, \&
  {Kohno}}]{Umehata2019}
{Umehata}, H., {Fumagalli}, M., {Smail}, I., {et~al.} 2019, Science, 366, 97,
  \dodoi{10.1126/science.aaw5949}

\bibitem[{{van Albada}(1961)}]{vanAlbada1961}
{van Albada}, G.~B. 1961, \aj, 66, 590, \dodoi{10.1086/108469}

\bibitem[{{van de Voort} {et~al.}(2011){van de Voort}, {Schaye}, {Booth},
  {Haas}, \& {Dalla Vecchia}}]{vandeVoort2011}
{van de Voort}, F., {Schaye}, J., {Booth}, C.~M., {Haas}, M.~R., \& {Dalla
  Vecchia}, C. 2011, \mnras, 414, 2458,
  \dodoi{10.1111/j.1365-2966.2011.18565.x}

\bibitem[{{van de Weygaert} \& {Bond}(2008)}]{vandeWeygaer2008}
{van de Weygaert}, R., \& {Bond}, J.~R. 2008, {Clusters and the Theory of the
  Cosmic Web}, ed. M.~{Plionis}, O.~{L{\'o}pez-Cruz}, \& D.~{Hughes}, Vol. 740,
  335, \dodoi{10.1007/978-1-4020-6941-3_10}

\bibitem[{{Venemans} {et~al.}(2002){Venemans}, {Kurk}, {Miley},
  {R{\"o}ttgering}, {van Breugel}, {Carilli}, {De Breuck}, {Ford}, {Heckman},
  {McCarthy}, \& {Pentericci}}]{Venemans2002}
{Venemans}, B.~P., {Kurk}, J.~D., {Miley}, G.~K., {et~al.} 2002, \apjl, 569,
  L11, \dodoi{10.1086/340563}

\bibitem[{{Venemans} {et~al.}(2004){Venemans}, {R{\"o}ttgering}, {Overzier},
  {Miley}, {De Breuck}, {Kurk}, {van Breugel}, {Carilli}, {Ford}, {Heckman},
  {McCarthy}, \& {Pentericci}}]{Venemans2004}
{Venemans}, B.~P., {R{\"o}ttgering}, H.~J.~A., {Overzier}, R.~A., {et~al.}
  2004, \aap, 424, L17, \dodoi{10.1051/0004-6361:200400041}

\bibitem[{{Venemans} {et~al.}(2005){Venemans}, {R{\"o}ttgering}, {Miley},
  {Kurk}, {De Breuck}, {Overzier}, {van Breugel}, {Carilli}, {Ford}, {Heckman},
  {Pentericci}, \& {McCarthy}}]{Venemans2005}
{Venemans}, B.~P., {R{\"o}ttgering}, H.~J.~A., {Miley}, G.~K., {et~al.} 2005,
  \aap, 431, 793, \dodoi{10.1051/0004-6361:20042038}

\bibitem[{{Venemans} {et~al.}(2007){Venemans}, {R{\"o}ttgering}, {Miley}, {van
  Breugel}, {de Breuck}, {Kurk}, {Pentericci}, {Stanford}, {Overzier}, {Croft},
  \& {Ford}}]{Venemans2007}
---. 2007, \aap, 461, 823, \dodoi{10.1051/0004-6361:20053941}

\bibitem[{{Visvanathan} \& {Sandage}(1977)}]{Visvanathan1977}
{Visvanathan}, N., \& {Sandage}, A. 1977, \apj, 216, 214,
  \dodoi{10.1086/155464}

\bibitem[{{Wang} {et~al.}(2016){Wang}, {Elbaz}, {Daddi}, {Finoguenov}, {Liu},
  {Schreiber}, {Mart{\'\i}n}, {Strazzullo}, {Valentino}, {van der Burg},
  {Zanella}, {Ciesla}, {Gobat}, {Le Brun}, {Pannella}, {Sargent}, {Shu}, {Tan},
  {Cappelluti}, \& {Li}}]{Wang2016}
{Wang}, T., {Elbaz}, D., {Daddi}, E., {et~al.} 2016, \apj, 828, 56,
  \dodoi{10.3847/0004-637X/828/1/56}

\bibitem[{{Whitaker} {et~al.}(2011){Whitaker}, {Labb{\'e}}, {van Dokkum},
  {Brammer}, {Kriek}, {Marchesini}, {Quadri}, {Franx}, {Muzzin}, {Williams},
  {Bezanson}, {Illingworth}, {Lee}, {Lundgren}, {Nelson}, {Rudnick}, {Tal}, \&
  {Wake}}]{Whitaker2011}
{Whitaker}, K.~E., {Labb{\'e}}, I., {van Dokkum}, P.~G., {et~al.} 2011, \apj,
  735, 86, \dodoi{10.1088/0004-637X/735/2/86}

\bibitem[{{Williams} {et~al.}(2009){Williams}, {Quadri}, {Franx}, {van Dokkum},
  \& {Labb{\'e}}}]{Williams2009}
{Williams}, R.~J., {Quadri}, R.~F., {Franx}, M., {van Dokkum}, P., \&
  {Labb{\'e}}, I. 2009, \apj, 691, 1879, \dodoi{10.1088/0004-637X/691/2/1879}

\bibitem[{{Zheng} {et~al.}(2021){Zheng}, {Cai}, {An}, {Fan}, \&
  {Shi}}]{Zheng2020}
{Zheng}, X.~Z., {Cai}, Z., {An}, F.~X., {Fan}, X., \& {Shi}, D.~D. 2021,
  \mnras, 500, 4354, \dodoi{10.1093/mnras/staa2882}

\end{thebibliography}
\bibliographystyle{aasjournal}

\end{document}